\documentclass[aps,prl,twocolumn,superscriptaddress,floatfix,letterpaper]{revtex4-1}

\usepackage{floatrow,subfig,amssymb}
\usepackage{mathtools}

\allowdisplaybreaks[2]  
\usepackage{pifont,dsfont}
\usepackage{bm} 
\usepackage{graphicx,tikz}
\usetikzlibrary{arrows,shapes.arrows,decorations.markings}
     \tikzset{>=triangle 90}
     \tikzstyle{bbc}=[draw,circle,fill=black,scale=.75]
     \tikzstyle{rc}=[circle,fill=red,scale=.6]
     \tikzstyle{wc}=[draw,circle,scale=.75]

\usepackage[colorlinks]{hyperref}
\hypersetup{
    colorlinks = true,
    linkcolor=blue,
    citecolor=magenta,
    linktocpage}

\def\bar{\overline}

\def\ket#1{{|{#1}\rangle}}

\def\^{\wedge}

\def\dim{{\rm dim}}

\def\Tr{{\rm Tr}}

\def\cC{{\mathcal C}}

\def\cM{{\mathcal M}}

\def\cO{{\mathcal O}}

\def\cT{{\mathcal T}}

\def\Mm{\cM_{M}[\cT]}

\def\Csc{\mathscr{C}}

\def\beq{\begin{equation}}
\def\eeq{\end{equation}}

\newcommand{\bpmat}{\begin{pmatrix}}
\newcommand{\epmat}{\end{pmatrix}}
\newcommand{\bsmat}{\begin{smallmatrix}}
\newcommand{\esmat}{\end{smallmatrix}}
\newfloatcommand{capbtabbox}{table}[][\FBwidth]

\def\bar{\overline}

\def\ket#1{{|{#1}\rangle}}

\def\^{\wedge}

\def\dim{{\rm dim}}

\def\Tr{{\rm Tr}}

\def\Mm{\cM_{M}[\cT]}

\def\cC{{\mathcal C}}

\def\cM{{\mathcal M}}

\def\cO{{\mathcal O}}

\def\cT{{\mathcal T}}

\def\Mm

\def\Csc{\mathscr{C}}

\def\beq{\begin{equation}}
\def\eeq{\end{equation}}

\usepackage{float}

\usepackage{braket}

\def\bM{\begin{matrix}}
\def\eM{\end{matrix}}
\def\bar{\overline}

\def\^{\wedge}

\def\bra#1{{\langle{#1}|}} 
\def\ket#1{{|{#1}\rangle}}

\def\Tr{{\rm Tr}}

\def\cC{{\mathcal C}}

\def\cM{{\mathcal M}}

\def\cO{{\mathcal O}}

\def\cT{{\mathcal T}}

\def\CH{{\mathcal{H}}}
\def\CO{{\mathcal{O}}}
\def\CB{{\mathcal{B}}}

\def\qd{{\mathsf{d}}}

\usepackage{tikzit}

\tikzstyle{junction}=[fill=black, draw=black, shape=circle, inner sep=.8pt]
\tikzstyle{boundarylines}=[draw=blue, thick]

\tikzstyle{boundaryrectangle}=[-, draw=none, fill={rgb,255: red,191; green,191; blue,191}]
\tikzstyle{boundarylines}=[-, draw=blue, thick]
\tikzstyle{bulklines}=[-, draw=red, thick]
\tikzstyle{dd}=[-, dashed]
\tikzstyle{line}=[-, thick]

\newcommand{\CC}{\mathcal{C}}

\newcommand{\bea}{\begin{eqnarray}}
\newcommand{\eea}{\end{eqnarray}}

\def\Tr{{\mathrm{Tr}}}

\usepackage{comment}
\includecomment{HaveLagrangian}
\includecomment{CiteFusion}

\usepackage{enumitem}
    
\begin{document}

\begin{titlepage}
    
\title{Noninvertible Symmetry-Resolved Affleck-Ludwig-Cardy Formula \\and Entanglement Entropy from the Boundary Tube Algebra}

\author{Yichul Choi}
\affiliation{School of Natural Sciences, Institute for Advanced Study, Princeton, NJ}
\affiliation{C.\ N.\ Yang Institute for Theoretical Physics, Stony Brook University, NY}
\affiliation{Simons Center for Geometry and Physics, Stony Brook University, NY}

\author{Brandon C. Rayhaun}
\affiliation{C.\ N.\ Yang Institute for Theoretical Physics, Stony Brook University, NY}

\author{Yunqin Zheng}
\affiliation{C.\ N.\ Yang Institute for Theoretical Physics, Stony Brook University, NY}

\begin{abstract}
We derive a refined version of the Affleck-Ludwig-Cardy formula for a 1+1d conformal field theory, which controls the asymptotic density of high energy states on an interval transforming under a given representation of a noninvertible global symmetry. 
We use this to determine the universal leading and sub-leading contributions to the noninvertible symmetry-resolved entanglement entropy of a single interval.
As a concrete example, we show that the ground state entanglement Hamiltonian for a single interval in the critical double Ising model enjoys a Kac-Paljutkin $H_8$ Hopf algebra symmetry when the boundary conditions at the entangling points are chosen to preserve the product of two Kramers-Wannier symmetries, and we present the corresponding symmetry-resolved entanglement entropies. Our analysis utilizes recent developments in symmetry topological field theories (SymTFTs).
\end{abstract}

\maketitle 

\end{titlepage}

\section{Introduction} 
It is a classic result of Cardy \cite{Cardy:1986ie} that the high-temperature (high-$T$) limit of the torus partition function of a 1+1d conformal field theory (CFT) is controlled purely by the central charge $c$ \footnote{We assume that the gravitational anomaly vanishes throughout, so that $c = \bar{c}$.}, 
\begin{equation} \label{eq:Z}
    Z_{T^2}(q) = \Tr_{\mathcal{H}_{S^1}} (q^{L_0-\frac{c}{24}} \bar{q}^{\bar{L}_0 - \frac{c}{24}}) \sim  e^{\frac{c \pi }{6} \frac{\ell}{\beta}} \,, \quad \beta \ll \ell \,. 
\end{equation}
Here, $q= e^{ - 2\pi \beta/\ell}$, where $\beta=1/T$ is the inverse temperature and $\ell$ is the circumference of the spatial circle.
From this universal behavior of the partition function, one can derive the asymptotic density of high energy states of the CFT on a circle, a result which is commonly referred to as the Cardy formula. By a slight abuse of terminology, we will also refer to \eqref{eq:Z} as the Cardy formula.

There is a parallel thread of developments in the context of \emph{boundary conformal field theory} (BCFT).
Indeed, the high-$T$ limit of the annulus partition function of a 1+1d CFT displays a similar universal behavior \cite{Affleck:1991tk, Cardy:2004hm, PhysRevLett.67.161,Friedan:2003yc,Casini:2016fgb}, 
\begin{equation}\label{eq:ZBB}
    Z_{B_1 B_2}(q) = \Tr_{\mathcal{H}_{B_1 B_2}} q^{L_0-\frac{c}{24}} \sim g_1 g_2 e^{\frac{c \pi }{6} \frac{\ell}{\beta}} \,, \quad  \beta \ll \ell \,,
\end{equation}
where $q= e^{ - \pi \beta/\ell}$ with $\ell$  the length of the spatial interval. Here, $\mathcal{H}_{B_1 B_2}$ is the Hilbert space of the CFT on an interval with the simple conformal boundary conditions $B_1$ and $B_2$ imposed at the two endpoints \footnote{A conformal boundary condition $B$ is simple if the CFT Hamiltonian on the interval Hilbert space $\mathcal{H}_{BB}$  has a unique ground state. More generally, an extended object is called simple if the only topological point operator supported on it is the identity operator. Various extended objects and boundary conditions in this Letter are always assumed to be simple.} , and $g_i=\braket{B_i|0}$ is the $g$-function associated with $B_i$. 
We will refer to \eqref{eq:ZBB} as the  \emph{Affleck-Ludwig-Cardy (ALC) formula}. 

The universal high-$T$ behavior of the annulus partition function \eqref{eq:ZBB} also determines the leading and sub-leading contributions to the ground state entanglement entropy of a CFT for a single interval \cite{Lauchli:2013jga,Ohmori:2014eia,Cardy:2016fqc} (see Supplemental Material for a review),
\begin{align} \label{eq:EEformula}
\begin{split}
    S_{\mathrm{EE}}&=\lim_{n\to 1}\frac{1}{1-n} \mathrm{log} \frac{Z_{B_1 B_2}(q^n)}{Z_{B_1 B_2}(q)^n}\\
    &\sim \frac{c}{3} \log \frac{L}{\varepsilon} + \log g_1  + \log g_2 \,,
\end{split}
\end{align}
where $L$ is the length of the entanglement subregion, while $\varepsilon$ is the size of entangling surface. These two quantities are related via a conformal mapping to $\ell$ and $\beta$ as $L/\varepsilon= e^{\pi \ell/ \beta}$ \cite{Ohmori:2014eia}, and the high-$T$ limit corresponds to $\varepsilon \ll L$.
Hence, $\log g_i$ is also called a \emph{boundary entropy}.

In recent years, it has been realized that 1+1d CFTs often enjoy a zoo of global symmetries which are beyond groups, and are often called noninvertible symmetries or fusion category symmetries \cite{Bhardwaj:2017xup,Chang:2018iay,Thorngren:2019iar, Thorngren:2021yso}. Among their many applications, global symmetries are useful for organizing operators and states into different multiplets, labeled by irreducible representations of the symmetry algebra. 
Hence, one can further refine the Cardy formula \eqref{eq:Z}, the ALC formula \eqref{eq:ZBB}, and the entanglement entropy formula \eqref{eq:EEformula} by computing  \emph{symmetry-resolved partition functions} which, for our purposes, involves restricting the trace in Equation \eqref{eq:ZBB} so that it receives contributions only from states within a given representation $\rho$.
Symmetry-resolved entanglement entropies have been measured in experiments \cite{islam2015measuring,Lukin:2019tkq}, which has led to a lot of theoretical interest in the subject.

In the case of the torus, this has been achieved in \cite{Pal:2020wwd} for ordinary symmetries and in \cite{Lin:2022dhv,Lu:2022ver} for finite noninvertible symmetries.
In the case of the annulus, the goal is to compute
\begin{equation}\label{eq:SRInterval}
    Z_{B_1 B_2}^{\rho}(q) = \Tr_{\mathcal{H}^\rho_{B_1 B_2}} q^{L_0-\frac{c}{24}}= \Tr_{\mathcal{H}_{B_1 B_2}} P_{B_1 B_2}^{\rho} q^{L_0-\frac{c}{24}} \,,
\end{equation}
where $\mathcal{H}^\rho_{B_1 B_2} \subset \mathcal{H}_{B_1 B_2}$ is the subspace transforming under the fixed representation $\rho$, and $P_{B_1 B_2}^\rho : \mathcal{H}_{B_1 B_2} \rightarrow \mathcal{H}^{\rho}_{B_1 B_2}$ is the operator which projects the full interval Hilbert space $\mathcal{H}_{B_1 B_2}$ onto this $\rho$-sector.

The purpose of this Letter is to derive the universal high-$T$ behavior of the annulus partition function \eqref{eq:SRInterval} in theories with a noninvertible global symmetry.
For  ordinary invertible symmetries, this question has been recently addressed in \cite{Casini:2019kex,Magan:2021myk,Kusuki:2023bsp},
where it was demonstrated that the high-$T$ limit of the symmetry-resolved annulus partition function receives an additional multiplicative contribution ${\qd_\rho^2}/{|G|}$ on top of \eqref{eq:ZBB}.
See also \cite{Harlow:2021trr, Cao:2021euf, Melia:2020pzd, Kang:2022orq, Mukhametzhanov:2020swe, Benjamin:2023qsc,Bianchi:2024aim, Karch:2024udk,DiGiulio:2022jjd,Northe:2023khz,Banerjee:2024ldl} for related discussions. Here, $\qd_\rho$ is the dimension of the representation $\rho$ and $|G|$ is the volume of the group $G$. 
An analogous result holds for torus partition functions \cite{Pal:2020wwd}.

Below, we show that, in the case of a noninvertible symmetry described by a fusion category $\mathcal{C}$, there is a qualitative difference between the torus and the annulus cases, due to the rich interplay between boundary conditions and noninvertible topological line defects \cite{Choi:2023xjw}.
To systematically address this, one needs to answer the following questions:
\begin{enumerate}[label=(\Alph*)]
    \item What is the algebra of symmetry operators acting on the interval Hilbert space? 
    \item What are the irreducible representations $\rho$ of this algebra? 
    \item How can one construct the projection operators $P_{B_1 B_2}^\rho$? 
\end{enumerate}
For the $S^1$ Hilbert space, analogous questions have been discussed \cite{Lin:2022dhv}: 
(A) The algebra is the \emph{tube algebra} $\mathrm{Tube}(\CC)$ \cite{ocneanu1994chirality}. (B) Its irreducible representations \cite{evans1995ocneanu,Izumi:2000qa,MUGER2003159} are in one-to-one correspondence with simple anyons (i.e.\ topological lines) of the 2+1d Turaev-Viro topological quantum field theory (TQFT) \cite{Turaev:1992hq,Barrett:1993ab} based on the fusion category $\CC$ \footnote{In fact, the representation category of $\mathrm{Tube}(\CC)$ naturally admits the structure of a braided tensor category which is equivalent to that on the Drinfeld center $Z(\CC)$ \cite{Liu:2023lgl}. Note that $Z(\CC)$ is the modular tensor category underlying the 2+1d Turaev-Viro topological field theory based on $\CC$.} (a.k.a.\  the SymTFT for the symmetry $\CC$ \cite{Gaiotto:2014kfa,Gaiotto:2020iye,Ji:2019jhk,Apruzzi:2021nmk,Freed:2022qnc, Kong:2015flk, Kong:2017hcw, Kong:2020cie, Kaidi:2022cpf, Antinucci:2022vyk}). (C) The projectors admit a closed-form expression in terms of \emph{generalized} half-linking numbers (reviewed in Supplemental Material) associated with $\CC$ \cite{Lin:2022dhv,Choi:2024tri}.

When the theory is quantized on an interval, the answers to the above questions are more involved, as we address in this Letter. 
Nevertheless, the SymTFT again turns out to be a  useful tool for studying the representations of the algebra of symmetry operators acting on the interval Hilbert space, and we leverage the understanding it affords to determine the universal behavior of \eqref{eq:SRInterval} in the high-$T$ limit, as well as the resulting leading and sub-leading contributions to the noninvertible \emph{symmetry-resolved entanglement entropy} (SREE) of a single interval. 

In a companion paper \cite{Choi:2024tri}, we  systematically  explore the representation theory of noninvertible symmetries in the presence of boundaries and interfaces, develop the formalism used in this work in more depth, and highlight different applications.

We note that this Letter is not the first work attempting to find a noninvertible symmetry-resolved ALC formula \footnote{We thank Javier Mag\'{a}n for bringing \cite{Benedetti:2024dku} to our attention, in which the authors study the noninvertible symmetry resolution of certain entanglement measures using methods from algebraic quantum field theory.}.
The universal high-$T$ limit of \eqref{eq:SRInterval}, and its application to SREE, was first treated in \cite{Saura-Bastida:2024yye} in the context of diagonal rational conformal field theories with noninvertible Verlinde line symmetries. However, our results do not agree with those reported in \cite{Saura-Bastida:2024yye}. We offer a potential explanation for the discrepancy.

\section{Boundary tube algebras}
To determine how a noninvertible symmetry $\CC$ acts on a state $\ket{\mathcal{O}}\in \CH_{B_1 B_2}$ on an interval with two conformal boundary conditions $B_{1}$ and $B_2$ imposed at the endpoints, it is useful to employ the state-operator correspondence to map the state $\ket{\mathcal{O}}$ to a boundary-changing operator $\mathcal{O}$, 
\begin{equation}
    {\scriptsize\begin{tikzpicture}
            \fill [gray, opacity=0.5] (0,-1) rectangle (1,1); 
            \fill [gray, opacity=0.5] (-3,-1) rectangle (-2,1); 
			\draw[thick,blue, decoration = {markings, mark=at position 0.6 with {\arrow[scale=1]{stealth[reversed]}}}, postaction=decorate] (0, 1) -- (0,-1);
            \draw[thick,blue, decoration = {markings, mark=at position 0.6 with {\arrow[scale=1]{stealth}}}, postaction=decorate] (-2, 1) -- (-2,-1);
			\node[right] at (0,-0.4) {$B_2$};
			\node[left] at (-2,-0.4) {$B_1$};
            \node[] at (-1, 0) {$\ket{\mathcal{O}}$};
\end{tikzpicture}} \quad \leftrightsquigarrow \quad \raisebox{-0.1in}{\scriptsize \begin{tikzpicture}
            \fill [gray, opacity=0.5] (-2,0) rectangle (2,1); 
			\draw[thick,blue, decoration = {markings, mark=at position 0.25 with {\arrow[scale=1]{stealth[reversed]}}, mark=at position 0.75 with {\arrow[scale=1]{stealth[reversed]}}}, postaction=decorate] (-2, 0) -- (2,0);
            \draw[thick, fill=black] (0, 0) circle (2pt);
			\node[above] at (-1,0) {$B_1$};
			\node[above] at (1,0) {$B_2$};
            \node[below] at (0, 0) {${\mathcal{O}}$};
\end{tikzpicture}} 
    \,.
\end{equation}
The symmetry $\CC$ then acts on the boundary-changing operator $\mathcal{O}$ via a so-called \emph{boundary lasso action} implemented by the \emph{boundary lasso operators} $\mathsf{H}^{C_1 C_2, y_1 y_2}_{B_1B_2, c}$, which are pictorially represented as
\begin{align}\label{eq:bdylasso}
\mathsf{H}^{C_1 C_2, y_1 y_2}_{B_1B_2, c} (\mathcal{O}):= ~\raisebox{-0.42in}{\scriptsize
	\begin{tikzpicture}[scale=0.6]
			\fill [gray, opacity=0.5] (0,-1.5) rectangle (1,1.5);
			\draw[thick, decoration = {markings, mark=at position 0.75 with {\arrow[scale=1]{stealth}}}, postaction=decorate]  (0,1) -- (0,-1) arc(270:90:1) --cycle;
			\node[left] at (0, 0) {{$\mathcal{O}$}};
			\draw[thick,blue, decoration = {markings, mark=at position 0.3 with {\arrow[scale=1]{stealth}}, mark=at position 0.1 with {\arrow[scale=1]{stealth}}, mark=at position 0.75 with {\arrow[scale=1]{stealth}}, mark=at position 0.95 with {\arrow[scale=1]{stealth}}}, postaction=decorate] (0, 1.5) -- (0,-1.5);
            \draw[thick, fill=black] (0, 0) circle (2pt);
			\node[right] at (0, 1.5) {$C_2$};
			\node[right] at (0,-1.5) {$C_1$};
			\node[right] at (0, 0.5) {$B_2$};
			\node[right] at (0,-0.5) {$B_1$};
			\node[right] at (-0, 1) {$y_2$};
			\node[right] at (-0, -1) {$\bar{y}_1$};
			\node[] at (-0.8, -0.1) {$c$};
	\end{tikzpicture}}
    ~~= 
    \raisebox{-0.35in}{\scriptsize
	\begin{tikzpicture}[scale=0.6]
			\fill [gray, opacity=0.5] (0,-1.5) rectangle (1,1.5);
            \draw[thick,blue, decoration = {markings, mark=at position 0.3 with {\arrow[scale=1]{stealth}}, mark=at position 0.75 with {\arrow[scale=1]{stealth}}}, postaction=decorate] (0, 1.5) -- (0,-1.5);
			\draw[thick, fill=black] (0, 0) circle (2pt);
			\node[left] at (0, 0) {{$\mathcal{O}'$}};
			\node[right] at (0, 0.5) {$C_2$};
			\node[right] at (0,-0.5) {$C_1$};
	\end{tikzpicture}}
	\,.
\end{align}
In this picture, the topological line $c\in \CC$ wraps around $\mathcal{O}$, and terminates on topological boundary-changing junctions $\bar{y}_1$ and $y_2$ connecting the boundary condition pairs $(B_1, C_1)$ and $(B_2, C_2)$, respectively. One then shrinks the lasso around $\mathcal{O}$ to produce another boundary-changing operator $\mathcal{O}'$ interpolating between a pair of (generically different) boundary conditions $C_1$ and $C_2$. Thus, boundary lasso operators act on the extended Hilbert space 
\begin{equation} \label{eq:extended_Hilbert}
    \CH_{\CB_1^\vee \CB_2} := \bigoplus_{B_1 \in \CB_1^\vee}\bigoplus_{B_2 \in \CB_2} \CH_{B_1 B_2} \,.
\end{equation}
Here, $\CB_1^\vee$ and $\CB_2$ are multiplets of conformal boundary conditions transforming into each other under the action of bulk topological lines, on the left and right ends of the interval, respectively.

The \emph{boundary tube algebra} $\mathrm{Tube}(\mathcal{B}_1^\vee\vert\mathcal{B}_2)$ is, by definition, the algebra which is generated by the boundary lasso operators in Equation \eqref{eq:bdylasso}. It is straightforward to derive the multiplication of boundary lasso operators, which takes the schematic form
\begin{equation}\label{eq:bdytubealgebra}
    \mathsf{H}_{X'}\times \mathsf{H}_{X} = \sum_{Y} t_{X'X}^{Y} \mathsf{H}_{Y}
\end{equation}
where e.g.\ $X$ represents the collection of indices of $\mathsf{H}^{C_1 C_2, y_1 y_2}_{B_1B_2, c}$, and the coefficients $t_{X'X}^{Y}$ are determined by the so-called $\widetilde{F}$-symbols of the multiplets $\mathcal{B}_1$ and $\mathcal{B}_2$. (See \cite{Kojita:2016jwe,Konechny:2019wff,Barter_2022,Konechny:2024ixa,Cordova:2024vsq, Copetti:2024dcz, Cordova:2024iti,Choi:2024tri} for concrete expressions for the coefficients $t_{X'X}^Y$.) Unlike the fusion coefficients of lines in $\CC$, the coefficients $t_{X'X}^Y$ appearing in Equation \eqref{eq:bdytubealgebra} are generally not integers.

The boundary tube algebra \eqref{eq:bdytubealgebra} has a rich mathematical structure. For example, when the two boundary multiplets $\mathcal{B}_1$ and $\mathcal{B}_2$ are the same \footnote{Mathematically, ``same'' here means that they define the same $\CC$-module category.}, it defines a $C^\ast$-weak Hopf algebra \cite{2012CMaPh.313..351K,Cordova:2024iti}.
At the end of this Letter, we discuss an explicit example of a Hopf algebra symmetry which arises in the doubled Ising model on an interval.

We turn to the representation theory of boundary tube algebras next. It turns out that the SymTFT allows one to painlessly extract the representation category of $\mathrm{Tube}(\mathcal{B}_1^\vee\vert\mathcal{B}_2)$ without having to deal with the details of its structure.

\section{Representation theory of boundary tube algebras from the SymTFT}
The main proposal of the SymTFT approach is that a 1+1d CFT $Q$ with symmetry $\CC$ can be expanded into a triple
\begin{equation}
    Q \leftrightharpoons (\mathcal{B}_{\mathrm{reg}}, \mathrm{TV}_{\CC}, \widetilde{Q})
\end{equation}
where $\mathrm{TV}_{\CC}$ (which is short for Turaev-Viro) is the 2+1d SymTFT, $\mathcal{B}_{\mathrm{reg}}$ (which is described mathematically by the regular module category of $\CC$) is its canonical Dirichlet topological boundary, and $\widetilde{Q}$ is a ``dynamical'' boundary which depends on the details of $Q$ \cite{Gaiotto:2014kfa,Gaiotto:2020iye,Ji:2019jhk,Apruzzi:2021nmk,Freed:2022qnc, Kong:2015flk, Kong:2017hcw, Kong:2020cie, Kaidi:2022cpf, Antinucci:2022vyk}.
The topological line defects generating the $\mathcal{C}$ symmetry are supported on the Dirichlet boundary.

We claim that this setup can be generalized to include boundary conditions of $Q$. Specifically, a conformal boundary condition $B$ of $Q$ can also be expanded into a triple,
\begin{equation}\label{eqn:tripleboundary}
    B \leftrightharpoons (\underline{B}, \mathcal{B}, \widetilde{B})
\end{equation}
where $\mathcal{B}$ is a topological boundary condition of the SymTFT, $\underline{B}$ is a topological line interface between $\mathcal{B}$ and $\mathcal{B}_{\mathrm{reg}}$, and $\widetilde{B}$ is a conformal line interface between $\mathcal{B}$ and $\widetilde{Q}$. See \cite{Huang:2023pyk, Copetti:2024dcz,Cvetic:2024dzu,Copetti:2024onh,Cordova:2024iti,Choi:2024tri}.

To describe which topological boundary condition $\mathcal{B}$ arises in Equation \eqref{eqn:tripleboundary}, recall that the $\CC$-multiplet to which a boundary condition $B$ of $Q$ belongs carries the structure of a $\CC$-module category \cite{Fuchs:2002cm,Huang:2021zvu,Choi:2023xjw}. On the other hand, topological boundary conditions of the SymTFT are also labeled by $\CC$-module categories \cite{2012CMaPh.313..351K}. In \cite{Choi:2024tri}, our proposal is that the topological boundary condition $\mathcal{B}$ arising in the SymTFT description of $B$ can be identified with the $\CC$-multiplet containing $B$. 

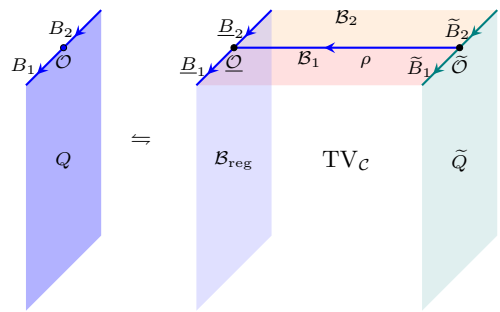
\begin{figure}[t]
	\centering
 {\scriptsize
    \raisebox{-63pt}{\begin{tikzpicture}
    \fill[fill=blue!50, opacity=.6](3,4) -- (3,1) -- (2,0) -- (2,3);
			\draw [color=blue, thick, decoration = {markings, mark=at position 0.5 with {\arrow[scale=1]{stealth[reversed]}}}, postaction=decorate] (2,3) -- (2.25,3.25) node[left]{\color{black}$B_1$} -- (2.5,3.5);
			\draw [color=blue, thick, decoration = {markings, mark=at position 0.5 with {\arrow[scale=1]{stealth[reversed]}}}, postaction=decorate] (2.5,3.5) -- (2.75,3.75) node[left]{\color{black}$B_2$} -- (3,4);
			\draw [fill=blue] (2.5,3.5) circle (0.04) node [below] {{\color{black}$\cO$}};			
			\draw (2.5,2) node{$Q$};
	\end{tikzpicture}}
    \quad $\leftrightharpoons$ ~
	\raisebox{-63pt}{\begin{tikzpicture}
             \fill[fill=red!40,opacity=.3] (0-1,3)  -- (3-1,3)  -- (4-1-0.5,4-0.5) -- (1-1-0.5,4-0.5) -- cycle;
             \fill[fill=orange!40,opacity=.3]  (4-1-0.5,4-0.5) -- (1-1-0.5,4-0.5)-- (1-1,4) -- (4-1,4) -- cycle;
             \fill[fill=teal!40,opacity=.3] (3,4) -- (3,1) -- (2,0) -- (2,3) -- cycle;
			\draw (1.4-.4,2) node{\small $\mathrm{TV}_\cC$};
			\fill[fill=blue!40, opacity=.3] (1-1,4) -- (1-1,1) -- (0-1,0) -- (0-1,3);
			\draw [color=blue, thick, decoration = {markings, mark=at position 0.5 with {\arrow[scale=1]{stealth[reversed]}}}, postaction=decorate] (0-1,3) -- (0.25-1,3.25) node[left]{\color{black}$\underline{B}_1$} -- (0.5-1,3.5);
			\draw [color=blue, thick, decoration = {markings, mark=at position 0.5 with {\arrow[scale=1]{stealth[reversed]}}}, postaction=decorate] (0.5-1,3.5) -- (0.75-1,3.75) node[left]{\color{black}$\underline{B}_2$} -- (1-1,4);
			\draw (-0.5,2) node{$\mathcal{B}_{\mathrm{reg}}$};
			\draw [color=blue, thick, decoration = {markings, mark=at position 0.6 with {\arrow[scale=1]{stealth}}}, postaction=decorate] (2.5,3.5) -- (1.5-.25,3.5) node[below]{\color{black}$\rho$} -- (0.5-1,3.5);
   \draw (.5,3.55) node[below]  {$\mathcal{B}_1$};
   \draw (1,4.1) node[below]  {$\mathcal{B}_2$};
			\draw [fill=black] (0.5-1,3.5) circle (0.04) node [below] {$\underline{\CO}$};
			\draw (2.5,2) node{ $\widetilde{Q}$};
			\draw [color=teal, thick, decoration = {markings, mark=at position 0.5 with {\arrow[scale=1]{stealth[reversed]}}}, postaction=decorate] (2,3) -- (2.25,3.25) node[left]{\color{black} $\widetilde{B}_1$} -- (2.5,3.5);
			\draw [color=teal, thick, decoration = {markings, mark=at position 0.5 with {\arrow[scale=1]{stealth[reversed]}}}, postaction=decorate] (2.5,3.5) -- (2.75,3.75) node[left]{\color{black} $\widetilde{B}_2$} -- (3,4);
			\draw [fill=black] (2.5,3.5) circle (0.04) node [below] {$\widetilde{\CO}$};
	\end{tikzpicture}}
 }
	\caption{
 The SymTFT picture of boundary conditions and  boundary-changing local operators. 
	}
	\label{fig:bdydefect.operatorsletter}
\end{figure}

A boundary-changing local operator $\CO$ between conformal boundary conditions $B_1$ and $B_2$ can be associated with a triple as well, 
\begin{equation}
    \CO \leftrightharpoons (\underline{\CO}, \rho, \widetilde{\mathcal{O}})\,.
\end{equation}
Here, $\rho$ is a topological line interface between $\CB_1$ and $\CB_2$. Mathematically, such line interfaces correspond to objects of the category $\mathrm{Fun}_{\mathcal{C}}(\mathcal{B}_1,\mathcal{B}_2)$ of $\mathcal{C}$-module functors from $\mathcal{B}_1$ to $\mathcal{B}_2$. Further, $\underline{\CO}$ is a topological junction between the three topological line interfaces $\rho, \underline{B}_1$ and $\underline{B}_2$, and we use $W_{\underline{B}_1 \underline{B}_2}^{\rho}$ to denote the space which is spanned by such topological junctions. Similarly, $\widetilde{\mathcal{O}}$ is a (non-topological) junction between $\rho, \widetilde{B}_1$ and $\widetilde{B}_2$, and we use $\mathcal{V}_\rho^{\widetilde{B}_1\widetilde{B}_2}$ to denote the space of such junctions. See Figure \ref{fig:bdydefect.operatorsletter} for a summary of the ingredients explained so far, and \cite{Choi:2024tri} for further details.

It follows from the discussion in the previous paragraph that the Hilbert space $\mathcal{H}_{B_1B_2}$ admits a decomposition of the form 
\begin{align}\label{eqn:HB1B2decomp}
\mathcal{H}_{B_1B_2} = \bigoplus_\rho \mathcal{H}^\rho_{B_1B_2}
\end{align}
where $\mathcal{H}^\rho_{B_1B_2}=W_{\underline{B}_1 \underline{B}_2}^{\rho} \otimes \mathcal{V}_{\rho}^{\widetilde{B}_1 \widetilde{B}_2}$ is the subspace of boundary-changing operators $\mathcal{O}$ which  inflate into the topological line interface $\rho$ in the SymTFT, i.e.\ correspond to triples of the form $(\ast,\rho,\ast)$. This decomposition can be interpreted in terms of the representation theory of the boundary tube algebra. 
In \cite{Choi:2024tri}, we explain that, given two multiplets $\mathcal{B}_1$ and $\mathcal{B}_2$ of conformal boundary conditions of $Q$, the irreducible representations of the corresponding boundary tube algebra $\mathrm{Tube}(\mathcal{B}_1^\vee\vert\mathcal{B}_2)$ are precisely in one-to-one correspondence with the different options for the simple topological line interface $\rho$ \footnote{Indeed, because the $\CC$ symmetry operators are only supported on the Dirichlet boundary $\mathcal{B}_{\mathrm{reg}}$, the boundary lasso operators \eqref{eq:bdylasso} only act on the topological junction $\underline{\CO}$, while keeping the topological boundary conditions $\mathcal{B}_1$ and $\CB_2$, as well as the topological line interface $\rho$ interpolating between them, unchanged.}. 
That is, $\mathcal{H}^\rho_{B_1B_2} \subset \mathcal{H}_{B_1B_2}$ is the subspace which transforms according to the representation $\rho$ of the boundary tube algebra. 

In \cite{Choi:2024tri}, the following closed form expression is obtained for the projector $P^\rho_{B_1B_2}:\mathcal{H}_{B_1B_2}\to\mathcal{H}^\rho_{B_1B_2}$ in terms of the boundary lasso operators in Equation \eqref{eq:bdylasso},

\begin{align}\label{eq:projfinal}
\begin{split}
    &P_{B_1 B_2}^{\rho} = \frac{\qd_{\rho}}{\qd_{\underline{B}_1} \qd_{\underline{B}_2}} \sum_{\substack{\mu x y\\ a x' z z'}} \sqrt{\frac{\qd_{a}}{S_{\mu 1}}} {^{\CB_1 \CB_2}}\widetilde{\Psi}^{11}_{\rho \rho(\mu x y)}\\
    &\ \ \ \ \ \  {^{\CB_1 \CB_{\mathrm{reg}}}}{\Psi}^{1 (az)}_{\underline{B}_1 \underline{B}_1 (\mu x x')} {^{\CB_2 \CB_{\mathrm{reg}}}}\widetilde{\Psi}^{1 (az')}_{\underline{B}_2 \underline{B}_2 (\mu y x')}\mathsf{H}^{B_1 B_2, z z'}_{B_1B_2, a}\,.
\end{split}
\end{align}
The derivation is also briefly discussed in End Matter. 
Here, $\qd_X$ denotes the quantum dimension of the topological line interface $X$ \cite{Diatlyk:2023fwf}, $S_{\mu\nu}$ is the modular S-matrix of the SymTFT, and the complex numbers ${^{\mathcal{B}_1\mathcal{B}_2}}\Psi^{(az)(bw)}_{\alpha\beta(\mu x y)}$ and ${^{\mathcal{B}_1\mathcal{B}_2}}\widetilde{\Psi}^{(az)(bw)}_{\alpha\beta(\mu x y)}$ are so-called ``generalized half-linking numbers'' which are defined in End Matter. We will make crucial use of this projector below.

\section{Noninvertible symmetry-resolved ALC formula}
We now return to the symmetry-resolved partition function \eqref{eq:SRInterval}. It is useful to define the \emph{charged moment} \cite{Kusuki:2023bsp, Goldstein:2017bua, Xavier:2018kqb, Calabrese:2021wvi, DiGiulio:2022jjd} as the annulus partition function with a line $a$ stretched along the interval direction and intersecting the two boundaries at  junctions $\bar{z}$ and $z'$ respectively, 
\begin{equation}
    Z_{B_1 B_2} ^{a z z'}(q) = \Tr_{\CH_{B_1 B_2}} (\mathsf{H}_{B_1 B_1, a}^{B_2 B_2, zz'} q^{L_0 - \frac{c}{24}}) = {\scriptsize \begin{tikzpicture}
            \fill [gray, opacity=0.5] (0,-0.6) rectangle (1,0.6); 
            \fill [gray, opacity=0.5] (-3,-0.6) rectangle (-2,0.6); 
			\draw[thick,blue, decoration = {markings, mark=at position 0.3 with {\arrow[scale=1]{stealth[reversed]}}, mark=at position 0.75 with {\arrow[scale=1]{stealth[reversed]}}}, postaction=decorate] (0, 0.6) -- (0,-0.6);
            \draw[thick,blue, decoration = {markings, mark=at position 0.3 with {\arrow[scale=1]{stealth}}, mark=at position 0.75 with {\arrow[scale=1]{stealth}}}, postaction=decorate] (-2, 0.6) -- (-2,-0.6);
			\node[right] at (0, 0.4) {$B_2$};
			\node[right] at (0,-0.4) {$B_2$};
            \node[left] at (-2, 0.4) {$B_1$};
			\node[left] at (-2,-0.4) {$B_1$};
            \draw[thick, black, decoration = {markings, mark=at position 0.5 with {\arrow[scale=1]{stealth}}}, postaction=decorate] (-2,0) -- (0,0);
            \draw[black,dashed] (-2,0.6) -- (0,0.6);
            \draw[black,dashed] (-2,-0.6) -- (0,-0.6);
            \node[above] at (-1,0) {$a$}; 
            \node[] at (-1.8,-0.2) {$\bar{z}$};
            \node[] at (-0.25,-0.2) {$z'$};
\end{tikzpicture}}\,.
\end{equation}
Substituting the projector \eqref{eq:projfinal} into \eqref{eq:SRInterval}, we see that the partition function $Z_{B_1 B_2}^\rho$ can be expressed as a linear combination of such charged moments,
\begin{equation}\label{eq:changebasis}
    \begin{split}
        Z_{B_1 B_2}^\rho(q) =&\frac{\qd_{\rho}}{\qd_{\underline{B}_1} \qd_{\underline{B}_2}} \sum_{\substack{\mu x y\\ a x' z z'}} \sqrt{\frac{\qd_{a}}{S_{\mu 1}}} {^{\CB_1 \CB_2}}\widetilde{\Psi}^{11}_{\rho \rho(\mu x y)} \\&\times {^{\CB_1 \CB_{\mathrm{reg}}}}{\Psi}^{1 (az)}_{\underline{B}_1 \underline{B}_1 (\mu x x')} {^{\CB_2 \CB_{\mathrm{reg}}}}\widetilde{\Psi}^{1 (az')}_{\underline{B}_2 \underline{B}_2 (\mu y x')} \\&\times {\scriptsize \begin{tikzpicture}
            \fill [gray, opacity=0.5] (0,-0.6) rectangle (1,0.6); 
            \fill [gray, opacity=0.5] (-3,-0.6) rectangle (-2,0.6); 
			\draw[thick,blue, decoration = {markings, mark=at position 0.3 with {\arrow[scale=1]{stealth[reversed]}}, mark=at position 0.75 with {\arrow[scale=1]{stealth[reversed]}}}, postaction=decorate] (0, 0.6) -- (0,-0.6);
            \draw[thick,blue, decoration = {markings, mark=at position 0.3 with {\arrow[scale=1]{stealth}}, mark=at position 0.75 with {\arrow[scale=1]{stealth}}}, postaction=decorate] (-2, 0.6) -- (-2,-0.6);
			\node[right] at (0, 0.4) {$B_2$};
			\node[right] at (0,-0.4) {$B_2$};
            \node[left] at (-2, 0.4) {$B_1$};
			\node[left] at (-2,-0.4) {$B_1$};
            \draw[thick, black, decoration = {markings, mark=at position 0.5 with {\arrow[scale=1]{stealth}}}, postaction=decorate] (-2,0) -- (0,0);
            \draw[black,dashed] (-2,0.6) -- (0,0.6);
            \draw[black,dashed] (-2,-0.6) -- (0,-0.6);
            \node[above] at (-1,0) {$a$}; 
            \node[] at (-1.8,-0.2) {$\bar{z}$};
            \node[] at (-0.25,-0.2) {$z'$};
\end{tikzpicture}}\,.
    \end{split}
\end{equation}

We then invoke a version of open-closed duality, which asserts that
\begin{equation}
    Z_{B_1 B_2} ^{a z z'}(q) = {_{az'}}\bra{B_2} \widetilde{q}^{\frac{1}{2}(L_0+ \bar{L}_0-\frac{c}{12})}  \ket{B_1}_{az}= \raisebox{0.2in}{\scriptsize \begin{tikzpicture}
            \fill [gray, opacity=0.5] (-0.6,0) rectangle (0.6,1); 
            \fill [gray, opacity=0.5] (-0.6,-3) rectangle (0.6, -2); 
			\draw[thick,blue, decoration = {markings, mark=at position 0.3 with {\arrow[scale=1]{stealth}}, mark=at position 0.75 with {\arrow[scale=1]{stealth}}}, postaction=decorate] (0.6,0) -- (-0.6,0);
            \draw[thick,blue, decoration = {markings, mark=at position 0.3 with {\arrow[scale=1]{stealth[reversed]}}, mark=at position 0.75 with {\arrow[scale=1]{stealth[reversed]}}}, postaction=decorate] (0.6,-2) -- (-0.6,-2);
			\node[above] at (0.4,0) {$B_2$};
			\node[above] at (-0.4,0) {$B_2$};
            \node[below] at (0.4,-2) {$B_1$};
			\node[below] at (-0.4,-2) {$B_1$};
            \draw[thick, black, decoration = {markings, mark=at position 0.5 with {\arrow[scale=1]{stealth}}}, postaction=decorate] (0,-2) -- (0,0);
            \draw[black,dashed] (0.6,-2) -- (0.6,0);
            \draw[black,dashed] (-0.6,-2) -- (-0.6,0);
            \node[left] at (0,-1) {$a$}; 
            \node[] at (-0.2,-1.8) {$\bar{z}$};
            \node[] at (-0.2, -0.25) {$z'$};
\end{tikzpicture}}
\end{equation}
for suitable twisted sector boundary states $\ket{B_i}_{az}$, where $\widetilde{q}= e^{-4\pi \ell/\beta}$ \cite{Choi:2024tri}. In the high-$T$ limit, $\ell \gg \beta$ and hence $\widetilde{q}\to 0$.
Assuming that the symmetry $\mathcal{C}$ acts faithfully,
the ground state contribution from the untwisted sector $a=1$ dominates the sum \eqref{eq:changebasis} in such a limit \cite{Lin:2022dhv}.
On the other hand, when $a=1$, the standard ALC formula, Equation \eqref{eq:ZBB}, asserts that $Z_{B_1 B_2} ^{a z z'}(q) \sim g_1 g_2 e^{\frac{c \pi }{6} \frac{\ell}{\beta}}$.
Hence,  we find the estimate
\begin{equation}
    \begin{split}
     &Z_{B_1 B_2}^\rho(q) \sim \frac{\qd_{\rho}}{\qd_{\underline{B}_1} \qd_{\underline{B}_2}} g_1 g_2 e^{\frac{c \pi }{6} \frac{\ell}{\beta}} \sum_{\substack{\mu x y x'}} \sqrt{\frac{1}{S_{\mu 1}}} \\
     & \ \ \ \ \   {^{\CB_1 \CB_2}}\widetilde{\Psi}^{11}_{\rho \rho(\mu x y)}  {^{\CB_1 \CB_{\mathrm{reg}}}}{\Psi}^{1 1}_{\underline{B}_1 \underline{B}_1 (\mu x x')} {^{\CB_2 \CB_{\mathrm{reg}}}}\widetilde{\Psi}^{1 1}_{\underline{B}_2 \underline{B}_2 (\mu y x')} \,.
    \end{split}
\end{equation}

Finally, we note that the sum on the right-hand side reduces to the dimension of the topological junction space $N_{\rho \underline{B}_2}^{\underline{B}_1}:= \dim W_{\underline{B}_1 \underline{B}_2}^{\rho}$, thanks to a generalized Verlinde formula derived in \cite[Section 5.2]{Choi:2024tri}.
Therefore, we arrive at the desired noninvertible symmetry-resolved ALC formula,
\begin{equation}\label{eq:main1}
    Z_{B_1 B_2}^\rho(q) \sim \frac{\qd_{\rho} N_{\rho \underline{B}_2}^{\underline{B}_1}}{\qd_{\underline{B}_1} \qd_{\underline{B}_2}} g_1 g_2 e^{\frac{c \pi }{6} \frac{\ell}{\beta}}, \quad \beta \ll \ell\,.
\end{equation}
The additional prefactor, compared to the original ALC formula \eqref{eq:ZBB}, is determined by the quantum dimensions of boundary-changing topological line interfaces $\rho$, $\underline{B}_1$, and $\underline{B}_2$ of the SymTFT and the fusion coefficient $N_{\rho \underline{B}_2}^{\underline{B}_1}$ between them.

As a consistency check, we note that, due to $\qd_{\underline{B}_1} \qd_{\underline{B}_2}= \sum_\rho N_{\rho \underline{B}_2}^{\underline{B}_1} \qd_\rho$ (which follows from the fusion rule of topological interfaces \cite{Diatlyk:2023fwf}), summing over all the representations $\rho$ gives back the standard ALC formula \eqref{eq:ZBB}. Another consistency check is that when we restrict to the special case of a non-anomalous invertible symmetry $\CC= \mathrm{Vec}_{G}$, with $B_1 = B_2 =B$ a $G$-symmtric boundary condition,
we have $\qd_{\underline{B}}=\sqrt{|G|}$ and $N_{\rho \underline{B}}^{\underline{B}} = \qd_{\rho}$, hence the prefactor $\qd_{\rho} N_{\rho \underline{B}}^{\underline{B}}/{\qd_{\underline{B}} \qd_{\underline{B}}}$ reproduces the known result $\qd_\rho^2/|G|$ \cite{Casini:2019kex,Magan:2021myk,Kusuki:2023bsp}.
In this special case, the quantum dimension $\qd_\rho$ coincides with the dimension of the representation $\rho$.

\section{Noninvertible symmetry-resolved entanglement entropy}
The noninvertible symmetry-resolved ALC formula \eqref{eq:main1} immediately implies the following leading and sub-leading contributions to the noninvertible symmetry-resolved entanglement entropy for the ground state of the 1+1d CFT defined on an infinite line, when the entanglement subregion is chosen to be a single interval: 
\begin{equation} \label{eq:SREE}
\begin{split}
    S^\rho_{\mathrm{EE}}&=\lim_{n\to 1} \frac{1}{1-n} \log \frac{Z^\rho_{B_1 B_2}(q^n)}{[Z^\rho_{B_1 B_2}(q)]^n}\\& \sim \frac{c}{3} \log \frac{L}{\varepsilon} + \log g_1  + \log g_2 + \log \frac{\qd_{\rho} N_{\rho \underline{B}_2}^{\underline{B}_1}}{\qd_{\underline{B}_1} \qd_{\underline{B}_2}} \,.
\end{split}
\end{equation}
The leading universal behavior, as well as the appearance of $g$-functions, is the same as the ordinary entanglement entropy.
On top of that, there is an additional sub-leading contribution depending on the representation $\rho$ of the boundary tube algebra, which is determined by the quantum dimensions and fusion coefficients of various topological boundary-changing line interfaces in the SymTFT.
Here, we have chosen the boundary conditions at the entangling surface (which consists of two points) to be $B_1$ and $B_2$, respectivley \cite{Ohmori:2014eia}. By taking the difference between the SREE and the ordinary entanglement entropy, one can also remove the dependence on the regularization scale $\varepsilon$,
\begin{equation}
 \lim_{\varepsilon \to 0} \left[ S^\rho_{\mathrm{EE}} - S_{\mathrm{EE}} \right]  = \log \frac{\qd_{\rho} N_{\rho \underline{B}_2}^{\underline{B}_1}}{\qd_{\underline{B}_1} \qd_{\underline{B}_2}} \,. 
\end{equation}
See Supplemental Material for a brief review of the BCFT approach to entanglement entropies.
Recently, special cases of noninvertible SREE have been numerically computed \cite{Heymann:2024vvf} in anyonic chains \cite{Feiguin:2006ydp,2013PhRvB..87w5120G,Buican:2017rxc}, and the results agree with our general formula \eqref{eq:SREE}.

When one considers the SREE, it is natural to choose the boundary conditions at the entangling surface to preserve the symmetries of the bulk CFT \cite{DiGiulio:2022jjd,Kusuki:2023bsp}.
On the other hand, defining the notion of a symmetric boundary condition under the action of a general fusion category symmetry requires care \cite{Choi:2023xjw}. 
We briefly comment on this in Supplemental Material.
As we explain more there, if one chooses the boundary conditions at the two entangling points to be the same ``strongly symmetric'' boundary condition (with $g$-function $g$), then \eqref{eq:SREE} reduces to
\begin{equation}
    S^\rho_{\mathrm{EE}} \sim \frac{c}{3} \log \frac{L}{\varepsilon} + \log g^2 + \log \frac{\qd_{\rho}^2 }{\mathrm{dim}(\mathcal{C})^2} \,,
\end{equation}
where $\mathrm{dim}(\mathcal{C})=\sqrt{\sum_{a\in\mathrm{Irr}(\mathcal{C})}\qd_a^2}$ is the total dimension of the fusion category $\mathcal{C}$.
We remark that, for the trivial representation $\rho =1$, the last term on the right-hand side coincides with the topological entanglement entropy of the SymTFT $\mathrm{TV}(\mathcal{C})$ \cite{Kitaev:2005dm,Levin:2006zz}. We further note that the same result can also be obtained by imposing the (non-simple) cloaking boundary condition introduced in \cite{Brehm:2021wev,Brehm:2024zun} at the entangling points, even when no symmetric boundary condition is known \footnote{The cloaking boundary condition is a non-simple boundary condition with a topological local operator on it whose defining property is that it is ``invisible'' to topological lines generating the symmetry. Such a boundary condition exists within any multiplet $\mathcal{B}$ of boundaries.}.

As mentioned at the end of the introduction, our results do not fully agree with those reported in an earlier paper \cite{Saura-Bastida:2024yye}. We believe that the discrepancy is due to the projectors used in \cite[Equation (17)]{Saura-Bastida:2024yye}. This expression first appeared in \cite[Equation (3.19)]{Lin:2022dhv} and, when used in tandem with the complementary projectors in \cite[Equation (3.24)]{Lin:2022dhv}, is appropriate for projecting the $a$-twisted $S^1$ Hilbert space of a diagonal rational CFT onto suitable irreducible representations of the \emph{ordinary} tube algebra. However, these projectors do not act on the interval Hilbert space, and hence are not suitable for analyzing the representations of the boundary tube algebra, which is the relevant one for the discussion of entanglement entropies. 

Finally, we note that an alternative derivation of \eqref{eq:SREE}, analogous to the ``3d interpretation'' of \cite{Lin:2022dhv}, can be obtained by leveraging the formulation of open-closed duality in \cite[Section 8.4]{Choi:2024tri}.

\section{Critical double Ising model}
We apply our results to the critical double Ising CFT (i.e.\ two decoupled copies of the Ising CFT).
The theory has a $\mathbb{Z}_2 \times \mathbb{Z}_2$ symmetry acting on Ising spins, as well as two copies of the noninvertible Kramers-Wannier symmetry \cite{Oshikawa:1996ww,Oshikawa:1996dj,Petkova:2000ip,Frohlich:2004ef,Frohlich:2006ch,Frohlich:2009gb,Aasen:2016dop,Chang:2018iay,Ji:2019ugf,Aasen:2020jwb,Seiberg:2023cdc}.
We denote the topological lines generating the $\mathbb{Z}_2 \times \mathbb{Z}_2$ symmetry as $\{1, a, b, ab \}$, and the \emph{product} of the two Kramers-Wannier duality lines as $\mathcal{N}$.
Together, these lines generate a $\mathbb{Z}_2 \times \mathbb{Z}_2$ Tambara-Yamagami fusion category symmetry of the double Ising CFT, which we denote as $\mathcal{C}_{\mathsf{TY}}$ \cite{Bhardwaj:2017xup,Thorngren:2021yso}.

In \cite{Choi:2023xjw}, it was shown that the double Ising CFT admits a conformal boundary condition which forms an irreducible multiplet by itself under the action of $\mathcal{C}_{\mathsf{TY}}$.
Namely, in the terminology used in that reference, the boundary condition is ``strongly'' symmetric under $\mathcal{C}_{\mathsf{TY}}$.
We denote this boundary condition as $B$ \footnote{The conformal boundary condition $B$ is obtained by folding a single copy of the Ising CFT along its Kramers-Wannier duality line \cite{Choi:2023xjw}.}.

\begin{figure}
\centering
    \includegraphics[width=0.8\textwidth]{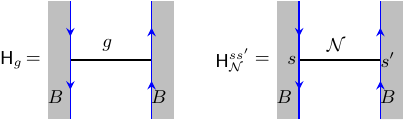}
    \caption{Boundary lasso opeartors in the double Ising CFT. Here, $g \in \mathbb{Z}_2 \times \mathbb{Z}_2$ and $s,s' = 0,1$ label topological point junctions. All the topological lines are self-dual and we do not draw arrows on them.} \label{fig:H8}
\end{figure}

We now consider the boundary tube algebra of the double Ising CFT defined on an interval with the $B$ boundary condition imposed at both ends.
The junction spaces between the $\mathbb{Z}_2 \times \mathbb{Z}_2$ topological lines and the boundary $B$ are all 1-dimensional, and hence each invertible topological line gives rise to a boundary lasso operator $\mathsf{H}_{BB,g}^{BB,11} \equiv \mathsf{H}_a$, where $g \in \mathbb{Z}_2 \times \mathbb{Z}_2$.
On the other hand, the noninvertible topological line $\mathcal{N}$ admits two linearly-independent topological junctions with $B$, and we distinguish them by a binary label $s= 0,1$ \cite{Choi:2023xjw}.
Therefore, the bulk topological line $\mathcal{N}$ defines 4 boundary lasso operators $\mathsf{H}_{BB,\mathcal{N}}^{BB,ss'} \equiv \mathsf{H}_\mathcal{N}^{s s'}$ acting on the interval Hilbert space.
See Figure \ref{fig:H8}.

Using the known boundary $\widetilde{F}$-symbols, we can compute the explicit boundary tube algebra.
This includes the multiplication relations
\begin{align} \label{eq:H8}
\begin{split}
    \mathsf{H}_g \times \mathsf{H}_h &= \mathsf{H}_h \times \mathsf{H}_g  = \mathsf{H}_{gh} \,, \\
    \mathsf{H}_{\mathcal{N}}^{ss'} \times  \mathsf{H}_{a}  &=  \mathsf{H}_{b} \times \mathsf{H}_{\mathcal{N}}^{ss'} = (-1)^s \mathsf{H}_{\mathcal{N}}^{s(s'+1)} \,, \\
    \mathsf{H}_{\mathcal{N}}^{ss'} \times  \mathsf{H}_{b} &= \mathsf{H}_{a} \times \mathsf{H}_{\mathcal{N}}^{ss'} = (-1)^{s'} \mathsf{H}_{\mathcal{N}}^{(s+1)s'} \,, \\
    \mathsf{H}_{\mathcal{N}}^{s_1 s_2} \times \mathsf{H}_{\mathcal{N}}^{s_3 s_4} &= 2 \sum_{m,n=0,1} (-1)^{f_{mn}(s_i)} \mathsf{H}_{a^m b^n}
    \,,
\end{split}
\end{align}
where $f_{mn}(s_i) \equiv s_1 s_3 + s_2 s_4 + m(s_2 + s_3) + n (s_1 + s_4) + mn$, and $g,h \in \mathbb{Z}_2 \times \mathbb{Z}_2$ are arbitrary group elements.
We explain this in more detail in Supplemental Material.
Interestingly, one recognizes \eqref{eq:H8} as the 8-dimensional Kac-Paljutkin Hopf algebra $H_8$ \cite{KP,masuoka1995semisimple}. 
$H_8$ was historically the first Hopf algebra discovered  which is neither commutative nor cocommutative (implying that its representation category is not equivalent to $\mathrm{Vec}_G$ or $\mathrm{Rep}(G)$ for any finite group $G$) \cite{Bhardwaj:2017xup}, and as such has a certain mathematical significance.
Our analysis shows that such an algebra appears in a physical system as familiar as (two copies of) the Ising model on an open interval.

The above discussion reveals that the ground state entanglement Hamiltonian of the critical double Ising model for a single interval possesses the $H_8$ Hopf algebra symmetry, when the boundary conditions at the two entangling points are chosen to be $B$.
Our general formula \eqref{eq:SREE} then allows us to compute the symmetry-resolved entanglement entropy for each representation sector of $H_8$.
The representation category of $H_8$ is well-understood, and in fact, it is isomorphic to the fusion category of bulk topological lines, $\mathrm{Rep}(H_8) \cong \mathcal{C}_{\mathsf{TY}}$ \cite{Bhardwaj:2017xup,Thorngren:2021yso}.
Therefore, we label the irreducible representations of $H_8$ by the corresponding topological lines in $\mathcal{C}_{\mathsf{TY}}$, and denote them as $\rho_g$ for $g \in \mathbb{Z}_2 \times \mathbb{Z}_2$ and $\rho_\mathcal{N}$.
The quantum dimensions and the fusion coefficients appearing in the general formula \eqref{eq:SREE} in this case are given by $\qd_{\rho_g} = 1$, $\qd_{\rho_\mathcal{N}} = 2$, $N_{\rho_g \underline{B}}^{\underline{B}} = 1$, $N_{\rho_\mathcal{N} \underline{B}}^{\underline{B}} = 2$, and $\qd_{\underline{B}} = \sqrt{8}$.
The corresponding symmetry-resolved entanglement entropies for the double Ising CFT are therefore
\begin{align}
\begin{split}
    S^{\rho_g}_{\mathrm{EE}} &\sim \frac{c}{3} \log \frac{L}{\varepsilon} + \log \braket{B|0}^2 + \log \frac{1}{8} \,,\\
    S^{\rho_\mathcal{N}}_{\mathrm{EE}} &\sim \frac{c}{3} \log \frac{L}{\varepsilon} + \log \braket{B|0}^2 + \log \frac{1}{2} \,,
\end{split}
\end{align}
where the central charge $c=1$ and the boundary $g$-function $\braket{B|0} = \sqrt{2}$.

\textit{Acknowledgements.}--We would like to thank Horacio Casini, Ryohei Kobayashi, Zohar Komargodski, Javier Mag\'{a}n, Shinsei Ryu, Sahand Seifnashri, Shu-Heng Shao, Nikita Sopenko, and Fei Yan for enlightening discussions. YC gratefully acknowledges funding provided by the Roger Dashen Member Fund and the Fund for Memberships in Natural Sciences at the Institute for Advanced Study. BR gratefully acknowledges NSF grant PHY-2210533.

\emph{Note Added:}
Near the completion of this work, the papers \cite{Copetti:2024onh,Cordova:2024iti,Copetti:2024dcz} appeared, where the SymTFT for boundary conditions were also studied with different perspectives and applications.
Recently, we were informed that related topics will be discussed in several upcoming papers  \cite{Heymann:2024vvf,Das:2024qdx,Bhardwaj:2024igy,GarciaEtxebarria:2024jfv} with which this paper and \cite{Choi:2024tri} were coordinated.

\section{End Matter}

Below, we review the definition of generalized half-linking numbers given in \cite{Choi:2024tri}, and sketch how they can be used to derive the projector defined in Equation \eqref{eq:projfinal}. 
A more systematic treatment is given in \cite{Choi:2024tri}, specifically Section 2.4 and Section 7.3.

We define the generalized half-linking numbers ${^{\mathcal{B}_1\mathcal{B}_2}}\Psi_{\alpha\beta(\mu xy)}^{(az)(bw)}$ and ${^{\mathcal{B}_1\mathcal{B}_2}}\widetilde{\Psi}_{\alpha\beta(\mu xy)}^{(az)(bw)}$ by the expectation value of the following configurations of various bulk, boundary, and boundary-changing topological line interfaces of the SymTFT $\mathrm{TV}_{\mathcal{C}}$:
\begin{align}\label{eqn:halflinkingdefn}
	\begin{split}
		{^{\mathcal{B}_1\mathcal{B}_2}}\Psi_{\alpha\beta(\mu xy)}^{(az)(bw)} &= \sqrt{\frac{S_{11}}{\qd_a \qd_b}}~{\scriptsize\begin{tikzpicture}[scale=0.7]
	\begin{pgfonlayer}{nodelayer}
		\node [style=none] (21) at (0, 1) {};
		\node [style=none] (22) at (0, 2.5) {};
		\node [style=none] (24) at (-1.25, -0.5) {};
		\node [style=none] (25) at (-1.5, -2) {};
		\node [style=none] (26) at (1.5, -2) {};
		\node [style=none] (27) at (1.5, 3) {};
		\node [style=none] (28) at (-1.5, 3) {};
		\node [style=none] (29) at (-1.75, -0.5) {$y$};
		\node [style=none] (30) at (1, -2) {$\bar x$};
		\node [style=none] (31) at (0, 0.5) {$w$};
		\node [style=none] (32) at (0, 3) {$\bar z$};
		\node [style=none] (33) at (-0.5, 1.75) {$\beta$};
		\node [style=none] (34) at (2.5, 1) {$a$};
		\node [style=none] (35) at (-2.5, 1) {$\alpha$};
		\node [style=none] (36) at (0, -1) {$\mu$};
		\node [style=none] (38) at (0, 2.25) {$\times$};
		\node [style=none] (39) at (0, 1.25) {$\times$};
		\node [style=none] (40) at (-1.5, 1.5) {$\mathcal{B}_2$};
		\node [style=none] (41) at (3, 2.75) {$\mathcal{B}_1$};
		\node [style=none] (42) at (1.25, -0.5) {};
		\node [style=none] (43) at (-1, 0.5) {$b$};
	\end{pgfonlayer}
	\begin{pgfonlayer}{edgelayer}
        \draw[decoration = {markings, mark=at position -0.45 with {\arrow[scale=1.5]{stealth}}}, postaction=decorate] [style=boundarylines, in=150, out=-165] (28.center) to (25.center);
		\draw [style=boundarylines, bend right=60] (25.center) to (42.center);
		\draw [style=boundarylines, bend right=45] (42.center) to (21.center);
  \draw[preaction={draw=white,line width=6pt},preaction={draw=white,line width=6pt},decoration = {markings, mark=at position -0.45 with {\arrow[scale=1.5]{stealth}}}, postaction=decorate] [style=bulklines, bend right=45] (26.center) to (24.center);
		\draw[decoration = {markings, mark=at position -0.4 with {\arrow[scale=1.5]{stealth}}}, postaction=decorate] [style=boundarylines] (21.center) to (22.center);
		\draw[thick] [bend left] (22.center) to (27.center);
		\draw[decoration = {markings, mark=at position -0.45 with {\arrow[scale=1.5]{stealth}}}, postaction=decorate][thick] [bend left=75] (27.center) to (26.center);
		\draw[decoration = {markings, mark=at position -0.45 with {\arrow[scale=1.5]{stealth}}}, postaction=decorate][thick] (24.center) to (21.center);
		
		\draw [style=boundarylines, bend right] (22.center) to (28.center);
	\end{pgfonlayer}
\end{tikzpicture}} \,, \\
		{^{\mathcal{B}_1\mathcal{B}_2}}\widetilde{\Psi}_{\alpha\beta(\mu xy)}^{(az)(bw)} &= \sqrt{\frac{S_{11}}{\qd_a \qd_b}}~{\scriptsize\begin{tikzpicture}[scale=0.7]
	\begin{pgfonlayer}{nodelayer}
		\node [style=none] (0) at (0, 1) {};
		\node [style=none] (1) at (0, 2.5) {};
		\node [style=none] (2) at (1.25, -0.5) {};
		\node [style=none] (3) at (-1.25, -0.5) {};
		\node [style=none] (4) at (-1.5, -2) {};
		\node [style=none] (5) at (1.5, -2) {};
		\node [style=none] (6) at (1.5, 3) {};
		\node [style=none] (7) at (-1.5, 3) {};
		\node [style=none] (8) at (-1, -2) {$\bar y$};
		\node [style=none] (9) at (1.75, -0.5) {$x$};
		\node [style=none] (10) at (0, 0.5) {$z$};
		\node [style=none] (11) at (0, 3) {$\bar w$};
		\node [style=none] (12) at (-0.5, 1.75) {$\beta$};
		\node [style=none] (13) at (2.5, 1) {$\alpha$};
		\node [style=none] (14) at (-2.5, 1) {$b$};
		\node [style=none] (15) at (0, -1) {$\mu$};
		\node [style=none] (16) at (-1.5, -2) {};
		\node [style=none] (19) at (0, 2.25) {$\times$};
		\node [style=none] (20) at (0, 1.25) {$\times$};
		\node [style=none] (21) at (1.5, 1.3) {$\mathcal{B}_1$};
		\node [style=none] (22) at (3.5, 2.5) {$\mathcal{B}_2$};
            \node [style=none] (23) at (.9, .7) {$a$};
	\end{pgfonlayer}
	\begin{pgfonlayer}{edgelayer}
		\draw[decoration = {markings, mark=at position -0.4 with {\arrow[scale=1.5]{stealth}}}, postaction=decorate] [style=boundarylines] (0.center) to (1.center);
  \draw [style=boundarylines, bend left] (1.center) to (6.center);
		\draw [style=boundarylines, bend right=300] (5.center) to (3.center);
		\draw [style=boundarylines, bend left=45] (3.center) to (0.center);
		\draw[preaction={draw=white,line width=6pt},decoration = {markings, mark=at position -0.45 with {\arrow[scale=1.5]{stealth}}}, postaction=decorate] [style=bulklines, bend left=45] (16.center) to (2.center);
		\draw[decoration = {markings, mark=at position -0.45 with {\arrow[scale=1.5]{stealth}}}, postaction=decorate][thick] (2.center) to (0.center);
  \draw[decoration = {markings, mark=at position -0.45 with {\arrow[scale=1.5]{stealth}}}, postaction=decorate] [style=boundarylines, in=30, out=-15] (6.center) to (5.center);
		\draw[thick] [bend right] (1.center) to (7.center);
		\draw[decoration = {markings, mark=at position -0.45 with {\arrow[scale=1.5]{stealth}}}, postaction=decorate][thick] [in=165, out=-165] (7.center) to (16.center);
	\end{pgfonlayer}
\end{tikzpicture}} \,.
	\end{split}
\end{align}
Here, $\CB_1$ and $\CB_2$ are two topological boundary conditions of $\mathrm{TV}_{\CC}$. The notation $\qd_X$ denotes the quantum dimension of a topological line interface $X$ \cite{Diatlyk:2023fwf}. The red line $\mu$ is a bulk topological line of  $\mathrm{TV}_{\CC}$, and $S_{\mu\nu}$ is the modular S-matrix of the SymTFT. The black lines $a$ and $b$ are topological lines supported on the topological boundaries $\CB_1$ and $\CB_2$, respectively. The blue lines $\alpha, \beta$ are topological boundary-changing line interfaces between $\CB_1$ and $\CB_2$. The $x,y,z,w$ are topological junctions between various lines (barred junctions are dual junctions). The $\times$ marks on the $\beta$ are there for technical purposes that we explain in \cite{Choi:2024tri}, but they will not be important for the purposes of this Letter.
These numbers generalize the ordinary half-linking numbers in \cite{Lin:2022dhv, Huang:2023pyk, PhysRevB.103.195155, Zhang:2023wlu}, which correspond to the special case that $a=b=1$.

The two generalized half-linking numbers $\Psi$ and $\widetilde{\Psi}$ are not completely independent, but they are inverses of each other in the sense that they satisfy the following orthogonality relations: 
\begin{align}\label{eqn:halflinkingorthogonality}
\begin{split}
    \sum_{\alpha\beta z w}{^{\mathcal{B}_1\mathcal{B}_2}}\Psi_{\alpha\beta(\mu x y)}^{(az)(bw)} {^{\mathcal{B}_1\mathcal{B}_2}}\widetilde{\Psi}_{\alpha\beta(\mu' x' y')}^{(az)(bw)} &=  \delta_{\mu\mu'} \delta_{xx'} \delta_{yy'} \,, \\ 
   \sum_{\mu x y}{^{\mathcal{B}_1\mathcal{B}_2}}\Psi_{\alpha\beta(\mu x y)}^{(az)(bw)} {^{\mathcal{B}_1\mathcal{B}_2}}\widetilde{\Psi}_{\alpha' \beta' (\mu x y)}^{(az')(bw')} &= \delta_{\alpha\alpha'}\delta_{\beta\beta'}\delta_{zz'}\delta_{ww'} \,.
\end{split}
\end{align}
From the definition \eqref{eqn:halflinkingdefn}, one can derive the relations
\begin{align}\label{eqn:halflinkingid1}
\begin{split}
    \begin{tikzpicture}
{\small
	\begin{pgfonlayer}{nodelayer}
		\node [style=none] (0) at (-2, 0) {};
		\node [style=none] (1) at (2, 0) {};
		\node [style=none] (4) at (0, 2.5) {};
		\node [style=none] (5) at (0, -2.5) {};
		\node [style=none] (6) at (0, 1) {};
		\node [style=none] (7) at (0, -1) {};
		\node [style=none] (8) at (1, 1) {$\mu$};
		\node [style=none] (9) at (-0.4, 1) {$\bar{x}$};
		\node [style=none] (10) at (-0.4, -1) {$y$};
		\node [style=none] (11) at (-1.5, 0.5) {$\alpha$};
		\node [style=none] (12) at (0.5, 1.75) {$a$};
		\node [style=none] (13) at (0.5, -1.75) {$b$};
        \node [style=none] (14) at (-1.5, 2) {$\mathcal{B}_1$};
        \node [style=none] (15) at (-1.5, -2) {$\mathcal{B}_2$};
	\end{pgfonlayer}
	\begin{pgfonlayer}{edgelayer}
		\draw[thick, blue, decoration = {markings, mark=at position -0.45 with {\arrow[scale=1.5]{stealth}}}, postaction=decorate] (1.center) to (0.center);
        \draw [thick, decoration = {markings, mark=at position .7 with {\arrow[scale=1.5]{stealth}}}, postaction=decorate] (4.center) to (6.center);
        \draw [thick, decoration = {markings, mark=at position .7 with {\arrow[scale=1.5]{stealth}}}, postaction=decorate] (7.center) to (5.center);
		\draw[thick, preaction={draw=white,line width=6pt},decoration = {markings, mark=at position .7 with {\arrow[scale=1.5]{stealth}}}, postaction=decorate] [style=bulklines, bend left=75, looseness=1.50] (6.center) to (7.center);
	\end{pgfonlayer}
 }
\end{tikzpicture} &= \sum_{\beta z w } \frac{{^{\mathcal{B}_1\mathcal{B}_2}}\Psi_{\alpha\beta(\mu x y)}^{(az)(bw)}}{\sqrt{S_{11} \qd_\alpha^2}} ~~ \begin{tikzpicture}
{\small
	\begin{pgfonlayer}{nodelayer}
		\node [style=none] (0) at (-2, 0) {};
		\node [style=none] (1) at (2, 0) {};
		\node [style=none] (2) at (0, 2.25) {};
		\node [style=none] (3) at (0, -2.25) {};
		\node [style=none] (4) at (1, 0) {};
		\node [style=none] (5) at (-1, 0) {};
		\node [style=none] (6) at (0, -0.6) {$\beta$};
		\node [style=none] (7) at (-1, 0.5) {$\bar{w}$};
		\node [style=none] (8) at (1, -0.5) {$z$};
		\node [style=none] (9) at (-1.75, -0.5) {$\alpha$};
		\node [style=none] (10) at (1.2, 1.25) {$a$};
		\node [style=none] (11) at (-1.2, -1.25) {$b$};
		\node [style=none] (12) at (1.75, -0.5) {$\alpha$};
  \node [style=none] (13) at (1.5, -2) {$\mathcal{B}_2$};
  \node [style=none] (14) at (-1.5, 2) {$\mathcal{B}_1$};
  \node [style=none] (15) at (0.7, 0) {$\times$};
		\node [style=none] (16) at (-0.7, 0) {$\times$};
	\end{pgfonlayer}
	\begin{pgfonlayer}{edgelayer}
		\draw[thick, blue, decoration = {markings,mark=at position -.85 with {\arrow[scale=1.5]{stealth}},mark=at position -.07 with {\arrow[scale=1.5]{stealth}}, mark=at position -0.45 with {\arrow[scale=1.5]{stealth}}}, postaction=decorate] (1.center) to (0.center);
        \draw [thick, decoration = {markings, mark=at position .7 with {\arrow[scale=1.5]{stealth}}}, postaction=decorate] (5) to (3.center);
        \draw [thick, decoration = {markings, mark=at position .7 with {\arrow[scale=1.5]{stealth}}}, postaction=decorate]  (2) to (4);
	\end{pgfonlayer}
 }
\end{tikzpicture} \,,\\
    \begin{tikzpicture}
{\small
	\begin{pgfonlayer}{nodelayer}
		\node [style=none] (0) at (-2, 0) {};
		\node [style=none] (1) at (2, 0) {};
		\node [style=none] (4) at (0, 2.5) {};
		\node [style=none] (5) at (0, -2.5) {};
		\node [style=none] (6) at (0, 1) {};
		\node [style=none] (7) at (0, -1) {};
		\node [style=none] (8) at (1, 1) {$\mu$};
		\node [style=none] (9) at (-0.4, 1) {$x$};
		\node [style=none] (10) at (-0.4, -1) {$\bar{y}$};
		\node [style=none] (11) at (-1.5, 0.5) {$\alpha$};
		\node [style=none] (12) at (0.5, 1.75) {$a$};
		\node [style=none] (13) at (0.5, -1.75) {$b$};
        \node [style=none] (14) at (-1.5, 2) {$\mathcal{B}_1$};
        \node [style=none] (15) at (-1.5, -2) {$\mathcal{B}_2$};
	\end{pgfonlayer}
	\begin{pgfonlayer}{edgelayer}
		\draw[thick, blue, decoration = {markings, mark=at position -0.45 with {\arrow[scale=1.5]{stealth}}}, postaction=decorate] (1.center) to (0.center);
        \draw [thick, decoration = {markings, mark=at position .7 with {\arrow[scale=1.5]{stealth[reversed]}}}, postaction=decorate] (4.center) to (6.center);
        \draw [thick, decoration = {markings, mark=at position .7 with {\arrow[scale=1.5]{stealth[reversed]}}}, postaction=decorate] (7.center) to (5.center);
		\draw[thick, preaction={draw=white,line width=6pt},decoration = {markings, mark=at position .7 with {\arrow[scale=1.5]{stealth[reversed]}}}, postaction=decorate] [style=bulklines, bend left=75, looseness=1.50] (6.center) to (7.center);
	\end{pgfonlayer}
 }
\end{tikzpicture} &= \sum_{\beta z w} \frac{{^{\mathcal{B}_1\mathcal{B}_2}}\widetilde{\Psi}_{\alpha\beta(\mu x y)}^{(az)(bw)}}{\sqrt{S_{11} \qd_\alpha^2}} ~~ \begin{tikzpicture}
{\small
	\begin{pgfonlayer}{nodelayer}
		\node [style=none] (0) at (-2, 0) {};
		\node [style=none] (1) at (2, 0) {};
		\node [style=none] (2) at (0, 2.25) {};
		\node [style=none] (3) at (0, -2.25) {};
		\node [style=none] (4) at (1, 0) {};
		\node [style=none] (5) at (-1, 0) {};
		\node [style=none] (6) at (0, -0.6) {$\beta$};
		\node [style=none] (7) at (-1.2, -0.5) {$\bar{z}$};
		\node [style=none] (8) at (1.2, 0.5) {$w$};
		\node [style=none] (9) at (-1.75, -0.5) {$\alpha$};
		\node [style=none] (10) at (-1.2, 1.25) {$a$};
		\node [style=none] (11) at (1.1, -1.25) {$b$};
		\node [style=none] (12) at (1.75, -0.5) {$\alpha$};
  \node [style=none] (13) at (1.5, -2) {$\mathcal{B}_2$};
  \node [style=none] (14) at (-1.5, 2) {$\mathcal{B}_1$};
  \node [style=none] (15) at (0.7, 0) {$\times$};
		\node [style=none] (16) at (-0.7, 0) {$\times$};
	\end{pgfonlayer}
	\begin{pgfonlayer}{edgelayer}
		\draw[thick, blue, decoration = {markings,mark=at position -.85 with {\arrow[scale=1.5]{stealth}},mark=at position -.07 with {\arrow[scale=1.5]{stealth}}, mark=at position -0.45 with {\arrow[scale=1.5]{stealth}}}, postaction=decorate] (1.center) to (0.center);
        \draw [thick, decoration = {markings, mark=at position .7 with {\arrow[scale=1.5]{stealth[reversed]}}}, postaction=decorate] (4) to (3.center);
        \draw [thick, decoration = {markings, mark=at position .7 with {\arrow[scale=1.5]{stealth[reversed]}}}, postaction=decorate]  (2) to (5);
	\end{pgfonlayer}
 }
\end{tikzpicture} \,,
\end{split}
\end{align}
which generalize Equation (C.5) of \cite{Lin:2022dhv}.
See \cite{Choi:2024tri} for details.

Combining \eqref{eqn:halflinkingid1} and \eqref{eqn:halflinkingorthogonality} and applying them within the SymTFT picture of boundary-changing local operators of a 1+1d CFT $Q$ (Figure \ref{fig:bdydefect.operatorsletter}), one obtains the following key identity,
\begin{equation}\label{eq:proj1}
\begin{split}
    &\sum_{\mu x y} \sqrt{S_{11} \qd_{\rho}^2 } {^{\CB_1 \CB_2}}\widetilde{\Psi}^{11}_{\rho \rho(\mu x y)}\begin{tikzpicture}[scale=.5]
{\scriptsize
	\begin{pgfonlayer}{nodelayer}
		\node [style=none] (0) at (-5, 1) {};
		\node [style=none] (1) at (-2, 4) {};
		\node [style=none] (2) at (2, 1) {};
		\node [style=none] (3) at (5, 4) {};
		\node [style=none] (4) at (-5, -3) {};
		\node [style=none] (5) at (2, -3) {};
		\node [style=none] (6) at (5, 0) {};
		\node [style=none] (7) at (-2, 0) {};
		\node [style=none] (8) at (-3.5, 2.5) {};
		\node [style=none] (9) at (3.5, 2.5) {};
		\node [style=none] (10) at (1.5, 3) {$\beta$};
		\node [style=none] (11) at (-1.25, 1.75) {};
		\node [style=none] (12) at (0.25, 3.25) {};
		\node [style=none] (17) at (-1.5, 1.25) {$\bar{x}$};
		\node [style=none] (18) at (0.75, 3.5) {$y$};
		\node [style=none] (21) at (0, 1.5) {$\mu$};
		\node [style=none] (22) at (-4.75, 2) {$\underline{B}_1$};
		\node [style=none] (23) at (-3.25, 3.5) {$\underline{B}_2$};
         \node [style=none] (26) at (-3.5, 2.5) {};
         \node [style=none] (27) at (3.5, 2.5) {};
	\end{pgfonlayer}
	\begin{pgfonlayer}{edgelayer}
  \draw[decoration = {markings, mark=at position -0.6 with {\arrow[scale=1]{stealth}}}, postaction=decorate] [style=bulklines, bend right=45, looseness=1.25] (11) to (12);
		\draw[preaction={draw=white,line width=4pt},decoration = {markings, mark=at position -0.65 with {\arrow[scale=1]{stealth}}}, postaction=decorate] [style=boundarylines] (9) to (8);
        \fill[fill=blue!40, opacity=.3] (4.center)--(0.center) -- (1.center) -- (7.center) -- cycle;
        \fill[fill=red!40,opacity=.3] (0.center)--(2.center) -- (9.center) -- (8.center) -- cycle;
        \fill[fill=orange!40,opacity=.3] (8.center)--(9.center) -- (3.center) -- (1.center) -- cycle;
        \fill[fill=teal!40,opacity=.3] (5.center)--(2.center) -- (3.center) -- (6.center) -- cycle;
        \draw[decoration = {markings, mark=at position -0.65 with {\arrow[scale=1]{stealth}}}, postaction=decorate] [style=boundarylines] (1.center) to (26.center);
        \draw[decoration = {markings, mark=at position -0.65 with {\arrow[scale=1]{stealth}}}, postaction=decorate] [style=boundarylines] (26.center) to (0.center);
        \draw[decoration = {markings, mark=at position -0.65 with {\arrow[scale=1]{stealth}}}, postaction=decorate] [style=boundarylines] [color=teal, thick] (3.center) to (27.center);
        \draw[decoration = {markings, mark=at position -0.65 with {\arrow[scale=1]{stealth}}}, postaction=decorate] [style=boundarylines] [color=teal, thick] (27.center) to (2.center);
	\end{pgfonlayer}
 }
\end{tikzpicture}\\&= \delta_{\rho \beta} \begin{tikzpicture}[scale=.5]
{\scriptsize
	\begin{pgfonlayer}{nodelayer}
		\node [style=none] (0) at (-5, 1) {};
		\node [style=none] (1) at (-2, 4) {};
		\node [style=none] (2) at (2, 1) {};
		\node [style=none] (3) at (5, 4) {};
		\node [style=none] (4) at (-5, -3) {};
		\node [style=none] (5) at (2, -3) {};
		\node [style=none] (6) at (5, 0) {};
		\node [style=none] (7) at (-2, 0) {};
		\node [style=none] (8) at (-3.5, 2.5) {};
		\node [style=none] (9) at (3.5, 2.5) {};
		\node [style=none] (10) at (1.5, 3) {$\rho$};
		\node [style=none] (11) at (-1.25, 1.75) {};
		\node [style=none] (12) at (0.25, 3.25) {};
		\node [style=none] (22) at (-4.75, 2) {$\underline{B}_1$};
		\node [style=none] (23) at (-3.25, 3.5) {$\underline{B}_2$};
         \node [style=none] (26) at (-3.5, 2.5) {};
         \node [style=none] (27) at (3.5, 2.5) {};
	\end{pgfonlayer}
	\begin{pgfonlayer}{edgelayer}
		\draw[preaction={draw=white,line width=4pt},decoration = {markings, mark=at position -0.65 with {\arrow[scale=1]{stealth}}}, postaction=decorate] [style=boundarylines] (9) to (8);
        \fill[fill=blue!40, opacity=.3] (4.center)--(0.center) -- (1.center) -- (7.center) -- cycle;
        \fill[fill=red!40,opacity=.3] (0.center)--(2.center) -- (9.center) -- (8.center) -- cycle;
        \fill[fill=orange!40,opacity=.3] (8.center)--(9.center) -- (3.center) -- (1.center) -- cycle;
        \fill[fill=teal!40,opacity=.3] (5.center)--(2.center) -- (3.center) -- (6.center) -- cycle;
        \draw[decoration = {markings, mark=at position -0.65 with {\arrow[scale=1]{stealth}}}, postaction=decorate] [style=boundarylines] (1.center) to (26.center);
        \draw[decoration = {markings, mark=at position -0.65 with {\arrow[scale=1]{stealth}}}, postaction=decorate] [style=boundarylines] (26.center) to (0.center);
        \draw[decoration = {markings, mark=at position -0.65 with {\arrow[scale=1]{stealth}}}, postaction=decorate] [style=boundarylines] [color=teal, thick] (3.center) to (27.center);
        \draw[decoration = {markings, mark=at position -0.65 with {\arrow[scale=1]{stealth}}}, postaction=decorate] [style=boundarylines] [color=teal, thick] (27.center) to (2.center);
	\end{pgfonlayer}
 }
\end{tikzpicture}\,.
\end{split}
\end{equation}
See Supplemental Material for further details.
We hasten to add that the quantum dimension $\qd_{\rho}$ generically differs from the dimension of the representation $\rho$. In particular, $\qd_{\rho}$ is not necessarily an integer.

Recall that the symmetry lines of $Q$ live on the topological Dirichlet boundary condition in the SymTFT picture. It follows that if we push the $\mu$ line appearing in this identity onto the Dirichlet topological boundary on the left, then we obtain a 2+1d presentation of (a linear combination of) lasso operators which act on the Hilbert space $\mathcal{H}_{B_1B_2}$.
Equation \eqref{eq:proj1} can then be read as saying that this combination of lassos \emph{is} the desired projector onto the representation labeled by $\rho$.

We can determine the coefficients of the linear combination obtained in this manner by using the completeness and orthogonality relations obeyed by bulk and boundary lines (and also special properties of generalized half-linking matrices, see below), to find that
\begin{widetext}
    \begin{equation}\label{eq:proj3}
    \begin{split}
        \begin{tikzpicture}[scale=.5]
{\scriptsize
	\begin{pgfonlayer}{nodelayer}
		\node [style=none] (0) at (-5, 1) {};
		\node [style=none] (1) at (-2, 4) {};
		\node [style=none] (2) at (2, 1) {};
		\node [style=none] (3) at (5, 4) {};
		\node [style=none] (4) at (-5, -3) {};
		\node [style=none] (5) at (2, -3) {};
		\node [style=none] (6) at (5, 0) {};
		\node [style=none] (7) at (-2, 0) {};
		\node [style=none] (8) at (-3.5, 2.5) {};
		\node [style=none] (9) at (3.5, 2.5) {};
		\node [style=none] (10) at (1.5, 3) {$\beta$};
		\node [style=none] (11) at (-1.25, 1.75) {};
		\node [style=none] (12) at (0.25, 3.25) {};
		\node [style=none] (17) at (-1.5, 1.25) {$\bar{x}$};
		\node [style=none] (18) at (0.75, 3.5) {$y$};
		\node [style=none] (21) at (0, 1.5) {$\mu$};
		\node [style=none] (22) at (-4.75, 2) {$\underline{B}_1$};
		\node [style=none] (23) at (-3.25, 3.5) {$\underline{B}_2$};
         \node [style=none] (26) at (-3.5, 2.5) {};
         \node [style=none] (27) at (3.5, 2.5) {};
	\end{pgfonlayer}
	\begin{pgfonlayer}{edgelayer}
  \draw[decoration = {markings, mark=at position -0.6 with {\arrow[scale=1]{stealth}}}, postaction=decorate] [style=bulklines, bend right=45, looseness=1.25] (11) to (12);
		\draw[preaction={draw=white,line width=4pt},decoration = {markings, mark=at position -0.65 with {\arrow[scale=1]{stealth}}}, postaction=decorate] [style=boundarylines] (9) to (8);
        \fill[fill=blue!40, opacity=.3] (4.center)--(0.center) -- (1.center) -- (7.center) -- cycle;
        \fill[fill=red!40,opacity=.3] (0.center)--(2.center) -- (9.center) -- (8.center) -- cycle;
        \fill[fill=orange!40,opacity=.3] (8.center)--(9.center) -- (3.center) -- (1.center) -- cycle;
        \fill[fill=teal!40,opacity=.3] (5.center)--(2.center) -- (3.center) -- (6.center) -- cycle;
        \draw[decoration = {markings, mark=at position -0.65 with {\arrow[scale=1]{stealth}}}, postaction=decorate] [style=boundarylines] (1.center) to (26.center);
        \draw[decoration = {markings, mark=at position -0.65 with {\arrow[scale=1]{stealth}}}, postaction=decorate] [style=boundarylines] (26.center) to (0.center);
        \draw[decoration = {markings, mark=at position -0.65 with {\arrow[scale=1]{stealth}}}, postaction=decorate] [style=boundarylines] [color=teal, thick] (3.center) to (27.center);
        \draw[decoration = {markings, mark=at position -0.65 with {\arrow[scale=1]{stealth}}}, postaction=decorate] [style=boundarylines] [color=teal, thick] (27.center) to (2.center);
	\end{pgfonlayer}
 }
\end{tikzpicture}= \sum_{a x' z z'} \frac{\sqrt{\qd_a} {^{\CB_1 \CB_{\mathrm{reg}}}}{\Psi}^{1 (az)}_{\underline{B}_1 \underline{B}_1 (\mu x x')} {^{\CB_2 \CB_{\mathrm{reg}}}}\widetilde{\Psi}^{1 (az')}_{\underline{B}_2 \underline{B}_2 (\mu y x')} }{\sqrt{\qd_{\mu}} \qd_{\underline{B}_1} \qd_{\underline{B}_2} S_{11} }   \begin{tikzpicture}[scale=.5]
{\scriptsize
	\begin{pgfonlayer}{nodelayer}
		\node [style=none] (0) at (-5, 1) {};
		\node [style=none] (1) at (-2, 4) {};
		\node [style=none] (2) at (2, 1) {};
		\node [style=none] (3) at (5, 4) {};
		\node [style=none] (4) at (-5, -3) {};
		\node [style=none] (5) at (2, -3) {};
		\node [style=none] (6) at (5, 0) {};
		\node [style=none] (7) at (-2, 0) {};
		\node [style=none] (8) at (-3.5, 2.5) {};
		\node [style=none] (9) at (3.5, 2.5) {};
		\node [style=none] (10) at (1.5, 3) {$\beta$};
		\node [style=none] (11) at (-1.25, 1.75) {};
		\node [style=none] (12) at (0.25, 3.25) {};
		\node [style=none] (22) at (-4.75, 2) {$\underline{B}_1$};
		\node [style=none] (23) at (-3.25, 3.5) {$\underline{B}_2$};
         \node [style=none] (26) at (-3.5, 2.5) {};
         \node [style=none] (27) at (3.5, 2.5) {};
         \node [style=none] (28) at (-4, 2) {};
		\node [style=none] (29) at (-3+0.2, 3+0.2) {};
        \node[right] at (29.center) {$z'$};
        \node[below] at (28.center) {$\bar{z}$};
        \node[right] at (-3.3,2) {$a$};
	\end{pgfonlayer}
	\begin{pgfonlayer}{edgelayer}
        \draw[decoration = {markings, mark=at position -0.5 with {\arrow[scale=1]{stealth}}}, postaction=decorate] [thick, bend right=75, looseness=1.50] (28.center) to (29.center);
		\draw[preaction={draw=white,line width=4pt},decoration = {markings, mark=at position -0.65 with {\arrow[scale=1]{stealth}}}, postaction=decorate] [style=boundarylines] (9) to (8);
        \fill[fill=blue!40, opacity=.3] (4.center)--(0.center) -- (1.center) -- (7.center) -- cycle;
        \fill[fill=red!40,opacity=.3] (0.center)--(2.center) -- (9.center) -- (8.center) -- cycle;
        \fill[fill=orange!40,opacity=.3] (8.center)--(9.center) -- (3.center) -- (1.center) -- cycle;
        \fill[fill=teal!40,opacity=.3] (5.center)--(2.center) -- (3.center) -- (6.center) -- cycle;
        \draw[decoration = {markings, mark=at position -0.65 with {\arrow[scale=1]{stealth}}}, postaction=decorate] [style=boundarylines] (1.center) to (26.center);
        \draw[decoration = {markings, mark=at position -0.65 with {\arrow[scale=1]{stealth}}}, postaction=decorate] [style=boundarylines] (26.center) to (0.center);
        \draw[decoration = {markings, mark=at position -0.65 with {\arrow[scale=1]{stealth}}}, postaction=decorate] [style=boundarylines] [color=teal, thick] (3.center) to (27.center);
        \draw[decoration = {markings, mark=at position -0.65 with {\arrow[scale=1]{stealth}}}, postaction=decorate] [style=boundarylines] [color=teal, thick] (27.center) to (2.center);
	\end{pgfonlayer}
 }
\end{tikzpicture}
    \end{split}
\end{equation}
Combining \eqref{eq:proj1} and \eqref{eq:proj3}, the projector $P^\rho_{B_1B_2}:\mathcal{H}_{B_1B_2}\to\mathcal{H}^\rho_{B_1B_2}$ takes the form in Equation \eqref{eq:projfinal}.
\end{widetext}

\appendix

\section{Supplemental Material}

\subsubsection{Review of symmetry-resolved entanglement entropy}
\label{app:EE}

In this appendix, we review the definition of symmetry-resolved entanglement entropy of a 1+1d CFT and its relation to the annulus partition function, following \cite{Ohmori:2014eia, Kusuki:2023bsp}. See also \cite{Calabrese:2009qy, Monkman:2023hup, Goldstein:2017bua, Xavier:2018kqb, Calabrese:2021wvi, DiGiulio:2022jjd}. 

Consider the ground state $\ket{\Omega}$ of the CFT on an infinite line. 
We wish to bipartition the space into two parts $A\cup \bar{A}$, and define the reduced density matrix $\rho_A$ by taking the partial trace of $\ket{\Omega}\bra{\Omega}$ over the Hilbert space on region $\bar{A}$. However, in continuum field theories, the Hilbert space does not factorize as the tensor product $\CH_A \otimes \CH_{\bar{A}}$ \cite{Witten:2018zxz}. 
Instead, one may define an explicit ``cutting'' operation \cite{Ohmori:2014eia},
\begin{equation}
    \iota : \mathcal{H} \rightarrow \CH_{A; B_1 B_2} \otimes \CH_{\bar{A}; B_1 B_2} \,,
\end{equation}
which maps the original unfactorized CFT Hilbert space to an explicitly factorized one.
The map $\iota$ is defined by a ``topology-changing'' Euclidean path integral where the (Euclidean) time slice evolves from an infinite line to a disconnected space where a finite interval $A$ is cut out from the rest of the space.
Here, $B_1$ and $B_2$ are the boundary conditions imposed at the entangling surface which explicitly enter in the geometry of this Euclidean path integral. 
The Hilbert space $\CH_{A; B_1 B_2}$ is identified with the interval Hilbert space $\CH_{B_1 B_2}$ of the CFT with the boundary conditions $B_1$ and $B_2$ imposed at the two endpoints of the interval.
See \cite[Figure 1]{Ohmori:2014eia} for the case where $B_1$ and $B_2$ are chosen to be the same boundary conditions (denoted as $a$ there).

With an explicit cutting map $\iota$ in hand, we define the reduced density matrix of the subregion $A$ by
\begin{equation}
    \varrho_{A} := \Tr_{\CH_{\bar{A}; B_1 B_2}} \iota \ket{\Omega} \bra{\Omega} \iota^\dagger \,,
\end{equation}
which is represented by a path integral on a plane, with two small disks and a thin strip connecting them excised.
The boundary conditions $B_1$ and $B_2$ are imposed on the boundary of the two disks, respectively.
By a conformal transformation, the reduced density matrix is given by 
\begin{equation}
    \varrho_A = \frac{q^{L_0-\frac{c}{24}}}{Z_{B_1 B_2}(q)}
\end{equation}
where $q= e^{-\pi^2/\log(L/\varepsilon)}$ \cite{Kusuki:2023bsp,Ohmori:2014eia}, and $Z_{B_1 B_2}(q)= \Tr_{\CH_{B_1 B_2}} q^{L_0-\frac{c}{24}}$. Here, $L$ is the length of the subregion $A$, and $\varepsilon$ is the size of the entangling surface (i.e.\ the excised disks). The limit $L/\varepsilon\gg 1$ precisely corresponds to $q\to 1$, i.e.\ the high-$T$ limit.

The entanglement Hamiltonian is defined as 
\begin{equation} \label{eq:EH}
    H_{\mathrm{EE}} = - \log \varrho_A\,,
\end{equation}
and hence the boundary lasso operators discussed in the main text, which commute with the Hamiltonian of the CFT on an interval, give rise to symmetries of the entanglement Hamiltonian.

The $n$-th Renyi entropy is defined as
\begin{equation}\label{eq:Renyi}
\begin{split}
    S_{n} &= \frac{1}{1-n} \log \Tr_{\CH_{A;B_1 B_2}} \varrho_A^n= \frac{1}{1-n} \log \frac{Z_{B_1 B_2}(q^n)}{[Z_{B_1 B_2}(q)]^n}\,.
\end{split}
\end{equation}
The entanglement entropy, defined as the von Neumann entropy of the reduced density matrix, is reproduced in the $n\to 1$ limit of $S_n$, 
\begin{equation}
    S_{\mathrm{EE}} = - \Tr_{\CH_{A;B_1 B_2}} (\varrho_{A} \log \varrho_{A})= \lim_{n\to 1} S_n\,.
\end{equation}

We proceed to discuss the symmetry-resolved entanglement entropy (SREE) and symmetry-resolved Renyi entropy (SRRE). 
The SRRE is defined by simply replacing the Hilbert space $\CH_{A; B_1 B_2}$ in \eqref{eq:Renyi} by the sub-Hilbert space $\CH_{A; B_1 B_2}^\rho$ transforming under a fixed representation $\rho$ of the boundary tube algebra.
That is,
\begin{equation}
    S_n^\rho= \frac{1}{1-n} \log \Tr_{\CH^\rho_{A;B_1 B_2}} \varrho_{A,\rho}^n= \frac{1}{1-n} \log \frac{Z^\rho_{B_1 B_2}(q^n)}{[Z^\rho_{B_1 B_2}(q)]^n} \,,
\end{equation}
where
\begin{equation}
    \varrho_{A,\rho} = \frac{\varrho_A |_{\CH_{A; B_1 B_2}^\rho}}{\Tr_{\CH^\rho_{A;B_1 B_2}}\varrho_A |_{\CH_{A; B_1 B_2}^\rho}}
\end{equation}
is the reduced density matrix restricted to the subspace $\CH_{A; B_1 B_2}^\rho$, appropriately normalized.
The SREE is given by the $n\to 1$ limit,
\begin{equation}
    S_{\mathrm{EE}}^\rho = \lim_{n\to 1} S_n^\rho\,.
\end{equation}

\subsubsection{Symmetric boundary conditions at the entangling surface} \label{app:bc}

Suppose the 1+1d CFT has an ordinary finite group symmetry $G$, and admits a $G$-symmetric conformal boundary condition.
To compute the SREE with respect to the symmetry $G$, one chooses the boundary condition at the entangling surface to be $G$-invariant \cite{Kusuki:2023bsp}.
Then, the entanglement Hamiltonian also possesses the $G$ symmetry, and the entanglement spectrum organizes into the representations of $G$ \footnote{If one imposes two distinct $G$-symmetric boundary conditions at the two entangling points, it can also happen that the entanglement spectrum organizes into projective representations of $G$. See, for instance, \cite{Cordova:2022lms}.}.
On the other hand, if one chooses the boundary condition to preserve only a subgroup $H \subset G$, then the entanglement Hamiltonian has a smaller symmetry $H$.

When computing SREE with respect to a noninvertible symmetry, there appears to be more flexibility in how one chooses the boundary conditions at the entangling surface. First, as we describe below, there are different notions of what it means for a boundary to be symmetric. Second, it is not clear that one is really required to impose any particular symmetry assumption on the boundaries at the entangling surface at all: indeed, within our formalism it is perfectly consistent to choose any boundary condition one wants, and content oneself with studying the SREE with respect to whichever representations $\rho$ of the appropriate boundary tube algebra that arise in the entanglement spectrum.

Nevertheless, let us focus on the cases that the entangling boundaries are symmetric in some sense. Following \cite{Choi:2023xjw}, we call a conformal boundary condition of a 1+1d CFT \emph{strongly symmetric} under the action of a fusion category symmetry $\mathcal{C}$ if the corresponding Cardy boundary state \cite{Cardy:2004hm} is an eigenstate of the symmetry operators in $\mathcal{C}$. 
On the other hand, we call it \emph{weakly symmetric} if every topological line in $\mathcal{C}$ can topologically end on the boundary.
A strongly symmetric boundary condition is necessarily weakly symmetric, but the converse is generally not true for noninvertible symmetries. 
See \cite{Choi:2023xjw} for more details.
For simplicity, below we focus on the case where the two boundary conditions $B_1$ and $B_2$ at the entangling surface, which consists of two points, are the same, $B_1 = B_2 = B$.

\emph{Strongly symmetric entangling surface.}
First, consider choosing the boundary condition $B$ to be strongly symmetric under $\mathcal{C}$.
The action of a topological line $c$ in $\mathcal{C}$ on the Cardy boundary state $\ket{B}$ is given by $c \ket{B} = \qd_c \ket{B}$, where $\ket{B}$ is an eigenstate with the eigenvalue given by the quantum dimension $\qd_c$ of $c$.
When a fusion category $\mathcal{C}$ admits a strongly symmetric boundary condition, the quantum dimension $\qd_c$ is a non-negative integer for all $c$, and moreover, it coincides with the number of linearly-independent topological point junction operators between $c$ and $B$ \cite{Choi:2023xjw}.

Therefore, each topological line $c$ leads to $\qd_c^2$ number of boundary lasso operators $\mathsf{H}_{BB,c}^{BB,y_1 y_2}$ where $y_1, y_2 = 1, 2, \cdots, \qd_c$.
The dimension of the boundary tube algebra (as a complex vector space) is then equal to the square of the total dimension of the fusion category, $\mathrm{dim}(\mathcal{C})^2= \sum_c \qd_c^2$.

\begin{figure*}
\centering
    \input{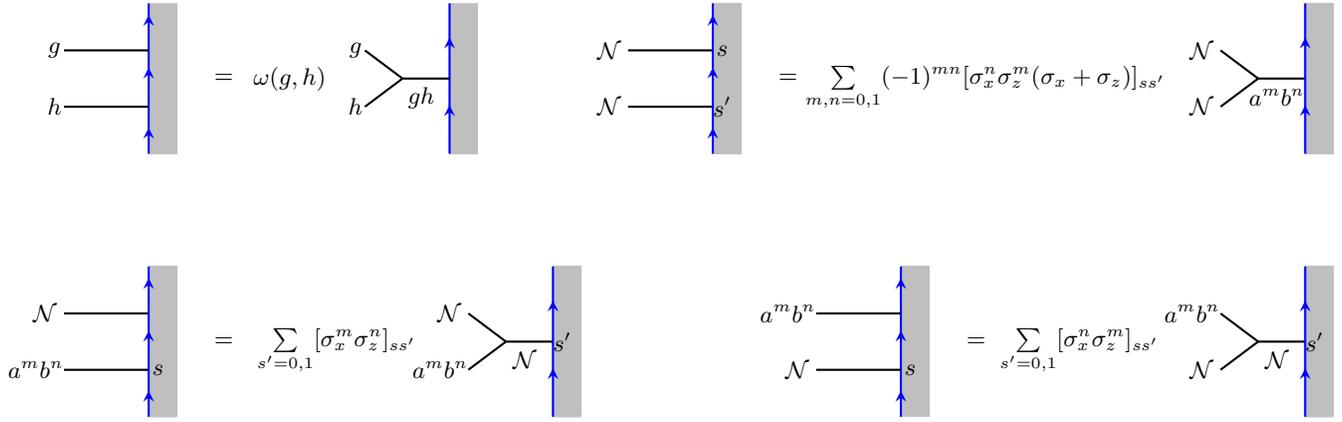}
    \caption{Boundary $\widetilde{F}$-symbols for the fusion category $\mathcal{C}_{\mathsf{TY}} \cong \mathrm{Rep}(H_8)$ acting on the (strongly) symmetric boundary condition $B$. Here, $\omega$ is a $\mathbb{Z}_2 \times \mathbb{Z}_2$ group 2-cocycle given by $\omega(a^{m_1} b^{n_1}, a^{m_2} b^{n_2}) = (-1)^{n_1 m_2}$, and $\sigma_i$'s are the Pauli matrices.}  \label{fig:H8F}
\end{figure*}

When a strongly symmetric boundary condition $B$ is imposed at the entangling surface, the entanglement spectrum satisfies the following ``completeness'' property.
In such a case, the fusion coefficient $N_{\rho\underline{B}}^{\underline{B}}$ appearing in the SREE formula (Equation (22) in the main text) is always non-negative, meaning that every representation $\rho$ of the boundary tube algebra appears in the entanglement spectrum.
More concretely, we have $N_{\rho\underline{B}}^{\underline{B}} = \qd_\rho > 0$.
Furthermore, $\qd_{\underline{B}} = \sqrt{\sum_c \qd_c^2} = \mathrm{dim}(\mathcal{C})$ \cite{Barter_2022,Choi:2024tri}.
The SREE in this case becomes
\begin{equation}
    S^\rho_{\mathrm{EE}} \sim \frac{c}{3} \log \frac{L}{\varepsilon} + \log g^2 + \log \frac{\qd_{\rho}^2 }{\mathrm{dim}(\mathcal{C})^2} \,,
\end{equation}
with $g$ the $g$-function of $B$.

\emph{Weakly symmetric entangling surface.} One may instead consider letting $B$ only weakly symmetric under the action of $\mathcal{C}$.
Then, it follows that every topological line in $\mathcal{C}$ leads to at least one boundary lasso operator acting on the interval Hilbert space $\mathcal{H}_{BB}$.
The collection of these lasso operators generally forms a subalgebra of the full boundary tube algebra acting on the extended Hilbert space (see \cite[Section 3.2]{Choi:2024tri}).
One may also perform symmetry resolution with respect to such a subalgebra, which we will not discuss in detail here.

For the case of a weakly symmetric $B$, it is not immediately obvious whether $N_{\rho\underline{B}}^{\underline{B}}$ is always nonzero so that every representation $\rho$ of the full boundary tube algebra appears in the entanglement spectrum at least once.
However, we note that, when the weakly symmetric boundary condition belongs to a regular module category, and furthermore the fusion rule of $\mathcal{C}$ is commutative, $N_{\rho\underline{B}}^{\underline{B}}$ is in fact always nonzero.
This is the case, for instance, for a boundary condition which is weakly symmetric under the Fibonacci fusion category symmetry.

\subsubsection{$H_8$ symmetry in the double Ising CFT} \label{app:H8}

Here, we derive the $H_8$ Hopf algebra symmetry of the double Ising CFT defined on an interval, with the boundary condition $B$.
We use the known module category structure of the multiplet formed by the boundary condition $B$ under the action of the bulk fusion category symmetry $\mathcal{C}_{\mathsf{TY}}$, and the associated $\widetilde{F}$-symbols.
For a detailed exposition on these concepts, we refer the readers to the companion paper \cite{Choi:2024tri}, or to \cite{Huang:2021zvu,Choi:2023xjw,Diatlyk:2023fwf}.

The nontrivial $\widetilde{F}$-symbols for the boundary condition $B$ are shown in Figure \ref{fig:H8F}.
They can be obtained from the data of the maximal algebra object in $\mathrm{Rep}(H_8)$, which was recently given in \cite{Choi:2023vgk,Diatlyk:2023fwf,Perez-Lona:2023djo}, by applying the (inverse) internal Hom construction \cite[Appendix A]{Choi:2023xjw}.
Using this, together with the standard $F$-symbols for $\mathcal{C}_{\mathsf{TY}} \cong \mathrm{Rep}(H_8)$ \cite{TAMBARA1998692,Bhardwaj:2017xup}, it is straightforward to compute the multiplication between the boundary lasso operators shown in Figure 2 in the main text.
We obtain \footnote{To be precise, the convention in Figure \ref{fig:H8F} gives us $\mathsf{H}_a \times \mathsf{H}_b = \mathsf{H}_b \times \mathsf{H}_a = - \mathsf{H}_{ab}$. 
By an appropriate gauge transformation of the topological junctions, we redefine $\mathsf{H}_{ab} \rightarrow - \mathsf{H}_{ab}$ and obtain \eqref{eq:H8app}. 
We thank the authors of \cite{Benini:2025lav} for pointing out an error in a previous version of Equation \eqref{eq:H8app}.}
\begin{align} \label{eq:H8app}
\begin{split}
    \mathsf{H}_g \times \mathsf{H}_h &= \mathsf{H}_h \times \mathsf{H}_g = \mathsf{H}_{gh} \,, \\
    \mathsf{H}_{\mathcal{N}}^{ss'} \times  \mathsf{H}_{a^m b^n} &= \sum_{s'' s'''}(-1)^{mn}[\sigma_x^n \sigma_z^m]_{s s''}   [\sigma_x^m \sigma_z^n]_{s' s'''} \mathsf{H}_{\mathcal{N}}^{s''  s'''} \,, \\
    \mathsf{H}_{a^m b^n} \times \mathsf{H}_{\mathcal{N}}^{ss'} &= \sum_{s'' s'''} (-1)^{mn}[\sigma_x^m \sigma_z^n]_{s s''} [\sigma_x^n \sigma_z^m]_{s' s'''}\mathsf{H}_{\mathcal{N}}^{s''  s'''} \,, \\
    \mathsf{H}_{\mathcal{N}}^{s_1 s_2} \times \mathsf{H}_{\mathcal{N}}^{s_3 s_4} &= 2 \sum_{m,n} (-1)^{f_{mn}(s_i)} \mathsf{H}_{a^m b^n} 
    \,,
\end{split}
\end{align}
where the $\sigma_i$ are the Pauli matrices, $g,h \in \mathbb{Z}_2 \times \mathbb{Z}_2 = \{1,a,b,ab \}$, and $f_{mn}(s_i) \equiv s_1 s_3 + s_2 s_4 + m(s_2 + s_3) + n (s_1 + s_4) + mn$.
The multiplication rules \eqref{eq:H8app} define the $H_8$ Hopf algebra symmetry of the double Ising CFT quantized on an interval with the strongly $\mathrm{Rep}(H_8)$-symmetric $B$ boundary condition imposed at the two endpoints of the interval.
For the full Hopf algebra structure involving comultiplication and antipode, see the general discussions in \cite{2012CMaPh.313..351K,Cordova:2024iti}.

The $H_8$ Hopf algebra is commonly presented in terms of three generators $x$, $y$, and $z$, satisfying the relations $x^2 = y^2=1$, $xy = yx$, $zx = yz$, $zy = xz$, and $z^2 = (1+x+y-xy)/2$ (see e.g.\ \cite[Section 4.2]{sage2012twisted}).
The precise relation between $x$, $y$, $z$ and the boundary lasso operators is
\begin{equation}
    x = \mathsf{H}_a \,, \quad y = \mathsf{H}_b\,, \quad z = \frac{1}{2} \mathsf{H}_{\mathcal{N}}^{00} \,.
\end{equation}

\bibliography{ref}

\begin{thebibliography}{114}%
\makeatletter
\providecommand \@ifxundefined [1]{%
 \@ifx{#1\undefined}
}%
\providecommand \@ifnum [1]{%
 \ifnum #1\expandafter \@firstoftwo
 \else \expandafter \@secondoftwo
 \fi
}%
\providecommand \@ifx [1]{%
 \ifx #1\expandafter \@firstoftwo
 \else \expandafter \@secondoftwo
 \fi
}%
\providecommand \natexlab [1]{#1}%
\providecommand \enquote  [1]{``#1''}%
\providecommand \bibnamefont  [1]{#1}%
\providecommand \bibfnamefont [1]{#1}%
\providecommand \citenamefont [1]{#1}%
\providecommand \href@noop [0]{\@secondoftwo}%
\providecommand \href [0]{\begingroup \@sanitize@url \@href}%
\providecommand \@href[1]{\@@startlink{#1}\@@href}%
\providecommand \@@href[1]{\endgroup#1\@@endlink}%
\providecommand \@sanitize@url [0]{\catcode `\\12\catcode `\$12\catcode `\&12\catcode `\#12\catcode `\^12\catcode `\_12\catcode `\%12\relax}%
\providecommand \@@startlink[1]{}%
\providecommand \@@endlink[0]{}%
\providecommand \url  [0]{\begingroup\@sanitize@url \@url }%
\providecommand \@url [1]{\endgroup\@href {#1}{\urlprefix }}%
\providecommand \urlprefix  [0]{URL }%
\providecommand \Eprint [0]{\href }%
\providecommand \doibase [0]{http://dx.doi.org/}%
\providecommand \selectlanguage [0]{\@gobble}%
\providecommand \bibinfo  [0]{\@secondoftwo}%
\providecommand \bibfield  [0]{\@secondoftwo}%
\providecommand \translation [1]{[#1]}%
\providecommand \BibitemOpen [0]{}%
\providecommand \bibitemStop [0]{}%
\providecommand \bibitemNoStop [0]{.\EOS\space}%
\providecommand \EOS [0]{\spacefactor3000\relax}%
\providecommand \BibitemShut  [1]{\csname bibitem#1\endcsname}%
\let\auto@bib@innerbib\@empty
\bibitem [{\citenamefont {Cardy}(1986)}]{Cardy:1986ie}%
  \BibitemOpen
  \bibfield  {author} {\bibinfo {author} {\bibfnamefont {J.~L.}\ \bibnamefont {Cardy}},\ }\href {\doibase 10.1016/0550-3213(86)90552-3} {\bibfield  {journal} {\bibinfo  {journal} {Nucl. Phys. B}\ }\textbf {\bibinfo {volume} {270}},\ \bibinfo {pages} {186} (\bibinfo {year} {1986})}\BibitemShut {NoStop}%
\bibitem [{Note1()}]{Note1}%
  \BibitemOpen
  \bibinfo {note} {We assume that the gravitational anomaly vanishes throughout, so that $c = \protect \overline {c}$.}\BibitemShut {Stop}%
\bibitem [{\citenamefont {Affleck}\ and\ \citenamefont {Ludwig}(1991{\natexlab{a}})}]{Affleck:1991tk}%
  \BibitemOpen
  \bibfield  {author} {\bibinfo {author} {\bibfnamefont {I.}~\bibnamefont {Affleck}}\ and\ \bibinfo {author} {\bibfnamefont {A.~W.~W.}\ \bibnamefont {Ludwig}},\ }\href {\doibase 10.1103/PhysRevLett.67.161} {\bibfield  {journal} {\bibinfo  {journal} {Phys. Rev. Lett.}\ }\textbf {\bibinfo {volume} {67}},\ \bibinfo {pages} {161} (\bibinfo {year} {1991}{\natexlab{a}})}\BibitemShut {NoStop}%
\bibitem [{\citenamefont {Cardy}(2004)}]{Cardy:2004hm}%
  \BibitemOpen
  \bibfield  {author} {\bibinfo {author} {\bibfnamefont {J.~L.}\ \bibnamefont {Cardy}},\ }\href@noop {} {\  (\bibinfo {year} {2004})},\ \Eprint {http://arxiv.org/abs/hep-th/0411189} {arXiv:hep-th/0411189} \BibitemShut {NoStop}%
\bibitem [{\citenamefont {Affleck}\ and\ \citenamefont {Ludwig}(1991{\natexlab{b}})}]{PhysRevLett.67.161}%
  \BibitemOpen
  \bibfield  {author} {\bibinfo {author} {\bibfnamefont {I.}~\bibnamefont {Affleck}}\ and\ \bibinfo {author} {\bibfnamefont {A.~W.~W.}\ \bibnamefont {Ludwig}},\ }\href {\doibase 10.1103/PhysRevLett.67.161} {\bibfield  {journal} {\bibinfo  {journal} {Phys. Rev. Lett.}\ }\textbf {\bibinfo {volume} {67}},\ \bibinfo {pages} {161} (\bibinfo {year} {1991}{\natexlab{b}})}\BibitemShut {NoStop}%
\bibitem [{\citenamefont {Friedan}\ and\ \citenamefont {Konechny}(2004)}]{Friedan:2003yc}%
  \BibitemOpen
  \bibfield  {author} {\bibinfo {author} {\bibfnamefont {D.}~\bibnamefont {Friedan}}\ and\ \bibinfo {author} {\bibfnamefont {A.}~\bibnamefont {Konechny}},\ }\href {\doibase 10.1103/PhysRevLett.93.030402} {\bibfield  {journal} {\bibinfo  {journal} {Phys. Rev. Lett.}\ }\textbf {\bibinfo {volume} {93}},\ \bibinfo {pages} {030402} (\bibinfo {year} {2004})},\ \Eprint {http://arxiv.org/abs/hep-th/0312197} {arXiv:hep-th/0312197} \BibitemShut {NoStop}%
\bibitem [{\citenamefont {Casini}\ \emph {et~al.}(2016)\citenamefont {Casini}, \citenamefont {Salazar~Landea},\ and\ \citenamefont {Torroba}}]{Casini:2016fgb}%
  \BibitemOpen
  \bibfield  {author} {\bibinfo {author} {\bibfnamefont {H.}~\bibnamefont {Casini}}, \bibinfo {author} {\bibfnamefont {I.}~\bibnamefont {Salazar~Landea}}, \ and\ \bibinfo {author} {\bibfnamefont {G.}~\bibnamefont {Torroba}},\ }\href {\doibase 10.1007/JHEP10(2016)140} {\bibfield  {journal} {\bibinfo  {journal} {JHEP}\ }\textbf {\bibinfo {volume} {10}},\ \bibinfo {pages} {140} (\bibinfo {year} {2016})},\ \Eprint {http://arxiv.org/abs/1607.00390} {arXiv:1607.00390 [hep-th]} \BibitemShut {NoStop}%
\bibitem [{Note2()}]{Note2}%
  \BibitemOpen
  \bibinfo {note} {A conformal boundary condition $B$ is simple if the CFT Hamiltonian on the interval Hilbert space $\protect \mathcal {H}_{BB}$ has a unique ground state. More generally, an extended object is called simple if the only topological point operator supported on it is the identity operator. Various extended objects and boundary conditions in this Letter are always assumed to be simple.}\BibitemShut {Stop}%
\bibitem [{\citenamefont {L\"auchli}(2013)}]{Lauchli:2013jga}%
  \BibitemOpen
  \bibfield  {author} {\bibinfo {author} {\bibfnamefont {A.~M.}\ \bibnamefont {L\"auchli}},\ }\href@noop {} {\  (\bibinfo {year} {2013})},\ \Eprint {http://arxiv.org/abs/1303.0741} {arXiv:1303.0741 [cond-mat.stat-mech]} \BibitemShut {NoStop}%
\bibitem [{\citenamefont {Ohmori}\ and\ \citenamefont {Tachikawa}(2015)}]{Ohmori:2014eia}%
  \BibitemOpen
  \bibfield  {author} {\bibinfo {author} {\bibfnamefont {K.}~\bibnamefont {Ohmori}}\ and\ \bibinfo {author} {\bibfnamefont {Y.}~\bibnamefont {Tachikawa}},\ }\href {\doibase 10.1088/1742-5468/2015/04/P04010} {\bibfield  {journal} {\bibinfo  {journal} {J. Stat. Mech.}\ }\textbf {\bibinfo {volume} {1504}},\ \bibinfo {pages} {P04010} (\bibinfo {year} {2015})},\ \Eprint {http://arxiv.org/abs/1406.4167} {arXiv:1406.4167 [hep-th]} \BibitemShut {NoStop}%
\bibitem [{\citenamefont {Cardy}\ and\ \citenamefont {Tonni}(2016)}]{Cardy:2016fqc}%
  \BibitemOpen
  \bibfield  {author} {\bibinfo {author} {\bibfnamefont {J.}~\bibnamefont {Cardy}}\ and\ \bibinfo {author} {\bibfnamefont {E.}~\bibnamefont {Tonni}},\ }\href {\doibase 10.1088/1742-5468/2016/12/123103} {\bibfield  {journal} {\bibinfo  {journal} {J. Stat. Mech.}\ }\textbf {\bibinfo {volume} {1612}},\ \bibinfo {pages} {123103} (\bibinfo {year} {2016})},\ \Eprint {http://arxiv.org/abs/1608.01283} {arXiv:1608.01283 [cond-mat.stat-mech]} \BibitemShut {NoStop}%
\bibitem [{\citenamefont {Bhardwaj}\ and\ \citenamefont {Tachikawa}(2018)}]{Bhardwaj:2017xup}%
  \BibitemOpen
  \bibfield  {author} {\bibinfo {author} {\bibfnamefont {L.}~\bibnamefont {Bhardwaj}}\ and\ \bibinfo {author} {\bibfnamefont {Y.}~\bibnamefont {Tachikawa}},\ }\href {\doibase 10.1007/JHEP03(2018)189} {\bibfield  {journal} {\bibinfo  {journal} {JHEP}\ }\textbf {\bibinfo {volume} {03}},\ \bibinfo {pages} {189} (\bibinfo {year} {2018})},\ \Eprint {http://arxiv.org/abs/1704.02330} {arXiv:1704.02330 [hep-th]} \BibitemShut {NoStop}%
\bibitem [{\citenamefont {Chang}\ \emph {et~al.}(2019)\citenamefont {Chang}, \citenamefont {Lin}, \citenamefont {Shao}, \citenamefont {Wang},\ and\ \citenamefont {Yin}}]{Chang:2018iay}%
  \BibitemOpen
  \bibfield  {author} {\bibinfo {author} {\bibfnamefont {C.-M.}\ \bibnamefont {Chang}}, \bibinfo {author} {\bibfnamefont {Y.-H.}\ \bibnamefont {Lin}}, \bibinfo {author} {\bibfnamefont {S.-H.}\ \bibnamefont {Shao}}, \bibinfo {author} {\bibfnamefont {Y.}~\bibnamefont {Wang}}, \ and\ \bibinfo {author} {\bibfnamefont {X.}~\bibnamefont {Yin}},\ }\href {\doibase 10.1007/JHEP01(2019)026} {\bibfield  {journal} {\bibinfo  {journal} {JHEP}\ }\textbf {\bibinfo {volume} {01}},\ \bibinfo {pages} {026} (\bibinfo {year} {2019})},\ \Eprint {http://arxiv.org/abs/1802.04445} {arXiv:1802.04445 [hep-th]} \BibitemShut {NoStop}%
\bibitem [{\citenamefont {Thorngren}\ and\ \citenamefont {Wang}(2019)}]{Thorngren:2019iar}%
  \BibitemOpen
  \bibfield  {author} {\bibinfo {author} {\bibfnamefont {R.}~\bibnamefont {Thorngren}}\ and\ \bibinfo {author} {\bibfnamefont {Y.}~\bibnamefont {Wang}},\ }\href@noop {} {\  (\bibinfo {year} {2019})},\ \Eprint {http://arxiv.org/abs/1912.02817} {arXiv:1912.02817 [hep-th]} \BibitemShut {NoStop}%
\bibitem [{\citenamefont {Thorngren}\ and\ \citenamefont {Wang}(2021)}]{Thorngren:2021yso}%
  \BibitemOpen
  \bibfield  {author} {\bibinfo {author} {\bibfnamefont {R.}~\bibnamefont {Thorngren}}\ and\ \bibinfo {author} {\bibfnamefont {Y.}~\bibnamefont {Wang}},\ }\href@noop {} {\  (\bibinfo {year} {2021})},\ \Eprint {http://arxiv.org/abs/2106.12577} {arXiv:2106.12577 [hep-th]} \BibitemShut {NoStop}%
\bibitem [{\citenamefont {Islam}\ \emph {et~al.}(2015)\citenamefont {Islam}, \citenamefont {Ma}, \citenamefont {Preiss}, \citenamefont {Eric~Tai}, \citenamefont {Lukin}, \citenamefont {Rispoli},\ and\ \citenamefont {Greiner}}]{islam2015measuring}%
  \BibitemOpen
  \bibfield  {author} {\bibinfo {author} {\bibfnamefont {R.}~\bibnamefont {Islam}}, \bibinfo {author} {\bibfnamefont {R.}~\bibnamefont {Ma}}, \bibinfo {author} {\bibfnamefont {P.~M.}\ \bibnamefont {Preiss}}, \bibinfo {author} {\bibfnamefont {M.}~\bibnamefont {Eric~Tai}}, \bibinfo {author} {\bibfnamefont {A.}~\bibnamefont {Lukin}}, \bibinfo {author} {\bibfnamefont {M.}~\bibnamefont {Rispoli}}, \ and\ \bibinfo {author} {\bibfnamefont {M.}~\bibnamefont {Greiner}},\ }\href@noop {} {\bibfield  {journal} {\bibinfo  {journal} {Nature}\ }\textbf {\bibinfo {volume} {528}},\ \bibinfo {pages} {77} (\bibinfo {year} {2015})}\BibitemShut {NoStop}%
\bibitem [{\citenamefont {Lukin}\ \emph {et~al.}(2019)\citenamefont {Lukin}, \citenamefont {Rispoli}, \citenamefont {Schittko}, \citenamefont {Tai}, \citenamefont {Kaufman}, \citenamefont {Choi}, \citenamefont {Khemani}, \citenamefont {L\'eonard},\ and\ \citenamefont {Greiner}}]{Lukin:2019tkq}%
  \BibitemOpen
  \bibfield  {author} {\bibinfo {author} {\bibfnamefont {A.}~\bibnamefont {Lukin}}, \bibinfo {author} {\bibfnamefont {M.}~\bibnamefont {Rispoli}}, \bibinfo {author} {\bibfnamefont {R.}~\bibnamefont {Schittko}}, \bibinfo {author} {\bibfnamefont {M.~E.}\ \bibnamefont {Tai}}, \bibinfo {author} {\bibfnamefont {A.~M.}\ \bibnamefont {Kaufman}}, \bibinfo {author} {\bibfnamefont {S.}~\bibnamefont {Choi}}, \bibinfo {author} {\bibfnamefont {V.}~\bibnamefont {Khemani}}, \bibinfo {author} {\bibfnamefont {J.}~\bibnamefont {L\'eonard}}, \ and\ \bibinfo {author} {\bibfnamefont {M.}~\bibnamefont {Greiner}},\ }\href {\doibase 10.1126/science.aau0818} {\bibfield  {journal} {\bibinfo  {journal} {Science}\ }\textbf {\bibinfo {volume} {364}},\ \bibinfo {pages} {aau0818} (\bibinfo {year} {2019})}\BibitemShut {NoStop}%
\bibitem [{\citenamefont {Pal}\ and\ \citenamefont {Sun}(2020)}]{Pal:2020wwd}%
  \BibitemOpen
  \bibfield  {author} {\bibinfo {author} {\bibfnamefont {S.}~\bibnamefont {Pal}}\ and\ \bibinfo {author} {\bibfnamefont {Z.}~\bibnamefont {Sun}},\ }\href {\doibase 10.1007/JHEP08(2020)064} {\bibfield  {journal} {\bibinfo  {journal} {JHEP}\ }\textbf {\bibinfo {volume} {08}},\ \bibinfo {pages} {064} (\bibinfo {year} {2020})},\ \Eprint {http://arxiv.org/abs/2004.12557} {arXiv:2004.12557 [hep-th]} \BibitemShut {NoStop}%
\bibitem [{\citenamefont {Lin}\ \emph {et~al.}(2022)\citenamefont {Lin}, \citenamefont {Okada}, \citenamefont {Seifnashri},\ and\ \citenamefont {Tachikawa}}]{Lin:2022dhv}%
  \BibitemOpen
  \bibfield  {author} {\bibinfo {author} {\bibfnamefont {Y.-H.}\ \bibnamefont {Lin}}, \bibinfo {author} {\bibfnamefont {M.}~\bibnamefont {Okada}}, \bibinfo {author} {\bibfnamefont {S.}~\bibnamefont {Seifnashri}}, \ and\ \bibinfo {author} {\bibfnamefont {Y.}~\bibnamefont {Tachikawa}},\ }\href@noop {} {\  (\bibinfo {year} {2022})},\ \Eprint {http://arxiv.org/abs/2208.05495} {arXiv:2208.05495 [hep-th]} \BibitemShut {NoStop}%
\bibitem [{\citenamefont {Lu}\ and\ \citenamefont {Sun}(2023)}]{Lu:2022ver}%
  \BibitemOpen
  \bibfield  {author} {\bibinfo {author} {\bibfnamefont {D.-C.}\ \bibnamefont {Lu}}\ and\ \bibinfo {author} {\bibfnamefont {Z.}~\bibnamefont {Sun}},\ }\href {\doibase 10.1007/JHEP02(2023)173} {\bibfield  {journal} {\bibinfo  {journal} {JHEP}\ }\textbf {\bibinfo {volume} {02}},\ \bibinfo {pages} {173} (\bibinfo {year} {2023})},\ \Eprint {http://arxiv.org/abs/2208.06077} {arXiv:2208.06077 [hep-th]} \BibitemShut {NoStop}%
\bibitem [{\citenamefont {Casini}\ \emph {et~al.}(2020)\citenamefont {Casini}, \citenamefont {Huerta}, \citenamefont {Mag\'an},\ and\ \citenamefont {Pontello}}]{Casini:2019kex}%
  \BibitemOpen
  \bibfield  {author} {\bibinfo {author} {\bibfnamefont {H.}~\bibnamefont {Casini}}, \bibinfo {author} {\bibfnamefont {M.}~\bibnamefont {Huerta}}, \bibinfo {author} {\bibfnamefont {J.~M.}\ \bibnamefont {Mag\'an}}, \ and\ \bibinfo {author} {\bibfnamefont {D.}~\bibnamefont {Pontello}},\ }\href {\doibase 10.1007/JHEP02(2020)014} {\bibfield  {journal} {\bibinfo  {journal} {JHEP}\ }\textbf {\bibinfo {volume} {02}},\ \bibinfo {pages} {014} (\bibinfo {year} {2020})},\ \Eprint {http://arxiv.org/abs/1905.10487} {arXiv:1905.10487 [hep-th]} \BibitemShut {NoStop}%
\bibitem [{\citenamefont {Mag\'{a}n}(2021)}]{Magan:2021myk}%
  \BibitemOpen
  \bibfield  {author} {\bibinfo {author} {\bibfnamefont {J.~M.}\ \bibnamefont {Mag\'{a}n}},\ }\href {\doibase 10.1007/JHEP12(2021)100} {\bibfield  {journal} {\bibinfo  {journal} {JHEP}\ }\textbf {\bibinfo {volume} {12}},\ \bibinfo {pages} {100} (\bibinfo {year} {2021})},\ \Eprint {http://arxiv.org/abs/2111.02418} {arXiv:2111.02418 [hep-th]} \BibitemShut {NoStop}%
\bibitem [{\citenamefont {Kusuki}\ \emph {et~al.}(2023)\citenamefont {Kusuki}, \citenamefont {Murciano}, \citenamefont {Ooguri},\ and\ \citenamefont {Pal}}]{Kusuki:2023bsp}%
  \BibitemOpen
  \bibfield  {author} {\bibinfo {author} {\bibfnamefont {Y.}~\bibnamefont {Kusuki}}, \bibinfo {author} {\bibfnamefont {S.}~\bibnamefont {Murciano}}, \bibinfo {author} {\bibfnamefont {H.}~\bibnamefont {Ooguri}}, \ and\ \bibinfo {author} {\bibfnamefont {S.}~\bibnamefont {Pal}},\ }\href {\doibase 10.1007/JHEP11(2023)216} {\bibfield  {journal} {\bibinfo  {journal} {JHEP}\ }\textbf {\bibinfo {volume} {11}},\ \bibinfo {pages} {216} (\bibinfo {year} {2023})},\ \Eprint {http://arxiv.org/abs/2309.03287} {arXiv:2309.03287 [hep-th]} \BibitemShut {NoStop}%
\bibitem [{\citenamefont {Harlow}\ and\ \citenamefont {Ooguri}(2022)}]{Harlow:2021trr}%
  \BibitemOpen
  \bibfield  {author} {\bibinfo {author} {\bibfnamefont {D.}~\bibnamefont {Harlow}}\ and\ \bibinfo {author} {\bibfnamefont {H.}~\bibnamefont {Ooguri}},\ }\href {\doibase 10.1088/1361-6382/ac5db2} {\bibfield  {journal} {\bibinfo  {journal} {Class. Quant. Grav.}\ }\textbf {\bibinfo {volume} {39}},\ \bibinfo {pages} {134003} (\bibinfo {year} {2022})},\ \Eprint {http://arxiv.org/abs/2109.03838} {arXiv:2109.03838 [hep-th]} \BibitemShut {NoStop}%
\bibitem [{\citenamefont {Cao}\ \emph {et~al.}(2024)\citenamefont {Cao}, \citenamefont {Melia},\ and\ \citenamefont {Pal}}]{Cao:2021euf}%
  \BibitemOpen
  \bibfield  {author} {\bibinfo {author} {\bibfnamefont {W.}~\bibnamefont {Cao}}, \bibinfo {author} {\bibfnamefont {T.}~\bibnamefont {Melia}}, \ and\ \bibinfo {author} {\bibfnamefont {S.}~\bibnamefont {Pal}},\ }\href {\doibase 10.1007/JHEP05(2024)031} {\bibfield  {journal} {\bibinfo  {journal} {JHEP}\ }\textbf {\bibinfo {volume} {05}},\ \bibinfo {pages} {031} (\bibinfo {year} {2024})},\ \Eprint {http://arxiv.org/abs/2111.04725} {arXiv:2111.04725 [hep-th]} \BibitemShut {NoStop}%
\bibitem [{\citenamefont {Melia}\ and\ \citenamefont {Pal}(2021)}]{Melia:2020pzd}%
  \BibitemOpen
  \bibfield  {author} {\bibinfo {author} {\bibfnamefont {T.}~\bibnamefont {Melia}}\ and\ \bibinfo {author} {\bibfnamefont {S.}~\bibnamefont {Pal}},\ }\href {\doibase 10.21468/SciPostPhys.10.5.104} {\bibfield  {journal} {\bibinfo  {journal} {SciPost Phys.}\ }\textbf {\bibinfo {volume} {10}},\ \bibinfo {pages} {104} (\bibinfo {year} {2021})},\ \Eprint {http://arxiv.org/abs/2010.08560} {arXiv:2010.08560 [hep-th]} \BibitemShut {NoStop}%
\bibitem [{\citenamefont {Kang}\ \emph {et~al.}(2023)\citenamefont {Kang}, \citenamefont {Lee},\ and\ \citenamefont {Ooguri}}]{Kang:2022orq}%
  \BibitemOpen
  \bibfield  {author} {\bibinfo {author} {\bibfnamefont {M.~J.}\ \bibnamefont {Kang}}, \bibinfo {author} {\bibfnamefont {J.}~\bibnamefont {Lee}}, \ and\ \bibinfo {author} {\bibfnamefont {H.}~\bibnamefont {Ooguri}},\ }\href {\doibase 10.1103/PhysRevD.107.026021} {\bibfield  {journal} {\bibinfo  {journal} {Phys. Rev. D}\ }\textbf {\bibinfo {volume} {107}},\ \bibinfo {pages} {026021} (\bibinfo {year} {2023})},\ \Eprint {http://arxiv.org/abs/2206.14814} {arXiv:2206.14814 [hep-th]} \BibitemShut {NoStop}%
\bibitem [{\citenamefont {Mukhametzhanov}\ and\ \citenamefont {Pal}(2020)}]{Mukhametzhanov:2020swe}%
  \BibitemOpen
  \bibfield  {author} {\bibinfo {author} {\bibfnamefont {B.}~\bibnamefont {Mukhametzhanov}}\ and\ \bibinfo {author} {\bibfnamefont {S.}~\bibnamefont {Pal}},\ }\href {\doibase 10.21468/SciPostPhys.8.6.088} {\bibfield  {journal} {\bibinfo  {journal} {SciPost Phys.}\ }\textbf {\bibinfo {volume} {8}},\ \bibinfo {pages} {088} (\bibinfo {year} {2020})},\ \Eprint {http://arxiv.org/abs/2003.14316} {arXiv:2003.14316 [hep-th]} \BibitemShut {NoStop}%
\bibitem [{\citenamefont {Benjamin}\ \emph {et~al.}(2024)\citenamefont {Benjamin}, \citenamefont {Lee}, \citenamefont {Ooguri},\ and\ \citenamefont {Simmons-Duffin}}]{Benjamin:2023qsc}%
  \BibitemOpen
  \bibfield  {author} {\bibinfo {author} {\bibfnamefont {N.}~\bibnamefont {Benjamin}}, \bibinfo {author} {\bibfnamefont {J.}~\bibnamefont {Lee}}, \bibinfo {author} {\bibfnamefont {H.}~\bibnamefont {Ooguri}}, \ and\ \bibinfo {author} {\bibfnamefont {D.}~\bibnamefont {Simmons-Duffin}},\ }\href {\doibase 10.1007/JHEP03(2024)115} {\bibfield  {journal} {\bibinfo  {journal} {JHEP}\ }\textbf {\bibinfo {volume} {03}},\ \bibinfo {pages} {115} (\bibinfo {year} {2024})},\ \Eprint {http://arxiv.org/abs/2306.08031} {arXiv:2306.08031 [hep-th]} \BibitemShut {NoStop}%
\bibitem [{\citenamefont {Bianchi}\ \emph {et~al.}(2024)\citenamefont {Bianchi}, \citenamefont {Dona},\ and\ \citenamefont {Kumar}}]{Bianchi:2024aim}%
  \BibitemOpen
  \bibfield  {author} {\bibinfo {author} {\bibfnamefont {E.}~\bibnamefont {Bianchi}}, \bibinfo {author} {\bibfnamefont {P.}~\bibnamefont {Dona}}, \ and\ \bibinfo {author} {\bibfnamefont {R.}~\bibnamefont {Kumar}},\ }\href@noop {} {\  (\bibinfo {year} {2024})},\ \Eprint {http://arxiv.org/abs/2405.00597} {arXiv:2405.00597 [quant-ph]} \BibitemShut {NoStop}%
\bibitem [{\citenamefont {Karch}\ \emph {et~al.}(2024)\citenamefont {Karch}, \citenamefont {Kusuki}, \citenamefont {Ooguri}, \citenamefont {Sun},\ and\ \citenamefont {Wang}}]{Karch:2024udk}%
  \BibitemOpen
  \bibfield  {author} {\bibinfo {author} {\bibfnamefont {A.}~\bibnamefont {Karch}}, \bibinfo {author} {\bibfnamefont {Y.}~\bibnamefont {Kusuki}}, \bibinfo {author} {\bibfnamefont {H.}~\bibnamefont {Ooguri}}, \bibinfo {author} {\bibfnamefont {H.-Y.}\ \bibnamefont {Sun}}, \ and\ \bibinfo {author} {\bibfnamefont {M.}~\bibnamefont {Wang}},\ }\href@noop {} {\  (\bibinfo {year} {2024})},\ \Eprint {http://arxiv.org/abs/2404.01515} {arXiv:2404.01515 [hep-th]} \BibitemShut {NoStop}%
\bibitem [{\citenamefont {Di~Giulio}\ \emph {et~al.}(2023)\citenamefont {Di~Giulio}, \citenamefont {Meyer}, \citenamefont {Northe}, \citenamefont {Scheppach},\ and\ \citenamefont {Zhao}}]{DiGiulio:2022jjd}%
  \BibitemOpen
  \bibfield  {author} {\bibinfo {author} {\bibfnamefont {G.}~\bibnamefont {Di~Giulio}}, \bibinfo {author} {\bibfnamefont {R.}~\bibnamefont {Meyer}}, \bibinfo {author} {\bibfnamefont {C.}~\bibnamefont {Northe}}, \bibinfo {author} {\bibfnamefont {H.}~\bibnamefont {Scheppach}}, \ and\ \bibinfo {author} {\bibfnamefont {S.}~\bibnamefont {Zhao}},\ }\href {\doibase 10.21468/SciPostPhysCore.6.3.049} {\bibfield  {journal} {\bibinfo  {journal} {SciPost Phys. Core}\ }\textbf {\bibinfo {volume} {6}},\ \bibinfo {pages} {049} (\bibinfo {year} {2023})},\ \Eprint {http://arxiv.org/abs/2212.09767} {arXiv:2212.09767 [hep-th]} \BibitemShut {NoStop}%
\bibitem [{\citenamefont {Northe}(2023)}]{Northe:2023khz}%
  \BibitemOpen
  \bibfield  {author} {\bibinfo {author} {\bibfnamefont {C.}~\bibnamefont {Northe}},\ }\href {\doibase 10.1103/PhysRevLett.131.151601} {\bibfield  {journal} {\bibinfo  {journal} {Phys. Rev. Lett.}\ }\textbf {\bibinfo {volume} {131}},\ \bibinfo {pages} {151601} (\bibinfo {year} {2023})},\ \Eprint {http://arxiv.org/abs/2303.07724} {arXiv:2303.07724 [hep-th]} \BibitemShut {NoStop}%
\bibitem [{\citenamefont {Banerjee}\ \emph {et~al.}(2024)\citenamefont {Banerjee}, \citenamefont {Basu}, \citenamefont {Bhattacharyya},\ and\ \citenamefont {Chakrabarti}}]{Banerjee:2024ldl}%
  \BibitemOpen
  \bibfield  {author} {\bibinfo {author} {\bibfnamefont {A.}~\bibnamefont {Banerjee}}, \bibinfo {author} {\bibfnamefont {R.}~\bibnamefont {Basu}}, \bibinfo {author} {\bibfnamefont {A.}~\bibnamefont {Bhattacharyya}}, \ and\ \bibinfo {author} {\bibfnamefont {N.}~\bibnamefont {Chakrabarti}},\ }\href {\doibase 10.1007/JHEP06(2024)121} {\bibfield  {journal} {\bibinfo  {journal} {JHEP}\ }\textbf {\bibinfo {volume} {06}},\ \bibinfo {pages} {121} (\bibinfo {year} {2024})},\ \Eprint {http://arxiv.org/abs/2404.02206} {arXiv:2404.02206 [hep-th]} \BibitemShut {NoStop}%
\bibitem [{\citenamefont {Choi}\ \emph {et~al.}(2023)\citenamefont {Choi}, \citenamefont {Rayhaun}, \citenamefont {Sanghavi},\ and\ \citenamefont {Shao}}]{Choi:2023xjw}%
  \BibitemOpen
  \bibfield  {author} {\bibinfo {author} {\bibfnamefont {Y.}~\bibnamefont {Choi}}, \bibinfo {author} {\bibfnamefont {B.~C.}\ \bibnamefont {Rayhaun}}, \bibinfo {author} {\bibfnamefont {Y.}~\bibnamefont {Sanghavi}}, \ and\ \bibinfo {author} {\bibfnamefont {S.-H.}\ \bibnamefont {Shao}},\ }\href {\doibase 10.1103/PhysRevD.108.125005} {\bibfield  {journal} {\bibinfo  {journal} {Phys. Rev. D}\ }\textbf {\bibinfo {volume} {108}},\ \bibinfo {pages} {125005} (\bibinfo {year} {2023})},\ \Eprint {http://arxiv.org/abs/2305.09713} {arXiv:2305.09713 [hep-th]} \BibitemShut {NoStop}%
\bibitem [{\citenamefont {Ocneanu}(1994)}]{ocneanu1994chirality}%
  \BibitemOpen
  \bibfield  {author} {\bibinfo {author} {\bibfnamefont {A.}~\bibnamefont {Ocneanu}},\ }\href@noop {} {\bibfield  {journal} {\bibinfo  {journal} {Subfactors (Kyuzeso, 1993)}\ }\textbf {\bibinfo {volume} {39}} (\bibinfo {year} {1994})}\BibitemShut {NoStop}%
\bibitem [{\citenamefont {Evans}\ and\ \citenamefont {Kawahigashi}(1995)}]{evans1995ocneanu}%
  \BibitemOpen
  \bibfield  {author} {\bibinfo {author} {\bibfnamefont {D.~E.}\ \bibnamefont {Evans}}\ and\ \bibinfo {author} {\bibfnamefont {Y.}~\bibnamefont {Kawahigashi}},\ }\href@noop {} {\bibfield  {journal} {\bibinfo  {journal} {International journal of mathematics}\ }\textbf {\bibinfo {volume} {6}},\ \bibinfo {pages} {205} (\bibinfo {year} {1995})}\BibitemShut {NoStop}%
\bibitem [{\citenamefont {Izumi}(2000)}]{Izumi:2000qa}%
  \BibitemOpen
  \bibfield  {author} {\bibinfo {author} {\bibfnamefont {M.}~\bibnamefont {Izumi}},\ }\href {\doibase 10.1007/s002200000234} {\bibfield  {journal} {\bibinfo  {journal} {Commun. Math. Phys.}\ }\textbf {\bibinfo {volume} {213}},\ \bibinfo {pages} {127} (\bibinfo {year} {2000})}\BibitemShut {NoStop}%
\bibitem [{\citenamefont {Müger}(2003)}]{MUGER2003159}%
  \BibitemOpen
  \bibfield  {author} {\bibinfo {author} {\bibfnamefont {M.}~\bibnamefont {Müger}},\ }\href {\doibase https://doi.org/10.1016/S0022-4049(02)00248-7} {\bibfield  {journal} {\bibinfo  {journal} {Journal of Pure and Applied Algebra}\ }\textbf {\bibinfo {volume} {180}},\ \bibinfo {pages} {159} (\bibinfo {year} {2003})},\ \Eprint {http://arxiv.org/abs/math/0111205} {arXiv:math/0111205} \BibitemShut {NoStop}%
\bibitem [{\citenamefont {Turaev}\ and\ \citenamefont {Viro}(1992)}]{Turaev:1992hq}%
  \BibitemOpen
  \bibfield  {author} {\bibinfo {author} {\bibfnamefont {V.~G.}\ \bibnamefont {Turaev}}\ and\ \bibinfo {author} {\bibfnamefont {O.~Y.}\ \bibnamefont {Viro}},\ }\href {\doibase 10.1016/0040-9383(92)90015-A} {\bibfield  {journal} {\bibinfo  {journal} {Topology}\ }\textbf {\bibinfo {volume} {31}},\ \bibinfo {pages} {865} (\bibinfo {year} {1992})}\BibitemShut {NoStop}%
\bibitem [{\citenamefont {Barrett}\ and\ \citenamefont {Westbury}(1996)}]{Barrett:1993ab}%
  \BibitemOpen
  \bibfield  {author} {\bibinfo {author} {\bibfnamefont {J.~W.}\ \bibnamefont {Barrett}}\ and\ \bibinfo {author} {\bibfnamefont {B.~W.}\ \bibnamefont {Westbury}},\ }\href {\doibase 10.1090/S0002-9947-96-01660-1} {\bibfield  {journal} {\bibinfo  {journal} {Trans. Am. Math. Soc.}\ }\textbf {\bibinfo {volume} {348}},\ \bibinfo {pages} {3997} (\bibinfo {year} {1996})},\ \Eprint {http://arxiv.org/abs/hep-th/9311155} {arXiv:hep-th/9311155} \BibitemShut {NoStop}%
\bibitem [{Note3()}]{Note3}%
  \BibitemOpen
  \bibinfo {note} {In fact, the representation category of $\protect \mathrm {Tube}(\protect \mathcal {C})$ naturally admits the structure of a braided tensor category which is equivalent to that on the Drinfeld center $Z(\protect \mathcal {C})$ \cite {Liu:2023lgl}. Note that $Z(\protect \mathcal {C})$ is the modular tensor category underlying the 2+1d Turaev-Viro topological field theory based on $\protect \mathcal {C}$.}\BibitemShut {Stop}%
\bibitem [{\citenamefont {Gaiotto}\ \emph {et~al.}(2015)\citenamefont {Gaiotto}, \citenamefont {Kapustin}, \citenamefont {Seiberg},\ and\ \citenamefont {Willett}}]{Gaiotto:2014kfa}%
  \BibitemOpen
  \bibfield  {author} {\bibinfo {author} {\bibfnamefont {D.}~\bibnamefont {Gaiotto}}, \bibinfo {author} {\bibfnamefont {A.}~\bibnamefont {Kapustin}}, \bibinfo {author} {\bibfnamefont {N.}~\bibnamefont {Seiberg}}, \ and\ \bibinfo {author} {\bibfnamefont {B.}~\bibnamefont {Willett}},\ }\href {\doibase 10.1007/JHEP02(2015)172} {\bibfield  {journal} {\bibinfo  {journal} {JHEP}\ }\textbf {\bibinfo {volume} {02}},\ \bibinfo {pages} {172} (\bibinfo {year} {2015})},\ \Eprint {http://arxiv.org/abs/1412.5148} {arXiv:1412.5148 [hep-th]} \BibitemShut {NoStop}%
\bibitem [{\citenamefont {Gaiotto}\ and\ \citenamefont {Kulp}(2021)}]{Gaiotto:2020iye}%
  \BibitemOpen
  \bibfield  {author} {\bibinfo {author} {\bibfnamefont {D.}~\bibnamefont {Gaiotto}}\ and\ \bibinfo {author} {\bibfnamefont {J.}~\bibnamefont {Kulp}},\ }\href {\doibase 10.1007/JHEP02(2021)132} {\bibfield  {journal} {\bibinfo  {journal} {JHEP}\ }\textbf {\bibinfo {volume} {02}},\ \bibinfo {pages} {132} (\bibinfo {year} {2021})},\ \Eprint {http://arxiv.org/abs/2008.05960} {arXiv:2008.05960 [hep-th]} \BibitemShut {NoStop}%
\bibitem [{\citenamefont {Ji}\ and\ \citenamefont {Wen}(2020)}]{Ji:2019jhk}%
  \BibitemOpen
  \bibfield  {author} {\bibinfo {author} {\bibfnamefont {W.}~\bibnamefont {Ji}}\ and\ \bibinfo {author} {\bibfnamefont {X.-G.}\ \bibnamefont {Wen}},\ }\href {\doibase 10.1103/PhysRevResearch.2.033417} {\bibfield  {journal} {\bibinfo  {journal} {Phys. Rev. Res.}\ }\textbf {\bibinfo {volume} {2}},\ \bibinfo {pages} {033417} (\bibinfo {year} {2020})},\ \Eprint {http://arxiv.org/abs/1912.13492} {arXiv:1912.13492 [cond-mat.str-el]} \BibitemShut {NoStop}%
\bibitem [{\citenamefont {Apruzzi}\ \emph {et~al.}(2021)\citenamefont {Apruzzi}, \citenamefont {Bonetti}, \citenamefont {Etxebarria}, \citenamefont {Hosseini},\ and\ \citenamefont {Schafer-Nameki}}]{Apruzzi:2021nmk}%
  \BibitemOpen
  \bibfield  {author} {\bibinfo {author} {\bibfnamefont {F.}~\bibnamefont {Apruzzi}}, \bibinfo {author} {\bibfnamefont {F.}~\bibnamefont {Bonetti}}, \bibinfo {author} {\bibfnamefont {I.~G.}\ \bibnamefont {Etxebarria}}, \bibinfo {author} {\bibfnamefont {S.~S.}\ \bibnamefont {Hosseini}}, \ and\ \bibinfo {author} {\bibfnamefont {S.}~\bibnamefont {Schafer-Nameki}},\ }\href@noop {} {\  (\bibinfo {year} {2021})},\ \Eprint {http://arxiv.org/abs/2112.02092} {arXiv:2112.02092 [hep-th]} \BibitemShut {NoStop}%
\bibitem [{\citenamefont {Freed}\ \emph {et~al.}(2022)\citenamefont {Freed}, \citenamefont {Moore},\ and\ \citenamefont {Teleman}}]{Freed:2022qnc}%
  \BibitemOpen
  \bibfield  {author} {\bibinfo {author} {\bibfnamefont {D.~S.}\ \bibnamefont {Freed}}, \bibinfo {author} {\bibfnamefont {G.~W.}\ \bibnamefont {Moore}}, \ and\ \bibinfo {author} {\bibfnamefont {C.}~\bibnamefont {Teleman}},\ }\href@noop {} {\  (\bibinfo {year} {2022})},\ \Eprint {http://arxiv.org/abs/2209.07471} {arXiv:2209.07471 [hep-th]} \BibitemShut {NoStop}%
\bibitem [{\citenamefont {Kong}\ \emph {et~al.}(2015)\citenamefont {Kong}, \citenamefont {Wen},\ and\ \citenamefont {Zheng}}]{Kong:2015flk}%
  \BibitemOpen
  \bibfield  {author} {\bibinfo {author} {\bibfnamefont {L.}~\bibnamefont {Kong}}, \bibinfo {author} {\bibfnamefont {X.-G.}\ \bibnamefont {Wen}}, \ and\ \bibinfo {author} {\bibfnamefont {H.}~\bibnamefont {Zheng}},\ }\href@noop {} {\  (\bibinfo {year} {2015})},\ \Eprint {http://arxiv.org/abs/1502.01690} {arXiv:1502.01690 [cond-mat.str-el]} \BibitemShut {NoStop}%
\bibitem [{\citenamefont {Kong}\ \emph {et~al.}(2017)\citenamefont {Kong}, \citenamefont {Wen},\ and\ \citenamefont {Zheng}}]{Kong:2017hcw}%
  \BibitemOpen
  \bibfield  {author} {\bibinfo {author} {\bibfnamefont {L.}~\bibnamefont {Kong}}, \bibinfo {author} {\bibfnamefont {X.-G.}\ \bibnamefont {Wen}}, \ and\ \bibinfo {author} {\bibfnamefont {H.}~\bibnamefont {Zheng}},\ }\href {\doibase 10.1016/j.nuclphysb.2017.06.023} {\bibfield  {journal} {\bibinfo  {journal} {Nucl. Phys. B}\ }\textbf {\bibinfo {volume} {922}},\ \bibinfo {pages} {62} (\bibinfo {year} {2017})},\ \Eprint {http://arxiv.org/abs/1702.00673} {arXiv:1702.00673 [cond-mat.str-el]} \BibitemShut {NoStop}%
\bibitem [{\citenamefont {Kong}\ \emph {et~al.}(2020)\citenamefont {Kong}, \citenamefont {Lan}, \citenamefont {Wen}, \citenamefont {Zhang},\ and\ \citenamefont {Zheng}}]{Kong:2020cie}%
  \BibitemOpen
  \bibfield  {author} {\bibinfo {author} {\bibfnamefont {L.}~\bibnamefont {Kong}}, \bibinfo {author} {\bibfnamefont {T.}~\bibnamefont {Lan}}, \bibinfo {author} {\bibfnamefont {X.-G.}\ \bibnamefont {Wen}}, \bibinfo {author} {\bibfnamefont {Z.-H.}\ \bibnamefont {Zhang}}, \ and\ \bibinfo {author} {\bibfnamefont {H.}~\bibnamefont {Zheng}},\ }\href {\doibase 10.1103/PhysRevResearch.2.043086} {\bibfield  {journal} {\bibinfo  {journal} {Phys. Rev. Res.}\ }\textbf {\bibinfo {volume} {2}},\ \bibinfo {pages} {043086} (\bibinfo {year} {2020})},\ \Eprint {http://arxiv.org/abs/2005.14178} {arXiv:2005.14178 [cond-mat.str-el]} \BibitemShut {NoStop}%
\bibitem [{\citenamefont {Kaidi}\ \emph {et~al.}(2022)\citenamefont {Kaidi}, \citenamefont {Ohmori},\ and\ \citenamefont {Zheng}}]{Kaidi:2022cpf}%
  \BibitemOpen
  \bibfield  {author} {\bibinfo {author} {\bibfnamefont {J.}~\bibnamefont {Kaidi}}, \bibinfo {author} {\bibfnamefont {K.}~\bibnamefont {Ohmori}}, \ and\ \bibinfo {author} {\bibfnamefont {Y.}~\bibnamefont {Zheng}},\ }\href@noop {} {\  (\bibinfo {year} {2022})},\ \Eprint {http://arxiv.org/abs/2209.11062} {arXiv:2209.11062 [hep-th]} \BibitemShut {NoStop}%
\bibitem [{\citenamefont {Antinucci}\ \emph {et~al.}(2022)\citenamefont {Antinucci}, \citenamefont {Benini}, \citenamefont {Copetti}, \citenamefont {Galati},\ and\ \citenamefont {Rizi}}]{Antinucci:2022vyk}%
  \BibitemOpen
  \bibfield  {author} {\bibinfo {author} {\bibfnamefont {A.}~\bibnamefont {Antinucci}}, \bibinfo {author} {\bibfnamefont {F.}~\bibnamefont {Benini}}, \bibinfo {author} {\bibfnamefont {C.}~\bibnamefont {Copetti}}, \bibinfo {author} {\bibfnamefont {G.}~\bibnamefont {Galati}}, \ and\ \bibinfo {author} {\bibfnamefont {G.}~\bibnamefont {Rizi}},\ }\href@noop {} {\  (\bibinfo {year} {2022})},\ \Eprint {http://arxiv.org/abs/2210.09146} {arXiv:2210.09146 [hep-th]} \BibitemShut {NoStop}%
\bibitem [{\citenamefont {Choi}\ \emph {et~al.}(2024{\natexlab{a}})\citenamefont {Choi}, \citenamefont {Rayhaun},\ and\ \citenamefont {Zheng}}]{Choi:2024tri}%
  \BibitemOpen
  \bibfield  {author} {\bibinfo {author} {\bibfnamefont {Y.}~\bibnamefont {Choi}}, \bibinfo {author} {\bibfnamefont {B.~C.}\ \bibnamefont {Rayhaun}}, \ and\ \bibinfo {author} {\bibfnamefont {Y.}~\bibnamefont {Zheng}},\ }\href@noop {} {\  (\bibinfo {year} {2024}{\natexlab{a}})},\ \Eprint {http://arxiv.org/abs/2409.02159} {arXiv:2409.02159 [hep-th]} \BibitemShut {NoStop}%
\bibitem [{Note4()}]{Note4}%
  \BibitemOpen
  \bibinfo {note} {We thank Javier Mag\'{a}n for bringing \cite {Benedetti:2024dku} to our attention, in which the authors study the noninvertible symmetry resolution of certain entanglement measures using methods from algebraic quantum field theory.}\BibitemShut {Stop}%
\bibitem [{\citenamefont {Saura-Bastida}\ \emph {et~al.}(2024)\citenamefont {Saura-Bastida}, \citenamefont {Das}, \citenamefont {Sierra},\ and\ \citenamefont {Molina-Vilaplana}}]{Saura-Bastida:2024yye}%
  \BibitemOpen
  \bibfield  {author} {\bibinfo {author} {\bibfnamefont {P.}~\bibnamefont {Saura-Bastida}}, \bibinfo {author} {\bibfnamefont {A.}~\bibnamefont {Das}}, \bibinfo {author} {\bibfnamefont {G.}~\bibnamefont {Sierra}}, \ and\ \bibinfo {author} {\bibfnamefont {J.}~\bibnamefont {Molina-Vilaplana}},\ }\href {\doibase 10.1103/PhysRevD.109.105026} {\bibfield  {journal} {\bibinfo  {journal} {Phys. Rev. D}\ }\textbf {\bibinfo {volume} {109}},\ \bibinfo {pages} {105026} (\bibinfo {year} {2024})},\ \Eprint {http://arxiv.org/abs/2402.06322} {arXiv:2402.06322 [hep-th]} \BibitemShut {NoStop}%
\bibitem [{\citenamefont {Kojita}\ \emph {et~al.}(2018)\citenamefont {Kojita}, \citenamefont {Maccaferri}, \citenamefont {Masuda},\ and\ \citenamefont {Schnabl}}]{Kojita:2016jwe}%
  \BibitemOpen
  \bibfield  {author} {\bibinfo {author} {\bibfnamefont {T.}~\bibnamefont {Kojita}}, \bibinfo {author} {\bibfnamefont {C.}~\bibnamefont {Maccaferri}}, \bibinfo {author} {\bibfnamefont {T.}~\bibnamefont {Masuda}}, \ and\ \bibinfo {author} {\bibfnamefont {M.}~\bibnamefont {Schnabl}},\ }\href {\doibase 10.1007/JHEP04(2018)057} {\bibfield  {journal} {\bibinfo  {journal} {JHEP}\ }\textbf {\bibinfo {volume} {04}},\ \bibinfo {pages} {057} (\bibinfo {year} {2018})},\ \Eprint {http://arxiv.org/abs/1612.01997} {arXiv:1612.01997 [hep-th]} \BibitemShut {NoStop}%
\bibitem [{\citenamefont {Konechny}(2020)}]{Konechny:2019wff}%
  \BibitemOpen
  \bibfield  {author} {\bibinfo {author} {\bibfnamefont {A.}~\bibnamefont {Konechny}},\ }\href {\doibase 10.1088/1751-8121/ab7c8b} {\bibfield  {journal} {\bibinfo  {journal} {J. Phys. A}\ }\textbf {\bibinfo {volume} {53}},\ \bibinfo {pages} {155401} (\bibinfo {year} {2020})},\ \Eprint {http://arxiv.org/abs/1911.06041} {arXiv:1911.06041 [hep-th]} \BibitemShut {NoStop}%
\bibitem [{\citenamefont {Barter}\ \emph {et~al.}(2022)\citenamefont {Barter}, \citenamefont {Bridgeman},\ and\ \citenamefont {Wolf}}]{Barter_2022}%
  \BibitemOpen
  \bibfield  {author} {\bibinfo {author} {\bibfnamefont {D.}~\bibnamefont {Barter}}, \bibinfo {author} {\bibfnamefont {J.}~\bibnamefont {Bridgeman}}, \ and\ \bibinfo {author} {\bibfnamefont {R.}~\bibnamefont {Wolf}},\ }\href {\doibase 10.21468/scipostphys.13.2.029} {\bibfield  {journal} {\bibinfo  {journal} {{SciPost} Physics}\ }\textbf {\bibinfo {volume} {13}} (\bibinfo {year} {2022}),\ 10.21468/scipostphys.13.2.029}\BibitemShut {NoStop}%
\bibitem [{\citenamefont {Konechny}\ and\ \citenamefont {Vergioglou}(2024)}]{Konechny:2024ixa}%
  \BibitemOpen
  \bibfield  {author} {\bibinfo {author} {\bibfnamefont {A.}~\bibnamefont {Konechny}}\ and\ \bibinfo {author} {\bibfnamefont {V.}~\bibnamefont {Vergioglou}},\ }\href@noop {} {\  (\bibinfo {year} {2024})},\ \Eprint {http://arxiv.org/abs/2405.10189} {arXiv:2405.10189 [hep-th]} \BibitemShut {NoStop}%
\bibitem [{\citenamefont {Cordova}\ \emph {et~al.}(2024{\natexlab{a}})\citenamefont {Cordova}, \citenamefont {Garc\'\i{}a-Sep\'ulveda},\ and\ \citenamefont {Holfester}}]{Cordova:2024vsq}%
  \BibitemOpen
  \bibfield  {author} {\bibinfo {author} {\bibfnamefont {C.}~\bibnamefont {Cordova}}, \bibinfo {author} {\bibfnamefont {D.}~\bibnamefont {Garc\'\i{}a-Sep\'ulveda}}, \ and\ \bibinfo {author} {\bibfnamefont {N.}~\bibnamefont {Holfester}},\ }\href {\doibase 10.1007/JHEP07(2024)154} {\bibfield  {journal} {\bibinfo  {journal} {JHEP}\ }\textbf {\bibinfo {volume} {07}},\ \bibinfo {pages} {154} (\bibinfo {year} {2024}{\natexlab{a}})},\ \Eprint {http://arxiv.org/abs/2403.08883} {arXiv:2403.08883 [hep-th]} \BibitemShut {NoStop}%
\bibitem [{\citenamefont {Copetti}\ \emph {et~al.}(2024)\citenamefont {Copetti}, \citenamefont {Cordova},\ and\ \citenamefont {Komatsu}}]{Copetti:2024dcz}%
  \BibitemOpen
  \bibfield  {author} {\bibinfo {author} {\bibfnamefont {C.}~\bibnamefont {Copetti}}, \bibinfo {author} {\bibfnamefont {L.}~\bibnamefont {Cordova}}, \ and\ \bibinfo {author} {\bibfnamefont {S.}~\bibnamefont {Komatsu}},\ }\href@noop {} {\  (\bibinfo {year} {2024})},\ \Eprint {http://arxiv.org/abs/2408.13132} {arXiv:2408.13132 [hep-th]} \BibitemShut {NoStop}%
\bibitem [{\citenamefont {Cordova}\ \emph {et~al.}(2024{\natexlab{b}})\citenamefont {Cordova}, \citenamefont {Holfester},\ and\ \citenamefont {Ohmori}}]{Cordova:2024iti}%
  \BibitemOpen
  \bibfield  {author} {\bibinfo {author} {\bibfnamefont {C.}~\bibnamefont {Cordova}}, \bibinfo {author} {\bibfnamefont {N.}~\bibnamefont {Holfester}}, \ and\ \bibinfo {author} {\bibfnamefont {K.}~\bibnamefont {Ohmori}},\ }\href@noop {} {\  (\bibinfo {year} {2024}{\natexlab{b}})},\ \Eprint {http://arxiv.org/abs/2408.11045} {arXiv:2408.11045 [hep-th]} \BibitemShut {NoStop}%
\bibitem [{Note5()}]{Note5}%
  \BibitemOpen
  \bibinfo {note} {Mathematically, ``same'' here means that they define the same $\protect \mathcal {C}$-module category.}\BibitemShut {Stop}%
\bibitem [{\citenamefont {{Kitaev}}\ and\ \citenamefont {{Kong}}(2012)}]{2012CMaPh.313..351K}%
  \BibitemOpen
  \bibfield  {author} {\bibinfo {author} {\bibfnamefont {A.}~\bibnamefont {{Kitaev}}}\ and\ \bibinfo {author} {\bibfnamefont {L.}~\bibnamefont {{Kong}}},\ }\href {\doibase 10.1007/s00220-012-1500-5} {\bibfield  {journal} {\bibinfo  {journal} {Communications in Mathematical Physics}\ }\textbf {\bibinfo {volume} {313}},\ \bibinfo {pages} {351} (\bibinfo {year} {2012})},\ \Eprint {http://arxiv.org/abs/1104.5047} {arXiv:1104.5047 [cond-mat.str-el]} \BibitemShut {NoStop}%
\bibitem [{\citenamefont {Huang}\ and\ \citenamefont {Cheng}(2023)}]{Huang:2023pyk}%
  \BibitemOpen
  \bibfield  {author} {\bibinfo {author} {\bibfnamefont {S.-J.}\ \bibnamefont {Huang}}\ and\ \bibinfo {author} {\bibfnamefont {M.}~\bibnamefont {Cheng}},\ }\href@noop {} {\  (\bibinfo {year} {2023})},\ \Eprint {http://arxiv.org/abs/2310.16878} {arXiv:2310.16878 [cond-mat.str-el]} \BibitemShut {NoStop}%
\bibitem [{\citenamefont {Cvetic}\ \emph {et~al.}(2024)\citenamefont {Cvetic}, \citenamefont {Donagi}, \citenamefont {Heckman}, \citenamefont {H\"ubner},\ and\ \citenamefont {Torres}}]{Cvetic:2024dzu}%
  \BibitemOpen
  \bibfield  {author} {\bibinfo {author} {\bibfnamefont {M.}~\bibnamefont {Cvetic}}, \bibinfo {author} {\bibfnamefont {R.}~\bibnamefont {Donagi}}, \bibinfo {author} {\bibfnamefont {J.~J.}\ \bibnamefont {Heckman}}, \bibinfo {author} {\bibfnamefont {M.}~\bibnamefont {H\"ubner}}, \ and\ \bibinfo {author} {\bibfnamefont {E.}~\bibnamefont {Torres}},\ }\href@noop {} {\  (\bibinfo {year} {2024})},\ \Eprint {http://arxiv.org/abs/2408.12600} {arXiv:2408.12600 [hep-th]} \BibitemShut {NoStop}%
\bibitem [{\citenamefont {Copetti}(2024)}]{Copetti:2024onh}%
  \BibitemOpen
  \bibfield  {author} {\bibinfo {author} {\bibfnamefont {C.}~\bibnamefont {Copetti}},\ }\href@noop {} {\  (\bibinfo {year} {2024})},\ \Eprint {http://arxiv.org/abs/2408.01490} {arXiv:2408.01490 [hep-th]} \BibitemShut {NoStop}%
\bibitem [{\citenamefont {Fuchs}\ \emph {et~al.}(2002)\citenamefont {Fuchs}, \citenamefont {Runkel},\ and\ \citenamefont {Schweigert}}]{Fuchs:2002cm}%
  \BibitemOpen
  \bibfield  {author} {\bibinfo {author} {\bibfnamefont {J.}~\bibnamefont {Fuchs}}, \bibinfo {author} {\bibfnamefont {I.}~\bibnamefont {Runkel}}, \ and\ \bibinfo {author} {\bibfnamefont {C.}~\bibnamefont {Schweigert}},\ }\href {\doibase 10.1016/S0550-3213(02)00744-7} {\bibfield  {journal} {\bibinfo  {journal} {Nucl. Phys. B}\ }\textbf {\bibinfo {volume} {646}},\ \bibinfo {pages} {353} (\bibinfo {year} {2002})},\ \Eprint {http://arxiv.org/abs/hep-th/0204148} {arXiv:hep-th/0204148} \BibitemShut {NoStop}%
\bibitem [{\citenamefont {Huang}\ \emph {et~al.}(2021)\citenamefont {Huang}, \citenamefont {Lin},\ and\ \citenamefont {Seifnashri}}]{Huang:2021zvu}%
  \BibitemOpen
  \bibfield  {author} {\bibinfo {author} {\bibfnamefont {T.-C.}\ \bibnamefont {Huang}}, \bibinfo {author} {\bibfnamefont {Y.-H.}\ \bibnamefont {Lin}}, \ and\ \bibinfo {author} {\bibfnamefont {S.}~\bibnamefont {Seifnashri}},\ }\href {\doibase 10.1007/JHEP12(2021)028} {\bibfield  {journal} {\bibinfo  {journal} {JHEP}\ }\textbf {\bibinfo {volume} {12}},\ \bibinfo {pages} {028} (\bibinfo {year} {2021})},\ \Eprint {http://arxiv.org/abs/2110.02958} {arXiv:2110.02958 [hep-th]} \BibitemShut {NoStop}%
\bibitem [{Note6()}]{Note6}%
  \BibitemOpen
  \bibinfo {note} {Indeed, because the $\protect \mathcal {C}$ symmetry operators are only supported on the Dirichlet boundary $\protect \mathcal {B}_{\protect \mathrm {reg}}$, the boundary lasso operators \protect \eqref {eq:bdylasso} only act on the topological junction $\protect \underline {{\protect \mathcal {O}}}$, while keeping the topological boundary conditions $\protect \mathcal {B}_1$ and ${\protect \mathcal {B}}_2$, as well as the topological line interface $\rho $ interpolating between them, unchanged.}\BibitemShut {Stop}%
\bibitem [{\citenamefont {Diatlyk}\ \emph {et~al.}(2024)\citenamefont {Diatlyk}, \citenamefont {Luo}, \citenamefont {Wang},\ and\ \citenamefont {Weller}}]{Diatlyk:2023fwf}%
  \BibitemOpen
  \bibfield  {author} {\bibinfo {author} {\bibfnamefont {O.}~\bibnamefont {Diatlyk}}, \bibinfo {author} {\bibfnamefont {C.}~\bibnamefont {Luo}}, \bibinfo {author} {\bibfnamefont {Y.}~\bibnamefont {Wang}}, \ and\ \bibinfo {author} {\bibfnamefont {Q.}~\bibnamefont {Weller}},\ }\href {\doibase 10.1007/JHEP03(2024)127} {\bibfield  {journal} {\bibinfo  {journal} {JHEP}\ }\textbf {\bibinfo {volume} {03}},\ \bibinfo {pages} {127} (\bibinfo {year} {2024})},\ \Eprint {http://arxiv.org/abs/2311.17044} {arXiv:2311.17044 [hep-th]} \BibitemShut {NoStop}%
\bibitem [{\citenamefont {Goldstein}\ and\ \citenamefont {Sela}(2018)}]{Goldstein:2017bua}%
  \BibitemOpen
  \bibfield  {author} {\bibinfo {author} {\bibfnamefont {M.}~\bibnamefont {Goldstein}}\ and\ \bibinfo {author} {\bibfnamefont {E.}~\bibnamefont {Sela}},\ }\href {\doibase 10.1103/PhysRevLett.120.200602} {\bibfield  {journal} {\bibinfo  {journal} {Phys. Rev. Lett.}\ }\textbf {\bibinfo {volume} {120}},\ \bibinfo {pages} {200602} (\bibinfo {year} {2018})},\ \Eprint {http://arxiv.org/abs/1711.09418} {arXiv:1711.09418 [cond-mat.stat-mech]} \BibitemShut {NoStop}%
\bibitem [{\citenamefont {Xavier}\ \emph {et~al.}(2018)\citenamefont {Xavier}, \citenamefont {Alcaraz},\ and\ \citenamefont {Sierra}}]{Xavier:2018kqb}%
  \BibitemOpen
  \bibfield  {author} {\bibinfo {author} {\bibfnamefont {J.~C.}\ \bibnamefont {Xavier}}, \bibinfo {author} {\bibfnamefont {F.~C.}\ \bibnamefont {Alcaraz}}, \ and\ \bibinfo {author} {\bibfnamefont {G.}~\bibnamefont {Sierra}},\ }\href {\doibase 10.1103/PhysRevB.98.041106} {\bibfield  {journal} {\bibinfo  {journal} {Phys. Rev. B}\ }\textbf {\bibinfo {volume} {98}},\ \bibinfo {pages} {041106} (\bibinfo {year} {2018})},\ \Eprint {http://arxiv.org/abs/1804.06357} {arXiv:1804.06357 [cond-mat.stat-mech]} \BibitemShut {NoStop}%
\bibitem [{\citenamefont {Calabrese}\ \emph {et~al.}(2021)\citenamefont {Calabrese}, \citenamefont {Dubail},\ and\ \citenamefont {Murciano}}]{Calabrese:2021wvi}%
  \BibitemOpen
  \bibfield  {author} {\bibinfo {author} {\bibfnamefont {P.}~\bibnamefont {Calabrese}}, \bibinfo {author} {\bibfnamefont {J.}~\bibnamefont {Dubail}}, \ and\ \bibinfo {author} {\bibfnamefont {S.}~\bibnamefont {Murciano}},\ }\href {\doibase 10.1007/JHEP10(2021)067} {\bibfield  {journal} {\bibinfo  {journal} {JHEP}\ }\textbf {\bibinfo {volume} {10}},\ \bibinfo {pages} {067} (\bibinfo {year} {2021})},\ \Eprint {http://arxiv.org/abs/2106.15946} {arXiv:2106.15946 [hep-th]} \BibitemShut {NoStop}%
\bibitem [{\citenamefont {Heymann}\ and\ \citenamefont {Quella}(2024)}]{Heymann:2024vvf}%
  \BibitemOpen
  \bibfield  {author} {\bibinfo {author} {\bibfnamefont {J.}~\bibnamefont {Heymann}}\ and\ \bibinfo {author} {\bibfnamefont {T.}~\bibnamefont {Quella}},\ }\href@noop {} {\  (\bibinfo {year} {2024})},\ \Eprint {http://arxiv.org/abs/2409.02315} {arXiv:2409.02315 [hep-th]} \BibitemShut {NoStop}%
\bibitem [{\citenamefont {Feiguin}\ \emph {et~al.}(2007)\citenamefont {Feiguin}, \citenamefont {Trebst}, \citenamefont {Ludwig}, \citenamefont {Troyer}, \citenamefont {Kitaev}, \citenamefont {Wang},\ and\ \citenamefont {Freedman}}]{Feiguin:2006ydp}%
  \BibitemOpen
  \bibfield  {author} {\bibinfo {author} {\bibfnamefont {A.}~\bibnamefont {Feiguin}}, \bibinfo {author} {\bibfnamefont {S.}~\bibnamefont {Trebst}}, \bibinfo {author} {\bibfnamefont {A.~W.~W.}\ \bibnamefont {Ludwig}}, \bibinfo {author} {\bibfnamefont {M.}~\bibnamefont {Troyer}}, \bibinfo {author} {\bibfnamefont {A.}~\bibnamefont {Kitaev}}, \bibinfo {author} {\bibfnamefont {Z.}~\bibnamefont {Wang}}, \ and\ \bibinfo {author} {\bibfnamefont {M.~H.}\ \bibnamefont {Freedman}},\ }\href {\doibase 10.1103/PhysRevLett.98.160409} {\bibfield  {journal} {\bibinfo  {journal} {Phys. Rev. Lett.}\ }\textbf {\bibinfo {volume} {98}},\ \bibinfo {pages} {160409} (\bibinfo {year} {2007})},\ \Eprint {http://arxiv.org/abs/cond-mat/0612341} {arXiv:cond-mat/0612341} \BibitemShut {NoStop}%
\bibitem [{\citenamefont {{Gils}}\ \emph {et~al.}(2013)\citenamefont {{Gils}}, \citenamefont {{Ardonne}}, \citenamefont {{Trebst}}, \citenamefont {{Huse}}, \citenamefont {{Ludwig}}, \citenamefont {{Troyer}},\ and\ \citenamefont {{Wang}}}]{2013PhRvB..87w5120G}%
  \BibitemOpen
  \bibfield  {author} {\bibinfo {author} {\bibfnamefont {C.}~\bibnamefont {{Gils}}}, \bibinfo {author} {\bibfnamefont {E.}~\bibnamefont {{Ardonne}}}, \bibinfo {author} {\bibfnamefont {S.}~\bibnamefont {{Trebst}}}, \bibinfo {author} {\bibfnamefont {D.~A.}\ \bibnamefont {{Huse}}}, \bibinfo {author} {\bibfnamefont {A.~W.~W.}\ \bibnamefont {{Ludwig}}}, \bibinfo {author} {\bibfnamefont {M.}~\bibnamefont {{Troyer}}}, \ and\ \bibinfo {author} {\bibfnamefont {Z.}~\bibnamefont {{Wang}}},\ }\href {\doibase 10.1103/PhysRevB.87.235120} {\bibfield  {journal} {\bibinfo  {journal} {Phys.Rev.B}\ }\textbf {\bibinfo {volume} {87}},\ \bibinfo {eid} {235120} (\bibinfo {year} {2013})},\ \Eprint {http://arxiv.org/abs/1303.4290} {arXiv:1303.4290 [cond-mat.str-el]} \BibitemShut {NoStop}%
\bibitem [{\citenamefont {Buican}\ and\ \citenamefont {Gromov}(2017)}]{Buican:2017rxc}%
  \BibitemOpen
  \bibfield  {author} {\bibinfo {author} {\bibfnamefont {M.}~\bibnamefont {Buican}}\ and\ \bibinfo {author} {\bibfnamefont {A.}~\bibnamefont {Gromov}},\ }\href {\doibase 10.1007/s00220-017-2995-6} {\bibfield  {journal} {\bibinfo  {journal} {Commun. Math. Phys.}\ }\textbf {\bibinfo {volume} {356}},\ \bibinfo {pages} {1017} (\bibinfo {year} {2017})},\ \Eprint {http://arxiv.org/abs/1701.02800} {arXiv:1701.02800 [hep-th]} \BibitemShut {NoStop}%
\bibitem [{\citenamefont {Kitaev}\ and\ \citenamefont {Preskill}(2006)}]{Kitaev:2005dm}%
  \BibitemOpen
  \bibfield  {author} {\bibinfo {author} {\bibfnamefont {A.}~\bibnamefont {Kitaev}}\ and\ \bibinfo {author} {\bibfnamefont {J.}~\bibnamefont {Preskill}},\ }\href {\doibase 10.1103/PhysRevLett.96.110404} {\bibfield  {journal} {\bibinfo  {journal} {Phys. Rev. Lett.}\ }\textbf {\bibinfo {volume} {96}},\ \bibinfo {pages} {110404} (\bibinfo {year} {2006})},\ \Eprint {http://arxiv.org/abs/hep-th/0510092} {arXiv:hep-th/0510092} \BibitemShut {NoStop}%
\bibitem [{\citenamefont {Levin}\ and\ \citenamefont {Wen}(2006)}]{Levin:2006zz}%
  \BibitemOpen
  \bibfield  {author} {\bibinfo {author} {\bibfnamefont {M.}~\bibnamefont {Levin}}\ and\ \bibinfo {author} {\bibfnamefont {X.-G.}\ \bibnamefont {Wen}},\ }\href {\doibase 10.1103/PhysRevLett.96.110405} {\bibfield  {journal} {\bibinfo  {journal} {Phys. Rev. Lett.}\ }\textbf {\bibinfo {volume} {96}},\ \bibinfo {pages} {110405} (\bibinfo {year} {2006})},\ \Eprint {http://arxiv.org/abs/cond-mat/0510613} {arXiv:cond-mat/0510613} \BibitemShut {NoStop}%
\bibitem [{\citenamefont {Brehm}\ and\ \citenamefont {Runkel}(2022)}]{Brehm:2021wev}%
  \BibitemOpen
  \bibfield  {author} {\bibinfo {author} {\bibfnamefont {E.~M.}\ \bibnamefont {Brehm}}\ and\ \bibinfo {author} {\bibfnamefont {I.}~\bibnamefont {Runkel}},\ }\href {\doibase 10.1088/1751-8121/ac6a91} {\bibfield  {journal} {\bibinfo  {journal} {J. Phys. A}\ }\textbf {\bibinfo {volume} {55}},\ \bibinfo {pages} {235001} (\bibinfo {year} {2022})},\ \Eprint {http://arxiv.org/abs/2112.01563} {arXiv:2112.01563 [cond-mat.stat-mech]} \BibitemShut {NoStop}%
\bibitem [{\citenamefont {Brehm}\ and\ \citenamefont {Runkel}(2024)}]{Brehm:2024zun}%
  \BibitemOpen
  \bibfield  {author} {\bibinfo {author} {\bibfnamefont {E.~M.}\ \bibnamefont {Brehm}}\ and\ \bibinfo {author} {\bibfnamefont {I.}~\bibnamefont {Runkel}},\ }\href@noop {} {\  (\bibinfo {year} {2024})},\ \Eprint {http://arxiv.org/abs/2410.19938} {arXiv:2410.19938 [math-ph]} \BibitemShut {NoStop}%
\bibitem [{Note7()}]{Note7}%
  \BibitemOpen
  \bibinfo {note} {The cloaking boundary condition is a non-simple boundary condition with a topological local operator on it whose defining property is that it is ``invisible'' to topological lines generating the symmetry. Such a boundary condition exists within any multiplet $\protect \mathcal {B}$ of boundaries.}\BibitemShut {Stop}%
\bibitem [{\citenamefont {Oshikawa}\ and\ \citenamefont {Affleck}(1996)}]{Oshikawa:1996ww}%
  \BibitemOpen
  \bibfield  {author} {\bibinfo {author} {\bibfnamefont {M.}~\bibnamefont {Oshikawa}}\ and\ \bibinfo {author} {\bibfnamefont {I.}~\bibnamefont {Affleck}},\ }\href {\doibase 10.1103/PhysRevLett.77.2604} {\bibfield  {journal} {\bibinfo  {journal} {Phys. Rev. Lett.}\ }\textbf {\bibinfo {volume} {77}},\ \bibinfo {pages} {2604} (\bibinfo {year} {1996})},\ \Eprint {http://arxiv.org/abs/hep-th/9606177} {arXiv:hep-th/9606177} \BibitemShut {NoStop}%
\bibitem [{\citenamefont {Oshikawa}\ and\ \citenamefont {Affleck}(1997)}]{Oshikawa:1996dj}%
  \BibitemOpen
  \bibfield  {author} {\bibinfo {author} {\bibfnamefont {M.}~\bibnamefont {Oshikawa}}\ and\ \bibinfo {author} {\bibfnamefont {I.}~\bibnamefont {Affleck}},\ }\href {\doibase 10.1016/S0550-3213(97)00219-8} {\bibfield  {journal} {\bibinfo  {journal} {Nucl. Phys. B}\ }\textbf {\bibinfo {volume} {495}},\ \bibinfo {pages} {533} (\bibinfo {year} {1997})},\ \Eprint {http://arxiv.org/abs/cond-mat/9612187} {arXiv:cond-mat/9612187} \BibitemShut {NoStop}%
\bibitem [{\citenamefont {Petkova}\ and\ \citenamefont {Zuber}(2001)}]{Petkova:2000ip}%
  \BibitemOpen
  \bibfield  {author} {\bibinfo {author} {\bibfnamefont {V.~B.}\ \bibnamefont {Petkova}}\ and\ \bibinfo {author} {\bibfnamefont {J.~B.}\ \bibnamefont {Zuber}},\ }\href {\doibase 10.1016/S0370-2693(01)00276-3} {\bibfield  {journal} {\bibinfo  {journal} {Phys. Lett. B}\ }\textbf {\bibinfo {volume} {504}},\ \bibinfo {pages} {157} (\bibinfo {year} {2001})},\ \Eprint {http://arxiv.org/abs/hep-th/0011021} {arXiv:hep-th/0011021} \BibitemShut {NoStop}%
\bibitem [{\citenamefont {Frohlich}\ \emph {et~al.}(2004)\citenamefont {Frohlich}, \citenamefont {Fuchs}, \citenamefont {Runkel},\ and\ \citenamefont {Schweigert}}]{Frohlich:2004ef}%
  \BibitemOpen
  \bibfield  {author} {\bibinfo {author} {\bibfnamefont {J.}~\bibnamefont {Frohlich}}, \bibinfo {author} {\bibfnamefont {J.}~\bibnamefont {Fuchs}}, \bibinfo {author} {\bibfnamefont {I.}~\bibnamefont {Runkel}}, \ and\ \bibinfo {author} {\bibfnamefont {C.}~\bibnamefont {Schweigert}},\ }\href {\doibase 10.1103/PhysRevLett.93.070601} {\bibfield  {journal} {\bibinfo  {journal} {Phys. Rev. Lett.}\ }\textbf {\bibinfo {volume} {93}},\ \bibinfo {pages} {070601} (\bibinfo {year} {2004})},\ \Eprint {http://arxiv.org/abs/cond-mat/0404051} {arXiv:cond-mat/0404051 [cond-mat]} \BibitemShut {NoStop}%
\bibitem [{\citenamefont {Frohlich}\ \emph {et~al.}(2007)\citenamefont {Frohlich}, \citenamefont {Fuchs}, \citenamefont {Runkel},\ and\ \citenamefont {Schweigert}}]{Frohlich:2006ch}%
  \BibitemOpen
  \bibfield  {author} {\bibinfo {author} {\bibfnamefont {J.}~\bibnamefont {Frohlich}}, \bibinfo {author} {\bibfnamefont {J.}~\bibnamefont {Fuchs}}, \bibinfo {author} {\bibfnamefont {I.}~\bibnamefont {Runkel}}, \ and\ \bibinfo {author} {\bibfnamefont {C.}~\bibnamefont {Schweigert}},\ }\href {\doibase 10.1016/j.nuclphysb.2006.11.017} {\bibfield  {journal} {\bibinfo  {journal} {Nucl. Phys. B}\ }\textbf {\bibinfo {volume} {763}},\ \bibinfo {pages} {354} (\bibinfo {year} {2007})},\ \Eprint {http://arxiv.org/abs/hep-th/0607247} {arXiv:hep-th/0607247} \BibitemShut {NoStop}%
\bibitem [{\citenamefont {Frohlich}\ \emph {et~al.}(2010)\citenamefont {Frohlich}, \citenamefont {Fuchs}, \citenamefont {Runkel},\ and\ \citenamefont {Schweigert}}]{Frohlich:2009gb}%
  \BibitemOpen
  \bibfield  {author} {\bibinfo {author} {\bibfnamefont {J.}~\bibnamefont {Frohlich}}, \bibinfo {author} {\bibfnamefont {J.}~\bibnamefont {Fuchs}}, \bibinfo {author} {\bibfnamefont {I.}~\bibnamefont {Runkel}}, \ and\ \bibinfo {author} {\bibfnamefont {C.}~\bibnamefont {Schweigert}},\ }in\ \href {\doibase 10.1142/9789814304634_0056} {\emph {\bibinfo {booktitle} {{16th International Congress on Mathematical Physics}}}}\ (\bibinfo {year} {2010})\ pp.\ \bibinfo {pages} {608--613},\ \Eprint {http://arxiv.org/abs/0909.5013} {arXiv:0909.5013 [math-ph]} \BibitemShut {NoStop}%
\bibitem [{\citenamefont {Aasen}\ \emph {et~al.}(2016)\citenamefont {Aasen}, \citenamefont {Mong},\ and\ \citenamefont {Fendley}}]{Aasen:2016dop}%
  \BibitemOpen
  \bibfield  {author} {\bibinfo {author} {\bibfnamefont {D.}~\bibnamefont {Aasen}}, \bibinfo {author} {\bibfnamefont {R.~S.~K.}\ \bibnamefont {Mong}}, \ and\ \bibinfo {author} {\bibfnamefont {P.}~\bibnamefont {Fendley}},\ }\href {\doibase 10.1088/1751-8113/49/35/354001} {\bibfield  {journal} {\bibinfo  {journal} {J. Phys. A}\ }\textbf {\bibinfo {volume} {49}},\ \bibinfo {pages} {354001} (\bibinfo {year} {2016})},\ \Eprint {http://arxiv.org/abs/1601.07185} {arXiv:1601.07185 [cond-mat.stat-mech]} \BibitemShut {NoStop}%
\bibitem [{\citenamefont {Ji}\ \emph {et~al.}(2020)\citenamefont {Ji}, \citenamefont {Shao},\ and\ \citenamefont {Wen}}]{Ji:2019ugf}%
  \BibitemOpen
  \bibfield  {author} {\bibinfo {author} {\bibfnamefont {W.}~\bibnamefont {Ji}}, \bibinfo {author} {\bibfnamefont {S.-H.}\ \bibnamefont {Shao}}, \ and\ \bibinfo {author} {\bibfnamefont {X.-G.}\ \bibnamefont {Wen}},\ }\href {\doibase 10.1103/PhysRevResearch.2.033317} {\bibfield  {journal} {\bibinfo  {journal} {Phys. Rev. Res.}\ }\textbf {\bibinfo {volume} {2}},\ \bibinfo {pages} {033317} (\bibinfo {year} {2020})},\ \Eprint {http://arxiv.org/abs/1909.01425} {arXiv:1909.01425 [cond-mat.str-el]} \BibitemShut {NoStop}%
\bibitem [{\citenamefont {Aasen}\ \emph {et~al.}(2020)\citenamefont {Aasen}, \citenamefont {Fendley},\ and\ \citenamefont {Mong}}]{Aasen:2020jwb}%
  \BibitemOpen
  \bibfield  {author} {\bibinfo {author} {\bibfnamefont {D.}~\bibnamefont {Aasen}}, \bibinfo {author} {\bibfnamefont {P.}~\bibnamefont {Fendley}}, \ and\ \bibinfo {author} {\bibfnamefont {R.~S.~K.}\ \bibnamefont {Mong}},\ }\href@noop {} {\  (\bibinfo {year} {2020})},\ \Eprint {http://arxiv.org/abs/2008.08598} {arXiv:2008.08598 [cond-mat.stat-mech]} \BibitemShut {NoStop}%
\bibitem [{\citenamefont {Seiberg}\ and\ \citenamefont {Shao}(2023)}]{Seiberg:2023cdc}%
  \BibitemOpen
  \bibfield  {author} {\bibinfo {author} {\bibfnamefont {N.}~\bibnamefont {Seiberg}}\ and\ \bibinfo {author} {\bibfnamefont {S.-H.}\ \bibnamefont {Shao}},\ }\href@noop {} {\  (\bibinfo {year} {2023})},\ \Eprint {http://arxiv.org/abs/2307.02534} {arXiv:2307.02534 [cond-mat.str-el]} \BibitemShut {NoStop}%
\bibitem [{Note8()}]{Note8}%
  \BibitemOpen
  \bibinfo {note} {The conformal boundary condition $B$ is obtained by folding a single copy of the Ising CFT along its Kramers-Wannier duality line \cite {Choi:2023xjw}.}\BibitemShut {Stop}%
\bibitem [{\citenamefont {Kats}\ and\ \citenamefont {Palyutkin}(1966)}]{KP}%
  \BibitemOpen
  \bibfield  {author} {\bibinfo {author} {\bibfnamefont {G.~I.}\ \bibnamefont {Kats}}\ and\ \bibinfo {author} {\bibfnamefont {V.~G.}\ \bibnamefont {Palyutkin}},\ }\href {\doibase https://www.mathnet.ru/php/archive.phtml?wshow=paper&jrnid=mmo&paperid=170&option_lang=eng#forwardlinks} {\bibfield  {journal} {\bibinfo  {journal} {Tr. Mosk. Mat. Obs.}\ }\textbf {\bibinfo {volume} {15}},\ \bibinfo {pages} {224} (\bibinfo {year} {1966})}\BibitemShut {NoStop}%
\bibitem [{\citenamefont {Masuoka}(1995)}]{masuoka1995semisimple}%
  \BibitemOpen
  \bibfield  {author} {\bibinfo {author} {\bibfnamefont {A.}~\bibnamefont {Masuoka}},\ }\href@noop {} {\bibfield  {journal} {\bibinfo  {journal} {Israel Journal of Mathematics}\ }\textbf {\bibinfo {volume} {92}},\ \bibinfo {pages} {361} (\bibinfo {year} {1995})}\BibitemShut {NoStop}%
\bibitem [{\citenamefont {Das}\ \emph {et~al.}(2024)\citenamefont {Das}, \citenamefont {Molina-Vilaplana},\ and\ \citenamefont {Saura-Bastida}}]{Das:2024qdx}%
  \BibitemOpen
  \bibfield  {author} {\bibinfo {author} {\bibfnamefont {A.}~\bibnamefont {Das}}, \bibinfo {author} {\bibfnamefont {J.}~\bibnamefont {Molina-Vilaplana}}, \ and\ \bibinfo {author} {\bibfnamefont {P.}~\bibnamefont {Saura-Bastida}},\ }\href@noop {} {\  (\bibinfo {year} {2024})},\ \Eprint {http://arxiv.org/abs/2409.02162} {arXiv:2409.02162 [hep-th]} \BibitemShut {NoStop}%
\bibitem [{\citenamefont {Bhardwaj}\ \emph {et~al.}(2024)\citenamefont {Bhardwaj}, \citenamefont {Copetti}, \citenamefont {Pajer},\ and\ \citenamefont {Schafer-Nameki}}]{Bhardwaj:2024igy}%
  \BibitemOpen
  \bibfield  {author} {\bibinfo {author} {\bibfnamefont {L.}~\bibnamefont {Bhardwaj}}, \bibinfo {author} {\bibfnamefont {C.}~\bibnamefont {Copetti}}, \bibinfo {author} {\bibfnamefont {D.}~\bibnamefont {Pajer}}, \ and\ \bibinfo {author} {\bibfnamefont {S.}~\bibnamefont {Schafer-Nameki}},\ }\href@noop {} {\  (\bibinfo {year} {2024})},\ \Eprint {http://arxiv.org/abs/2409.02166} {arXiv:2409.02166 [hep-th]} \BibitemShut {NoStop}%
\bibitem [{\citenamefont {Garc\'\i{}a~Etxebarria}\ \emph {et~al.}(2024)\citenamefont {Garc\'\i{}a~Etxebarria}, \citenamefont {Huertas},\ and\ \citenamefont {Uranga}}]{GarciaEtxebarria:2024jfv}%
  \BibitemOpen
  \bibfield  {author} {\bibinfo {author} {\bibfnamefont {I.~n.}\ \bibnamefont {Garc\'\i{}a~Etxebarria}}, \bibinfo {author} {\bibfnamefont {J.}~\bibnamefont {Huertas}}, \ and\ \bibinfo {author} {\bibfnamefont {A.~M.}\ \bibnamefont {Uranga}},\ }\href@noop {} {\  (\bibinfo {year} {2024})},\ \Eprint {http://arxiv.org/abs/2409.02156} {arXiv:2409.02156 [hep-th]} \BibitemShut {NoStop}%
\bibitem [{\citenamefont {Lin}\ \emph {et~al.}(2021)\citenamefont {Lin}, \citenamefont {Levin},\ and\ \citenamefont {Burnell}}]{PhysRevB.103.195155}%
  \BibitemOpen
  \bibfield  {author} {\bibinfo {author} {\bibfnamefont {C.-H.}\ \bibnamefont {Lin}}, \bibinfo {author} {\bibfnamefont {M.}~\bibnamefont {Levin}}, \ and\ \bibinfo {author} {\bibfnamefont {F.~J.}\ \bibnamefont {Burnell}},\ }\href {\doibase 10.1103/PhysRevB.103.195155} {\bibfield  {journal} {\bibinfo  {journal} {Phys. Rev. B}\ }\textbf {\bibinfo {volume} {103}},\ \bibinfo {pages} {195155} (\bibinfo {year} {2021})}\BibitemShut {NoStop}%
\bibitem [{\citenamefont {Zhang}\ and\ \citenamefont {C\'ordova}(2023)}]{Zhang:2023wlu}%
  \BibitemOpen
  \bibfield  {author} {\bibinfo {author} {\bibfnamefont {C.}~\bibnamefont {Zhang}}\ and\ \bibinfo {author} {\bibfnamefont {C.}~\bibnamefont {C\'ordova}},\ }\href@noop {} {\  (\bibinfo {year} {2023})},\ \Eprint {http://arxiv.org/abs/2304.01262} {arXiv:2304.01262 [cond-mat.str-el]} \BibitemShut {NoStop}%
\bibitem [{\citenamefont {Calabrese}\ and\ \citenamefont {Cardy}(2009)}]{Calabrese:2009qy}%
  \BibitemOpen
  \bibfield  {author} {\bibinfo {author} {\bibfnamefont {P.}~\bibnamefont {Calabrese}}\ and\ \bibinfo {author} {\bibfnamefont {J.}~\bibnamefont {Cardy}},\ }\href {\doibase 10.1088/1751-8113/42/50/504005} {\bibfield  {journal} {\bibinfo  {journal} {J. Phys. A}\ }\textbf {\bibinfo {volume} {42}},\ \bibinfo {pages} {504005} (\bibinfo {year} {2009})},\ \Eprint {http://arxiv.org/abs/0905.4013} {arXiv:0905.4013 [cond-mat.stat-mech]} \BibitemShut {NoStop}%
\bibitem [{\citenamefont {Monkman}\ and\ \citenamefont {Sirker}(2023)}]{Monkman:2023hup}%
  \BibitemOpen
  \bibfield  {author} {\bibinfo {author} {\bibfnamefont {K.}~\bibnamefont {Monkman}}\ and\ \bibinfo {author} {\bibfnamefont {J.}~\bibnamefont {Sirker}},\ }\href {\doibase 10.1088/1751-8121/ad086d} {\bibfield  {journal} {\bibinfo  {journal} {J. Phys. A}\ }\textbf {\bibinfo {volume} {56}},\ \bibinfo {pages} {495001} (\bibinfo {year} {2023})},\ \Eprint {http://arxiv.org/abs/2307.05820} {arXiv:2307.05820 [cond-mat.stat-mech]} \BibitemShut {NoStop}%
\bibitem [{\citenamefont {Witten}(2018)}]{Witten:2018zxz}%
  \BibitemOpen
  \bibfield  {author} {\bibinfo {author} {\bibfnamefont {E.}~\bibnamefont {Witten}},\ }\href {\doibase 10.1103/RevModPhys.90.045003} {\bibfield  {journal} {\bibinfo  {journal} {Rev. Mod. Phys.}\ }\textbf {\bibinfo {volume} {90}},\ \bibinfo {pages} {045003} (\bibinfo {year} {2018})},\ \Eprint {http://arxiv.org/abs/1803.04993} {arXiv:1803.04993 [hep-th]} \BibitemShut {NoStop}%
\bibitem [{Note9()}]{Note9}%
  \BibitemOpen
  \bibinfo {note} {If one imposes two distinct $G$-symmetric boundary conditions at the two entangling points, it can also happen that the entanglement spectrum organizes into projective representations of $G$. See, for instance, \cite {Cordova:2022lms}.}\BibitemShut {Stop}%
\bibitem [{\citenamefont {Choi}\ \emph {et~al.}(2024{\natexlab{b}})\citenamefont {Choi}, \citenamefont {Lu},\ and\ \citenamefont {Sun}}]{Choi:2023vgk}%
  \BibitemOpen
  \bibfield  {author} {\bibinfo {author} {\bibfnamefont {Y.}~\bibnamefont {Choi}}, \bibinfo {author} {\bibfnamefont {D.-C.}\ \bibnamefont {Lu}}, \ and\ \bibinfo {author} {\bibfnamefont {Z.}~\bibnamefont {Sun}},\ }\href {\doibase 10.1007/JHEP01(2024)142} {\bibfield  {journal} {\bibinfo  {journal} {JHEP}\ }\textbf {\bibinfo {volume} {01}},\ \bibinfo {pages} {142} (\bibinfo {year} {2024}{\natexlab{b}})},\ \Eprint {http://arxiv.org/abs/2310.19867} {arXiv:2310.19867 [hep-th]} \BibitemShut {NoStop}%
\bibitem [{\citenamefont {Perez-Lona}\ \emph {et~al.}(2024)\citenamefont {Perez-Lona}, \citenamefont {Robbins}, \citenamefont {Sharpe}, \citenamefont {Vandermeulen},\ and\ \citenamefont {Yu}}]{Perez-Lona:2023djo}%
  \BibitemOpen
  \bibfield  {author} {\bibinfo {author} {\bibfnamefont {A.}~\bibnamefont {Perez-Lona}}, \bibinfo {author} {\bibfnamefont {D.}~\bibnamefont {Robbins}}, \bibinfo {author} {\bibfnamefont {E.}~\bibnamefont {Sharpe}}, \bibinfo {author} {\bibfnamefont {T.}~\bibnamefont {Vandermeulen}}, \ and\ \bibinfo {author} {\bibfnamefont {X.}~\bibnamefont {Yu}},\ }\href {\doibase 10.1007/JHEP02(2024)154} {\bibfield  {journal} {\bibinfo  {journal} {JHEP}\ }\textbf {\bibinfo {volume} {02}},\ \bibinfo {pages} {154} (\bibinfo {year} {2024})},\ \Eprint {http://arxiv.org/abs/2311.16230} {arXiv:2311.16230 [hep-th]} \BibitemShut {NoStop}%
\bibitem [{\citenamefont {Tambara}\ and\ \citenamefont {Yamagami}(1998)}]{TAMBARA1998692}%
  \BibitemOpen
  \bibfield  {author} {\bibinfo {author} {\bibfnamefont {D.}~\bibnamefont {Tambara}}\ and\ \bibinfo {author} {\bibfnamefont {S.}~\bibnamefont {Yamagami}},\ }\href {\doibase https://doi.org/10.1006/jabr.1998.7558} {\bibfield  {journal} {\bibinfo  {journal} {Journal of Algebra}\ }\textbf {\bibinfo {volume} {209}},\ \bibinfo {pages} {692} (\bibinfo {year} {1998})}\BibitemShut {NoStop}%
\bibitem [{Note10()}]{Note10}%
  \BibitemOpen
  \bibinfo {note} {To be precise, the convention in Figure \ref {fig:H8F} gives us $\protect \mathsf {H}_a \times \protect \mathsf {H}_b = \protect \mathsf {H}_b \times \protect \mathsf {H}_a = - \protect \mathsf {H}_{ab}$. By an appropriate gauge transformation of the topological junctions, we redefine $\protect \mathsf {H}_{ab} \rightarrow - \protect \mathsf {H}_{ab}$ and obtain \protect \eqref {eq:H8app}. We thank the authors of \cite {Benini:2025lav} for pointing out an error in a previous version of Equation \protect \eqref {eq:H8app}.}\BibitemShut {Stop}%
\bibitem [{\citenamefont {Sage}\ and\ \citenamefont {Vega}(2012)}]{sage2012twisted}%
  \BibitemOpen
  \bibfield  {author} {\bibinfo {author} {\bibfnamefont {D.~S.}\ \bibnamefont {Sage}}\ and\ \bibinfo {author} {\bibfnamefont {M.~D.}\ \bibnamefont {Vega}},\ }\href@noop {} {\bibfield  {journal} {\bibinfo  {journal} {Journal of Algebra}\ }\textbf {\bibinfo {volume} {354}},\ \bibinfo {pages} {136} (\bibinfo {year} {2012})}\BibitemShut {NoStop}%
\bibitem [{\citenamefont {Liu}\ \emph {et~al.}(2023)\citenamefont {Liu}, \citenamefont {Ming}, \citenamefont {Wang},\ and\ \citenamefont {Wu}}]{Liu:2023lgl}%
  \BibitemOpen
  \bibfield  {author} {\bibinfo {author} {\bibfnamefont {Z.}~\bibnamefont {Liu}}, \bibinfo {author} {\bibfnamefont {S.}~\bibnamefont {Ming}}, \bibinfo {author} {\bibfnamefont {Y.}~\bibnamefont {Wang}}, \ and\ \bibinfo {author} {\bibfnamefont {J.}~\bibnamefont {Wu}},\ }\href@noop {} {\  (\bibinfo {year} {2023})},\ \Eprint {http://arxiv.org/abs/2307.12284} {arXiv:2307.12284 [math.QA]} \BibitemShut {NoStop}%
\bibitem [{\citenamefont {Benedetti}\ \emph {et~al.}(2024)\citenamefont {Benedetti}, \citenamefont {Casini}, \citenamefont {Kawahigashi}, \citenamefont {Longo},\ and\ \citenamefont {Mag\'{a}n}}]{Benedetti:2024dku}%
  \BibitemOpen
  \bibfield  {author} {\bibinfo {author} {\bibfnamefont {V.}~\bibnamefont {Benedetti}}, \bibinfo {author} {\bibfnamefont {H.}~\bibnamefont {Casini}}, \bibinfo {author} {\bibfnamefont {Y.}~\bibnamefont {Kawahigashi}}, \bibinfo {author} {\bibfnamefont {R.}~\bibnamefont {Longo}}, \ and\ \bibinfo {author} {\bibfnamefont {J.~M.}\ \bibnamefont {Mag\'{a}n}},\ }\href@noop {} {\  (\bibinfo {year} {2024})},\ \Eprint {http://arxiv.org/abs/2408.04011} {arXiv:2408.04011 [hep-th]} \BibitemShut {NoStop}%
\bibitem [{\citenamefont {Cordova}\ and\ \citenamefont {Garc\'\i{}a-Sep\'ulveda}(2022)}]{Cordova:2022lms}%
  \BibitemOpen
  \bibfield  {author} {\bibinfo {author} {\bibfnamefont {C.}~\bibnamefont {Cordova}}\ and\ \bibinfo {author} {\bibfnamefont {D.}~\bibnamefont {Garc\'\i{}a-Sep\'ulveda}},\ }\href@noop {} {\  (\bibinfo {year} {2022})},\ \Eprint {http://arxiv.org/abs/2210.01135} {arXiv:2210.01135 [hep-th]} \BibitemShut {NoStop}%
\bibitem [{\citenamefont {Benini}\ \emph {et~al.}(2025)\citenamefont {Benini}, \citenamefont {Calabrese}, \citenamefont {Fossati}, \citenamefont {Singh},\ and\ \citenamefont {Venuti}}]{Benini:2025lav}%
  \BibitemOpen
  \bibfield  {author} {\bibinfo {author} {\bibfnamefont {F.}~\bibnamefont {Benini}}, \bibinfo {author} {\bibfnamefont {P.}~\bibnamefont {Calabrese}}, \bibinfo {author} {\bibfnamefont {M.}~\bibnamefont {Fossati}}, \bibinfo {author} {\bibfnamefont {A.~H.}\ \bibnamefont {Singh}}, \ and\ \bibinfo {author} {\bibfnamefont {M.}~\bibnamefont {Venuti}},\ }\href@noop {} {\  (\bibinfo {year} {2025})},\ \Eprint {http://arxiv.org/abs/2509.16311} {arXiv:2509.16311 [hep-th]} \BibitemShut {NoStop}%
\end{thebibliography}%

\end{document}